\documentclass{statsoc}
\usepackage[utf8]{inputenc}
\usepackage{textcomp}

\usepackage{amsthm}
\usepackage{amsmath}
\usepackage{amssymb}
\usepackage{courier}
\usepackage{enumerate}
\usepackage{float}
\usepackage{graphicx}
\usepackage{comment}
\usepackage[authoryear]{natbib}
\makeatletter
\theoremstyle{remark}
\newtheorem{remark}{Remark}
\newtheorem{thm}{Theorem}
\newtheorem*{thm1}{Theorem 1}
\newtheorem*{thm2}{Theorem 2}
\newtheorem*{thmA.1}{Theorem A.1}
\newtheorem{definition}{Definition}
\newtheorem{example}{Example}
\usepackage[a4paper,left=2.5cm,right=2.5cm,top=2.5cm,bottom=2.5cm]{geometry}
\usepackage[title]{appendix}

\title[Visualizing Supervised Learning Models]{Visualizing the Effects of Predictor Variables in Black Box Supervised Learning Models}
\author[D. Apley and J. Zhu]{Daniel W. Apley\footnote{Address for correspondence: Daniel W. Apley, Department of Industrial Engineering \& Management Sciences, Northwestern University, Evanston, IL 60208, USA.\\ E-mail: apley@northwestern.edu} and Jingyu Zhu}
\address{Northwestern University, USA}
\begin{document}
\maketitle
\begin{abstract}
In many supervised learning applications, understanding and visualizing the effects of the predictor variables on the predicted response is of paramount importance. A shortcoming of black box supervised learning models (e.g., complex trees, neural networks, boosted trees, random forests, nearest neighbors, local kernel-weighted methods, support vector regression, etc.) in this regard is their lack of interpretability or transparency. Partial dependence (PD) plots, which are the most popular approach for visualizing the effects of the predictors with black box supervised learning models, can produce erroneous results if the predictors are strongly correlated, because they require extrapolation of the response at predictor values that are far outside the multivariate envelope of the training data. As an alternative to PD plots, we present a new visualization approach that we term accumulated local effects (ALE) plots, which do not require this unreliable extrapolation with correlated predictors. Moreover, ALE plots are far less computationally expensive than PD plots. 
\end{abstract}
\section{Introduction}\label{introduction}
\par 
With the proliferation of larger and richer data sets in many predictive modeling application domains, black box supervised learning models (e.g., complex trees, neural networks, boosted trees, random forests, nearest neighbors, local kernel-weighted methods, support vector regression, etc.) are increasingly commonly used in place of more transparent linear and logistic regression models to capture nonlinear phenomena. However, one shortcoming of black box supervised learning models is that they are difficult to interpret in terms of understanding the effects of the predictor variables (aka predictors) on the predicted response. For many applications, understanding the effects of the predictors is critically important. This is obviously the case if the purpose of the predictive modeling is explanatory, such as identifying new disease risk factors from electronic medical record databases. Even if the purpose is purely predictive, understanding the effects of the predictors may still be quite important. If the effect of a predictor violates intuition (e.g., if it appears from the supervised learning model that the risk of experiencing a cardiac event \emph{decreases} as patients age), then this is either an indication that the fitted model is unreliable or that a surprising new phenomenon has been discovered. In addition, predictive models must be transparent in many regulatory environments, e.g., to demonstrate to regulators that consumer credit risk models do not penalize credit applicants based on age, race, etc.  
\par 
To be more concrete, suppose we have fit a supervised learning model for approximating $\mathbb{E}[Y|\mathbf{X}=\mathbf{x}]\approx f(\mathbf{x})$, where $Y$ is a scalar response variable, $\mathbf{X}=(X_1,X_2,\ldots,X_d )$ is a vector of $d$ predictors, and $f(\cdot)$ is the fitted model that predicts $Y$ (or the probability that $Y$ falls into a particular class, in the classification setting) as a function of $\mathbf{X}$. To simplify notation, we omit any $ \quad\hat{}\quad$ symbol over $f$, with the understanding that it is fitted from data. The training data to which the model is fit consists of $n$ $(d+1)$-variate observations $\{y_i,\mathbf{x}_i=(x_{i,1},x_{i,2},\ldots,x_{i,d}):  i=1,2,\ldots,n\}$. Throughout, we use upper case to denote a random variable and lower case to denote specific or observed values of the random variable. 
\par 
Our objective is to visualize and understand the ``main effects'' dependence of $f(\mathbf{x})=f(x_1,x_2,\ldots,x_d)$ on each of the individual predictors, as well as the low-order ``interaction effects'' among pairs of predictors. Throughout the introduction we illustrate concepts for the simple $d=2$ case. The most popular approach for visualizing the effects of the predictors is partial dependence (PD) plots, introduced by \cite{Friedman}. To understand the effect of one predictor (say $X_1$) on the predicted response, a PD plot is a plot of the function
\begin{equation}\label{PD def} f_{1,PD}(x_1)\equiv \mathbb{E}[f(x_1,X_2 )]=\int{p_2(x_2)f(x_1,x_2)dx_2}\end{equation} versus $x_1$, where $p_2 (\cdot)$ denotes the marginal distribution of $X_2$. We use $p(\cdot)$ to denote the full joint probability density of $\mathbf{X}$, and use $p_{\cdot} (\cdot)$, $p_{\cdot|\cdot}(\cdot|\cdot)$, and $p_{\cdot,\cdot}(\cdot,\cdot)$ to respectively denote the marginal, conditional, and joint probability density functions of various elements of $\mathbf{X}$, with the subscripts indicating which elements. An estimate of \eqref{PD def}, calculated pointwise in $x_1$ for a range of $x_1$ values, is \begin{equation}\label{PD estimate}\hat{f}_{1,PD} (x_1)\equiv \frac{1}{n} \sum_{i=1}^{n}f(x_1,x_{i,2}).\end{equation}
\par
\begin{figure}
    \centering
    \includegraphics[width=\textwidth]{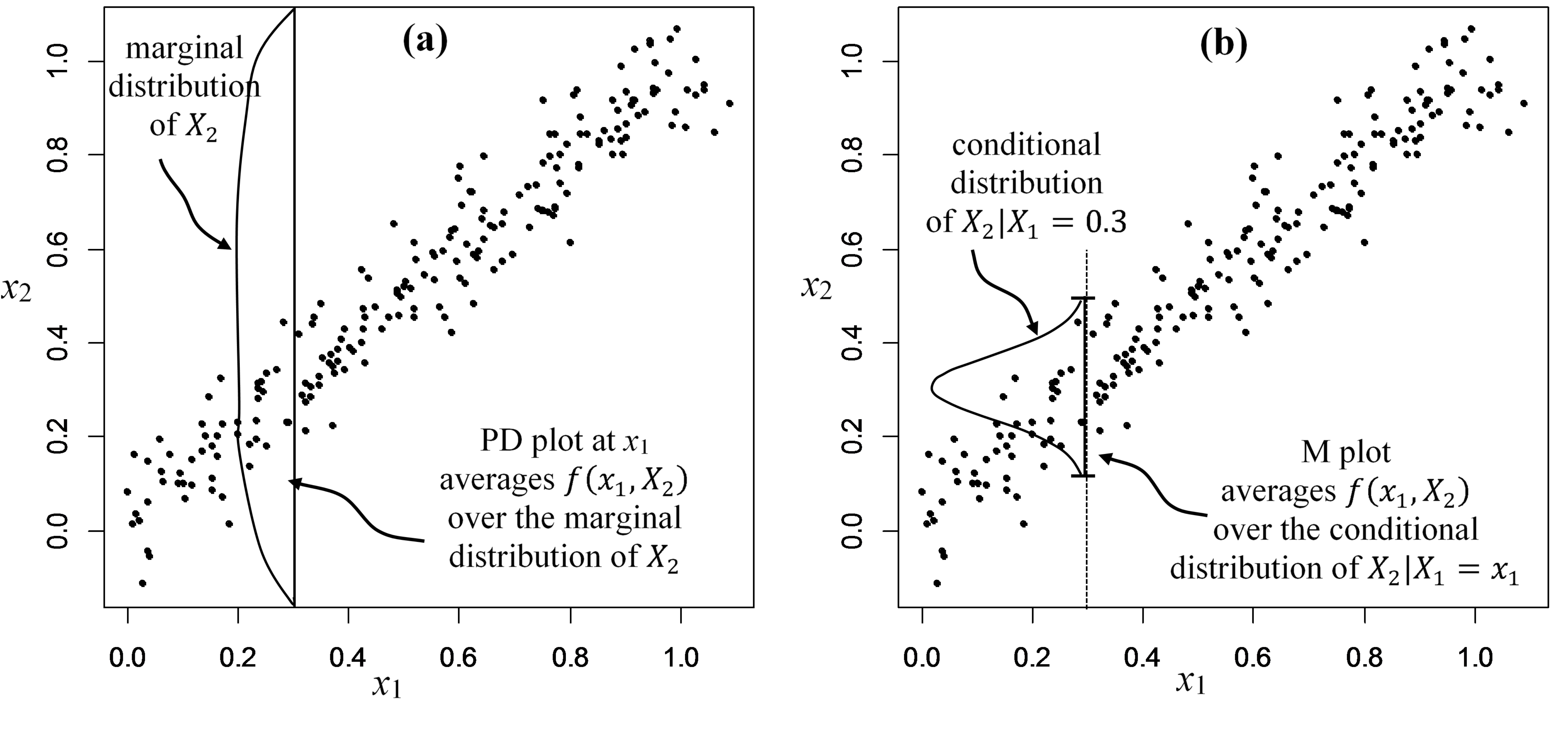}
    \caption{Illustration of the differences between the computation of (a) $f_{1,PD}(x_1)$ and (b) $f_{1,M}(x_1)$ at $x_1=0.3$. }
    \label{PD and M difference}
\end{figure}

 
Figure \ref{PD and M difference}(a) illustrates how $f_{1,PD}(x_1)$ is computed at a specific value $x_1=0.3$ for a toy example with $n=200$ observations of $(X_1,X_2)$ following a uniform distribution along the line segment $x_2=x_1$ but with independent $N(0,0.05^2)$ variables added to both predictors (see \cite{Hooker}, for a similar example demonstrating the adverse consequences of extrapolation in PD plots). Although we ignore the response variable for now, we return to this example in Section \ref{toy examples ale vs pd} and fit various models $f(\mathbf{x})$ to these data. The salient point in Figure \ref{PD and M difference}(a), which illustrates the problem with PD plots, is that the integral in \eqref{PD def} is the weighted average of $f(x_1,X_2)$ as $X_2$ varies over its marginal distribution. This integral is over the entire vertical line segment in Figure \ref{PD and M difference}(a) and requires rather severe extrapolation beyond the envelope of the training data. If one were to fit a simple parametric model of the correct form (e.g., $f(\mathbf{x})=\beta_0+\beta_1x_1+\beta_2 x_2^2$), then this extrapolation might be reliable. However, by nature of its flexibility, a nonparametric supervised learning model like a regression tree cannot be expected to extrapolate reliably. As we demonstrate later (see Figures \ref{1st8split}—\ref{50reps}), this renders the PD plot an unreliable indicator of the effect of $x_1$.
 
\par 
The extrapolation in Figure \ref{PD and M difference}(a) that is required to calculate $f_{1,PD}(x_1)$ occurs because the marginal density $p_2(x_2)$ is much less concentrated around the data than the conditional density $p_{2|1} (x_2|x_1)$, due to the strong dependence between $X_2$ and $X_1$. Marginal plots (M plots) are alternatives to PD plots that avoid such extrapolation by using the conditional density in place of the marginal density. As illustrated in Figure \ref{PD and M difference}(b), an M plot of the effect of $X_1$ is a plot of the function \begin{equation}\label{M def}f_{1,M} (x_1)\equiv \mathbb{E}[f(X_1,X_2)|X_1=x_1]=\int{p_{2|1}(x_2|x_1)f(x_1,x_2)dx_2}\end{equation} versus $x_1$. A crude estimate of $f_{1,M}(x_1)$ is \begin{equation}\label{M estimate}\hat{f}_{1,M}(x_1)\equiv \frac{1}{n(x_1)}\sum_{i\in N(x_1)}{f(x_1,x_{i,2})}, \end{equation} where $N(x_1)\subset\{1,2,\ldots,n\}$ is the subset of row indices $i$ for which $x_{i,1}$ falls into some small, appropriately selected neighborhood of $x_1$, and $n(x_1)$ is the number of observations in the neighborhood. Although more sophisticated kernel smoothing methods are typically used to estimate $f_{1,M}(x_1)$, we do not consider them here, because there is a more serious problem with using $f_{1,M}(x_1)$ to visualize the main effect of $X_1$ when $X_1$ and $X_2$ are dependent. Namely, using $f_{1,M}(x_1)$ is like regressing $Y$ onto $X_1$ while ignoring (i.e., marginalizing\footnote{Regarding the terminology, plots of an estimate of $f_{1,M}(x_1)$ are often referred to as ``marginal plots", because ignoring $X_2$ in this manner is equivalent to working with the joint distribution of $(Y,X_1)$ after marginalizing across $X_2$. Unfortunately, plots of $\hat{f}_{1,PD}(x_1)$ are also sometimes referred to as ``marginal plots" (e.g., in the \textbf{\texttt{gbm}} package for fitting boosted trees in \texttt{R}), presumably because the integral in \eqref{PD def} is with respect to the marginal distribution $p_2(x_2)$. In this paper, marginal plots will refer to how we have defined them above.}  over) the nuisance variable $X_2$. Consequently, if $Y$ depends on $X_1$ and $X_2$, $f_{1,M}(x_1)$ will reflect both of their effects, a consequence of the omitted variable bias phenomenon in regression. 
\par 
The main objective of this paper is to introduce a new method of assessing the main and interaction effects of the predictors in black box supervised learning models that avoids the foregoing problems with PD plots and M plots. We refer to the approach as accumulated local effects (ALE) plots. For the case that $d=2$ and $f(\cdot)$ is differentiable (the more general definition is deferred until Section \ref{def of main and second}), we define the ALE main-effect of $X_1$ as \begin{equation}\label{ALE 1 main def} 
\begin{split} f_{1,ALE}(x_1) &\equiv \int_{x_{\min,1}}^{x_1}\mathbb{E}[f^1(X_1,X_2)|X_1=z_1]dz_1-\text{constant} \\
			&=\int_{x_{\min,1}}^{x_1}\int{p_{2|1}(x_2|z_1) f^1 (z_1,x_2)dx_2} dz_1-\text{constant},\end{split}\end{equation}
where $f^1(x_1,x_2 )\equiv \frac{\partial{f(x_1,x_2)}}{\partial{x_1}}$ represents the \emph{local effect} of $x_1$ on $f(\cdot)$ at $(x_1,x_2)$, and $x_{\min,1}$ is some value chosen near the lower bound of the effective support of $p_1(\cdot)$, e.g., just below the smallest observation $\min\{x_{i,1}: i=1,2,\ldots,n\}$. Choice of $x_{\min,1}$ is not important, as it only affects the vertical translation of the ALE plot of $f_{1,ALE}(x_1)$ versus $x_1$, and the constant in \eqref{ALE 1 main def} will be chosen to vertically center the plot.
\par
\begin{figure}
   \centering
    \includegraphics[width=\textwidth]{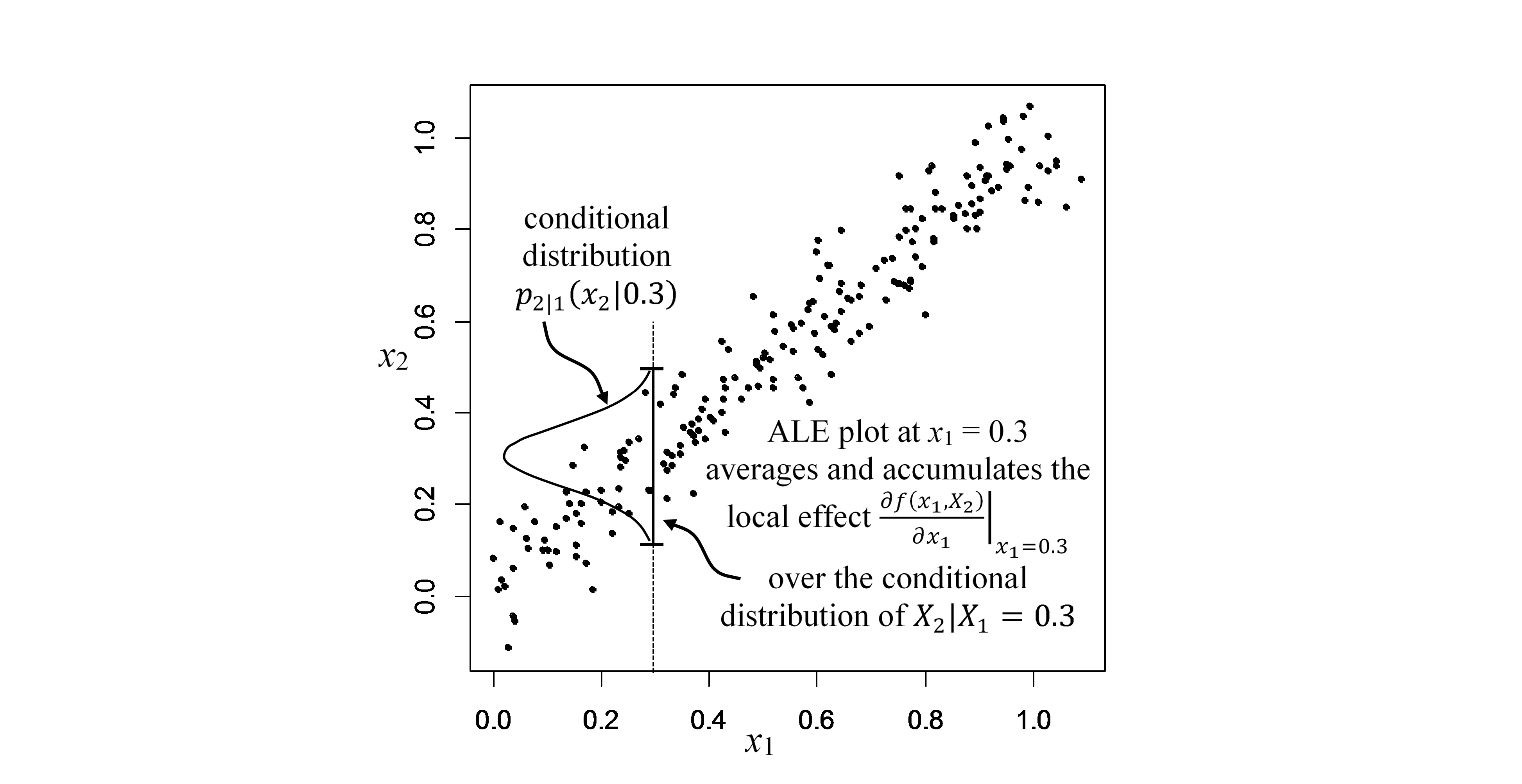}
    \caption{Illustration of the computation of $f_{1,ALE}(x_1)$ at $x_1=0.3$}
    \label{Illustration of ALE Comp}
\end{figure}

 
The function $f_{1,ALE}(x_1)$ can be interpreted as the \emph{accumulated local effects} of $x_1$ in the following sense. In \eqref{ALE 1 main def}, we calculate the local effect $f^1(x_1,x_2)$ of $x_1$ at $(x_1=z_1,x_2)$, then average this local effect across all values of $x_2$ with weight $p_{2|1}(x_2|z_1)$, and then finally accumulate/integrate this averaged local effect over all values of $z_1$ up to $x_1$. As illustrated in Figure \ref{Illustration of ALE Comp}, when averaging the local effect $f^1(x_1,x_2)$ across $x_2$, the use of the conditional density $p_{2|1}(x_2|x_1)$, instead of the marginal density $p_2(x_2)$, avoids the extrapolation required in PD plots. The avoidance of extrapolation is similar to M plots, which also use the conditional density $p_{2|1} (x_2|x_1)$. However, by averaging (across $x_2$) and accumulating (up to $x_1$) the local effects via \eqref{ALE 1 main def}, as opposed to directly averaging $f(\cdot)$ via \eqref{M def}, ALE plots avoid the omitted nuisance variable bias that renders M plots of little use for assessing the main effects of the predictors. This relates closely to the use of paired differences to block out nuisance factors in more general statistical settings, which we discuss in Section \ref{paired diff and add recovery}.
\par 
Methods also exist for visualizing the effects of predictors by plotting a collection of curves, rather than a single curve that represents some aggregate effect. Consider the effect of a single predictor $X_j$, and let $\mathbf{X}_{\backslash j}$ denote the other predictors. Conditioning plots (coplots) (\cite{Chambers}; \cite{Cleveland}), conditional response (CORE) plots (\cite{Cook}), and individual conditional expectation (ICE) plots (\cite{Goldstein}) plot quantities like $f(x_j,\mathbf{x}_{\backslash j})$ vs.\ $x_j$ for a collection of discrete values of $\mathbf{x}_{\backslash j}$ (CORE and ICE plots), or similarly they plot $\mathbb{E}[f(x_j,\mathbf{X}_{\backslash j} )|\mathbf{X}_{\backslash j} \in S_k ]$ vs.\ $x_j$ for each set $S_k$ in some partition $\{S_k: k=1,2,\ldots\}$ of the space of $\mathbf{X}_{\backslash j}$. Such a collection of curves have more in common with interaction effect plots (as in Figure \ref{bs-interaction}, later) than with main effect plots, for which one desires, by definition, a single aggregated curve.
\par
The format of the remainder of the paper is as follows. In Section \ref{def of main and second}, we define the ALE main effects for individual predictors and the ALE second-order interaction effects for pairs of predictors. In Section \ref{estimation of main and second} we present estimators of the ALE main and second-order interaction effects, which are conceptually straightforward and computationally efficient (much more efficient than PD plots), and we prove their consistency. We focus on main and second-order interaction effects and discuss general higher-order effects and their estimation in the appendices. In Section \ref{toy examples ale vs pd} we give examples that illustrate the ALE plots and, in particular, how they can produce correct results when PD plots are corrupted due to their reliance on extrapolation. In Section \ref{discussion}, we discuss interpretation of ALE plots and a number of their desirable properties and computational advantages, and we illustrate with a real data example. We also discuss their relation to functional ANOVA decompositions for dependent variables (e.g., \cite{Hooker}) that have been developed to avoid the same extrapolation problem highlighted in Figure \ref{PD and M difference}(a). ALE plots are far more computationally efficient and systematic to compute than functional ANOVA decompositions; and they yield a fundamentally different decomposition of $f(\mathbf{x})$ that is better suited for visualization of the effects of the predictors. Section 6 concludes the paper. We also provide as supplementary material an \texttt{R} package \textbf{\texttt{ALEPlot}} to implement ALE plots.
\section{Definition of ALE Main and Second-Order Effects}\label{def of main and second}
\par 
In this section we define the ALE main effect functions for each predictor (Eq. \eqref{ALE 1 main def} is a special case for $d=2$ and differentiable $f(\cdot)$) and the ALE second-order effect functions for each pair of predictors. ALE plots are plots of estimates of these functions, and the estimators are defined in Section \ref{estimation of main and second}. We do not envision ALE plots being commonly used to visualize third- and high-order effects, since high-order effects are difficult to interpret and usually not as predominant as main and second-order effects. For this reason, and to simplify notation, we focus on main and second-order effects and relegate the definition of higher-order ALE effects to the appendices.
\par 
Throughout this section, we assume that $p$ has compact support $\mathcal{S}$, and the support of $p_j$ is the interval $\mathcal{S}_j=[x_{\min,j},x_{\max,j}]$ for each $j \in \{1,2,\ldots,d\}$. For each $K=1,2,\ldots,$ and $j\in \{1, 2, \ldots, d\}$, let $\mathcal{P}_{j}^K\equiv \{z_{k,j}^K: k = 0,1,\ldots,K\}$ be a partition of $\mathcal{S}_j$ into $K$ intervals with $z_{0,j}^K=x_{\min,j}$ and $z_{K,j}^K=x_{\max,j}$. Define $\delta_{j,K}\equiv \max\{|z_{k,j}^K-z_{k-1,j}^K|: k=1,2,\ldots,K\}$, which represents the fineness of the partition. For any $x \in \mathcal{S}_j$, define $k_j^K(x)$ to be the index of the interval of $\mathcal{P}_{j}^K$ into which $x$ falls, i.e., $x\in(z_{k-1,j}^K,z_{k,j}^K]$ for $k = k_j^K(x)$. Let $\mathbf{X}_{\backslash j}$ denote the subset of $d-1$ predictors excluding $X_j$, i.e., $\mathbf{X}_{\backslash j} = (X_k:  k=1,2,\ldots,d; k\neq j)$. The following definition is a generalization of \eqref{ALE 1 main def} for a function $f(\cdot)$ that is not necessarily differentiable and for any $d \geq 2$. The generalization essentially replaces the derivative and integral in \eqref{ALE 1 main def} with limiting forms of finite differences and summations, respectively.\\
\begin{definition} [\textbf{Uncentered ALE Main Effect}] \label{def as limit}
Consider any $j\in \{1, 2, \ldots, d\}$, and suppose the sequence of partitions $\{\mathcal{P}_{j}^K:K=1,2,\ldots\}$ is such that $\lim_{K\to\infty} \delta_{j,K} = 0$.  When $f(\cdot)$ and $p$ are such that the following limit exists and is independent of the particular sequence of partitions $\{\mathcal{P}_{j}^K: K = 1, 2, \ldots\}$ (see Theorem A.1 in Appendix \ref{statement and proofs of theorems} for sufficient conditions on the existence and uniqueness of the limit), we define the uncentered ALE main (aka first-order) effect function of $X_j$ as (for $x_j \in \mathcal{S}_j$)
\begin{equation}\label{ALE uncentered main def as limit}
    g_{j,ALE}(x_j)\equiv \lim_{K\to\infty} \sum_{k=1}^{k_j^K(x_j)} \mathbb{E}[f(z_{k,j}^K,\mathbf{X}_{\backslash j})-f(z_{k-1,j}^K,\mathbf{X}_{\backslash j} )|X_j\in (z_{k-1,j}^K,z_{k,j}^K]].	       
\end{equation}
\end{definition}
\par The following theorem, the proof of which is in Appendix \ref{statement and proofs of theorems}, states that for differentiable $f(\cdot)$, the uncentered ALE main effect of $X_j$ in \eqref{ALE uncentered main def as limit} has an equivalent but more revealing form that is analogous to \eqref{ALE 1 main def}. \\
\begin{thm1} [\textbf{Uncentered ALE Main Effect for differentiable $f(\cdot)$}] Let $f^j(x_j,\mathbf{x}_{\backslash j}) \equiv \frac{\partial{f(x_j,\mathbf{x}_{\backslash j})}} {\partial{x_j}}$ denote the partial derivative of $f(\mathbf{x})$ with respect to $x_j$ when the derivative exists. In Definition \ref{def as limit}, suppose 
\begin{enumerate} [(i)]
    \item $f(x_j,\mathbf{x}_{\backslash j})$ is differentiable in $x_j$ on $\mathcal{S}$,
    \item $f^j(x_j,\mathbf{x}_{\backslash j})$ is continuous in $(x_j,\mathbf{x}_{\backslash j})$ on $\mathcal{S}$, and
    \item  $\mathbb{E}[f^j(X_j,\mathbf{X}_{\backslash j})|X_j=z_j]$ is continuous in $z_j$ on $\mathcal{S}_j$.
\end{enumerate} Then, for each $x_j \in \mathcal{S}_j$, 
\begin{equation}\label{ALE uncentered main def as integral}
    	g_{j,ALE}(x_j)=\int_{x_{\min,j}}^{x_j} \mathbb{E}[f^j(X_j,\mathbf{X}_{\backslash j} )|X_j=z_j]dz_j.	
\end{equation}
\begin{flushright}
\textbf{(End of Theorem 1)}
\end{flushright}
\end{thm1}

\par 
The (centered) ALE main effect of $X_j$, denoted by $f_{j,ALE}(x_j )$, is defined the same as $g_{j,ALE}(x_j)$ but centered so that $f_{j,ALE}(X_j)$ has a mean of zero with respect to the marginal distribution of $X_j$. That is, 
\begin{equation} \label{ALE centered main def} 
\begin{split}
	f_{j,ALE}(x_j) &\equiv g_{j,ALE}(x_j)- \mathbb{E}[g_{j,ALE}(X_j )] \\
			&=g_{j,ALE}(x_j)-\int p_j(z_j) g_{j,ALE}(z_j)dz_j.				
\end{split}\end{equation}
\begin{remark}\label{Remark: explanations additive recovery}
The ALE plot function $f_{j,ALE}(x_j)$ attempts to quantify something quite similar to the PD plot function $f_{j,PD}(x_j)$ in \eqref{PD def} and can be interpreted in the same manner. For example, ALE plots and PD plots both have a desirable \textit{additive recovery} property. That is, if $f(\mathbf{x})=\sum_{j=1}^d f_j(x_j)$ is additive, then both $f_{j,ALE}(x_j)$ and $f_{j,PD}(x_j)$ are equal to the desired true effect $f_j(x_j)$, up to an additive constant. Hence, a plot of $f_{j,ALE}(x_j)$ vs. $x_j$ correctly reveals the true effect of $X_j$ on $f$, no matter how black-box the function $f$ is. If second-order interaction effects are present in $f$, a similar additive recovery property holds for the ALE second-order interaction effects that we define next (see Section \ref{paired diff and add recovery} for a more general additive recovery property that applies to interactions of any order). In spite of the similarities in the characteristics of $f$ that they are designed to extract, the differences in $f_{j,ALE}(x_j)$ and $f_{j,PD}(x_j)$ lead to very different methods of estimation. As will be demonstrated in the later sections, the ALE plot functions are estimated in a far more computationally efficient manner that also avoids the extrapolation problem that renders PD plots unreliable with highly correlated predictors.\\
\end{remark}
\par 
We next define the ALE second-order effects. For each pair of indices $\{j,l\}\subseteq \{1,2,\ldots,d\}$, let $\mathbf{X}_{\backslash \{j,l\}}$ denote the subset of $d-2$ predictors excluding $\{X_j,X_l\}$, i.e., $\mathbf{X}_{\backslash\{j,l\}}= (X_k:  k=1,2,\ldots,d;k\neq j; k\neq l)$. For general $f(\cdot)$, the uncentered ALE second-order effect of $(X_j,X_l)$ is defined similarly to \eqref{ALE uncentered main def as limit}, except that we replace the 1-D finite-differences by 2-D second-order finite differences on the 2-D grid that is the Cartesian product of the 1-D partitions of $\mathcal{S}_j$ and $\mathcal{S}_l$, and the summation is over this 2-D grid. \\
\begin{definition} [\textbf{Uncentered ALE Second-Order Effect}] \label{def 2nd as limit}
Consider any pair $\{j,l\}\subseteq \{1, \ldots, d\}$ and corresponding sequences of partitions $\{\mathcal{P}_{j}^K:K=1,2,\ldots\}$ and $\{\mathcal{P}_{l}^K:K=1,2,\ldots\}$ such that $\lim_{K\to\infty} \delta_{j,K} = \lim_{K\to\infty} \delta_{l,K} = 0$. When $f(\cdot)$ and $p$ are such that the following limit exists and is independent of the particular sequences of partitions, we define the uncentered ALE second-order effect function of $(X_j, X_l)$ as (for $(x_j, x_l) \in \mathcal{S}_j \times \mathcal{S}_l$)
\begin{equation}\label{ALE uncentered second def as limit}
    h_{\{j,l\},ALE}(x_j, x_l)\equiv \lim_{K\to\infty} \sum_{k=1}^{k_j^K(x_j)}\sum_{m=1}^{k_l^K(x_l)} \mathbb{E}[\Delta_f^{\{j,l\}}(K, k, m;\mathbf{X}_{\backslash\{j,l\}})|X_j\in (z_{k-1,j}^K,z_{k,j}^K], X_l\in (z_{m-1,l}^K,z_{m,l}^K]],
    \end{equation}
where 
\begin{equation}\label{second order finte diff}\begin{split} \Delta_f^{\{j,l\}}(K,k,m; \mathbf{x}_{\backslash\{j,l\}})=[f(z_{k,j}^K,z_{m,l}^K,\mathbf{x}_{\backslash\{j,l\}})-f(z_{k-1,j}^K,z_{m,l}^K,\mathbf{x}_{\backslash\{j,l\}})]\\
		-[f(z_{k,j}^K,z_{m-1,l}^K,\mathbf{x}_{\backslash\{j,l\}})-f(z_{k-1,j}^K,z_{m-1,l}^K,\mathbf{x}_{\backslash\{j,l\}})]\end{split}\end{equation}
is the second-order finite difference of $f(\mathbf{x})=f(x_j,x_l,\mathbf{x}_{\backslash\{j,l\}})$ with respect to $(x_j,x_l)$ across cell $(z_{k-1,j}^K,z_{k,j}^K]\times(z_{m-1,l}^K,z_{m,l}^K]$ of the 2-D grid that is the Cartesian product of $\mathcal{P}_{j}^K$ and $\mathcal{P}_{l}^K$.\\ 
\end{definition}
\par Analogous to Theorem \ref{Theorem1}, Theorem \ref{Theorem2} (proved in Appendix \ref{statement and proofs of theorems}) states that for differentiable $f(\cdot)$, the uncentered ALE second-order effect of $(X_j, X_l)$ in \eqref{ALE uncentered second def as limit} has an equivalent integral form. \\
\begin{thm2}
 [\textbf{Uncentered ALE Second-Order Effect for differentiable $f(\cdot)$}] Let\\ $f^{\{j,l\}}(x_j,x_l,\mathbf{x}_{\backslash\{j,l\}})\equiv \frac{\partial^2 f(x_j,x_l,\mathbf{x}_{\backslash\{j,l\}})}{\partial x_j \partial x_l}$ denote the second-order partial derivative of $f(\mathbf{x})$ with respect to $x_j$ and $x_l$ when the derivative exists. In Definition \ref{def 2nd as limit}, suppose 
\begin{enumerate} [(i)]
    \item $f(x_j, x_l, \mathbf{x}_{\backslash\{j,l\}})$ is differentiable in $(x_j, x_l)$ on $\mathcal{S}$,
    \item $f^{\{j,l\}}(x_j, x_l, \mathbf{x}_{\backslash \{j,l\}})$ is continuous in $(x_j, x_l, \mathbf{x}_{\backslash\{j,l\}})$ on $\mathcal{S}$, and
    \item  $\mathbb{E}[f^{\{j,l\}}(X_j,X_l,\mathbf{X}_{\backslash \{j,l\}})|X_j=z_j, X_l = z_l]$ is continuous in $(z_j, z_l)$ on $\mathcal{S}_j\times\mathcal{S}_l$.
\end{enumerate} Then, for each $(x_j, x_l) \in \mathcal{S}_j\times \mathcal{S}_l$, 
\begin{equation} \label{ALE uncentered second def as integral} 
h_{\{j,l\},ALE}(x_j,x_l)\equiv \int_{x_{\min,l}}^{x_l} \int_{x_{\min,j}}^{x_j} \mathbb{E}[f^{\{j,l\}}(X_j,X_l,\mathbf{X}_{\backslash\{j,l\}})|X_j=z_j,X_l=z_l]dz_j dz_l.
\end{equation}
\begin{flushright}
\textbf{(End of Theorem 2)}
\end{flushright}
\end{thm2}

\par 
The ALE second-order effect of $(X_j,X_l)$, denoted by $f_{\{j,l\},ALE} (x_j,x_l)$, is defined the same as $h_{\{j,l\},ALE}(x_j,x_l)$ but ``doubly-centered" so that $f_{\{j,l\},ALE}(X_j,X_l)$ has a mean of zero with respect to the marginal distribution of $(X_j,X_l)$ \emph{and} so that the ALE main effects of $X_j$ and $X_l$ on $f_{\{j,l\},ALE}(X_j,X_l)$ are both zero. The latter centering is accomplished by subtracting from $h_{\{j,l\},ALE}(x_j,x_l)$ its uncentered ALE main effects via 
\begin{equation} \label{ALE second centering 1 def} \begin{split}
g_{\{j,l\},ALE} (x_j,x_l) &\equiv h_{\{j,l\},ALE}(x_j,x_l )\\&-\lim_{K\to\infty} \sum_{k=1}^{k_j^K(x_j)} \mathbb{E}[h_{\{j,l\},ALE}(z_{k,j}^K,X_l)-h_{\{j,l\},ALE}(z_{k-1,j}^K,X_l)|X_j\in (z_{k-1,j}^K,z_{k,j}^K]]\\ &-\lim_{K\to\infty} \sum_{k=1}^{k_l^K(x_l)} \mathbb{E}[h_{\{j,l\},ALE}(X_j,z_{k,l}^K)-h_{\{j,l\},ALE}(X_j,z_{k-1,l}^K)|X_l\in (z_{k-1,l}^K,z_{k,l}^K]].
\end{split} \end{equation}
By Theorem \ref{Theorem1}, for differentiable $f$, \eqref{ALE second centering 1 def} is equivalent to 
\begin{equation} \label{ALE second centering 1 def differentiable} \begin{split}
g_{\{j,l\},ALE} (x_j,x_l) &\equiv h_{\{j,l\},ALE}(x_j,x_l )-\int_{x_{\min, j}}^{x_j} \mathbb{E}[\frac{\partial h_{\{j,l\},ALE} (X_j,X_l )}{\partial X_j}|X_j=z_j ]dz_j\\ &-\int_{x_{\min, l}}^{x_l} \mathbb{E}[\frac{\partial h_{\{j,l\},ALE} (X_j,X_l )}{\partial X_l}|X_l=z_l ]dz_l \\	
&=h_{\{j,l\},ALE} (x_j,x_l )-\int_{x_{\min, j}}^{x_j}\int p_{l|j} (z_l |z_j )\frac{\partial h_{\{j,l\},ALE}(z_j,z_l)}{\partial z_j} dz_l dz_j\\ &-\int_{x_{\min, l}}^{x_l} \int p_{j|l}(z_j |z_l ) \frac{ \partial h_{\{j,l\},ALE}(z_j,z_l)}{\partial z_l} dz_j dz_l.
\end{split} \end{equation}
The final centering is accomplished by taking
\begin{equation}\label{ALE centered second def} \begin{split}
f_{\{j,l\},ALE}(x_j,x_l) &\equiv g_{\{j,l\},ALE} (x_j,x_l )-\mathbb{E}[g_{\{j,l\},ALE}(X_j,X_l )]\\
& =g_{\{j,l\},ALE}(x_j,x_l)-\int \int p_{\{j,l\}}(z_j,z_l ) g_{\{j,l\},ALE}(z_j,z_l)dz_j dz_l.\end{split}\end{equation}
It can be verified that $f_{\{j,l\},ALE}(x_j,x_l)$ is centered in the sense that the ALE main effects of $X_j$ and $X_l$ on $f_{\{j,l\},ALE} (x_j,x_l)$ are both zero (see Appendix \ref{ale decomposition theorem and properties of L} for a formal proof of a related but more general result).
\par 
If we define the zero-order effect for any function of $\mathbf{X}$ as its expected value with respect to $p$, then we can view the ALE first-order effect of $X_j$ as being obtained by first calculating its uncentered first-order effect \eqref{ALE uncentered main def as limit}, and then for the resulting function, subtracting its zero-order effect. Likewise, the ALE second-order effect of $(X_j,X_l)$ is obtained by first calculating the uncentered second-order effect \eqref{ALE uncentered second def as limit}, then for the resulting function, subtracting both of its first-order effects of $X_j$ and of $X_l$, and then for this resulting function, subtracting its zero-order effect. The ALE higher-order effects are defined analogously in Appendix \ref{def of higher-order effects}. The uncentered higher-order effect is first calculated, and then all lower-order effects are sequentially calculated and subtracted one order at a time, until the final result has all lower-order effects that are identically zero. \\
\begin{remark}\label{Remark: Appendices B and C}
In Appendix \ref{def of higher-order effects} we define ALE higher-order effect functions $f_{J,ALE}(\mathbf{x}_J)$ for $|J|>2$, where $|J|$ denotes the cardinality of the set of predictor indices $J$. Appendix \ref{ale decomposition theorem and properties of L} shows that this leads to a functional-ANOVA-like decomposition of $f$ via  \[f(\mathbf{x})=\sum_{j=1}^d f_{j,ALE}(x_j)+\sum_{j=1}^d \sum_{l=j+1}^d f_{\{j,l\},ALE}(x_j,x_l)+\sum_{J\subseteq \{1,2,\ldots,d\},|J|\geq3} f_{J,ALE}(\mathbf{x}_J).\]This ALE decomposition has a certain orthogonality-like property, which we contrast with other functional ANOVA decompositions in Section \ref{relation to functional anova}. \\
\end{remark}
\begin{remark}\label{Remark: discrete measure}
 The ALE function definitions in this section apply to predictor distributions $p_j$ that are continuous numerical with compact support. For discrete $p_j$, one could consider modifying \eqref{ALE uncentered main def as limit} and \eqref{ALE uncentered second def as limit} by using a fixed finite partition whose interval endpoints coincide with the support of $p_j$. We do not develop this, however, because our focus is on \textit{estimation} and interpretation of the ALE effects, and the estimators in the following section are meaningful for either continuous or discrete $p_j$. In the case that $X_j$ is a nominal categorical predictor, one must decide how to order its categories prior to estimating its ALE effect (which requires differencing $f$ across neighboring categories of $X_j$). In Appendix \ref{implementation details}, we discuss a strategy for this that we have found to work well in practice.
\end{remark}
\section{Estimation of $f_{j,ALE}(x_j)$ and $f_{\{j,l\},ALE}(x_j,x_l)$}\label{estimation of main and second}
\par 
In Appendix \ref{estimation of higher-order effects} we briefly describe how to estimate the ALE higher-order effect $f_{J,ALE}(\mathbf{x}_J)$ for a general subset $J\subseteq \{1,2,\ldots,d\}$ of predictor indices. Our focus in this section is on estimating the first-order $(|J|=1)$ and second-order $(|J|=2)$ effects, since these are the most common and useful (i.e., interpretable). 
\par 
As an overview, the estimate $\hat{f}_{J,ALE}$ is obtained by computing estimates of the quantities in Eqs. \eqref{ALE uncentered main def as limit}—\eqref{ALE centered second def} for $J=j$ (a single index) or for $J=\{j,l\}$ (a pair of indices). For the estimates we (\textit{i}) replace the sequence of partitions in \eqref{ALE uncentered main def as limit} (or \eqref{ALE uncentered second def as limit}) by some appropriate fixed partition of the sample range of $\{\mathbf{x}_{i,J}: i = 1, \ldots, n\}$ and (\textit{ii}) replace the conditional expectations in \eqref{ALE uncentered main def as limit} (or \eqref{ALE uncentered second def as limit}) by sample averages across $\{\mathbf{x}_{i,\backslash J}: i=1,2,\ldots,n)\}$, conditioned on $\mathbf{x}_{i,J}$ falling into the corresponding interval/cell of the partition. In the preceding, $\mathbf{x}_{i,J}= (x_{i,j}: j \in J)$ and $\mathbf{x}_{i,\backslash J}=(x_{i,j}:  j=1,2,\ldots,d; j \not\in J)$ denote the $i$th observation of the subsets of predictors $\mathbf{X}_J$ and $\mathbf{X}_{\backslash J}$, respectively. 
\begin{figure}[!ht]
    \centering
    \includegraphics[width=\textwidth]{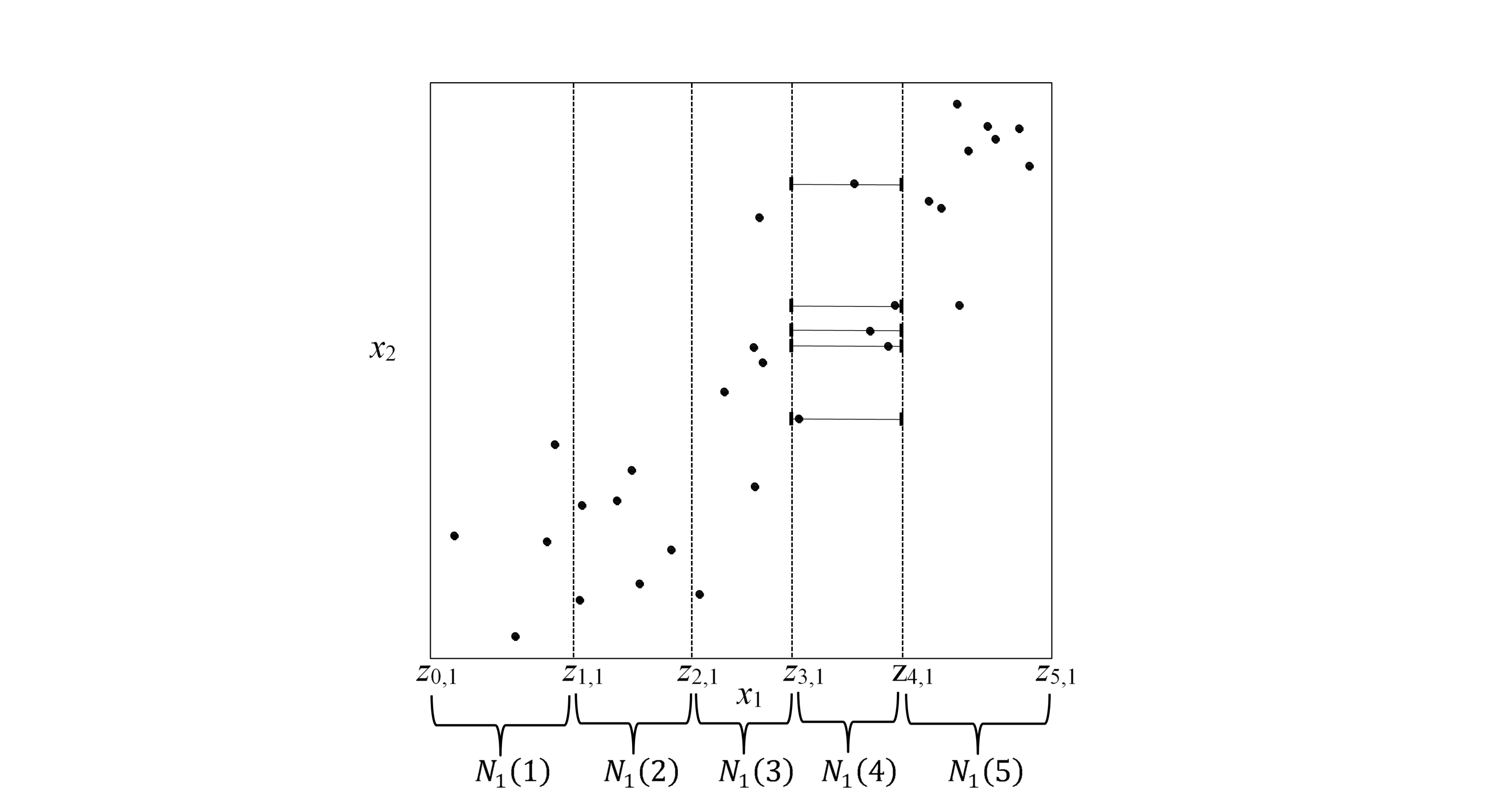}
    \caption{Illustration of the notation and concepts in computing the ALE main effect estimator $\hat{f}_{j,ALE}(x_j)$ for $j=1$ with $d=2$ predictors. The bullets are a scatterplot of $\{(x_{i,1},x_{i,2}): i=1,2,\ldots,n\}$ for $n=30$ training observations. The range of $\{x_{i,1}: i=1,2,\ldots,n\}$ is partitioned into $K=5$ intervals $\{N_1(k)=(z_{k-1,1},z_{k,1}]:k=1,2,\ldots,5\}$ (in practice, $K$ should usually be chosen much larger than $5$). The numbers of training observations falling into the $5$ intervals are $n_1(1)=4$, $n_1(2)=6$, $n_1(3)=6$, $n_1(4)=5$, and $n_1(5)=9$. The horizontal line segments shown in the $N_1(4)$ region are the segments across which the finite differences $f(z_{4,j},\mathbf{x}_{i,\backslash j} )-f(z_{3,j},\mathbf{x}_{i,\backslash j})$ are calculated and then averaged in the inner summand of Eq. \eqref{ALE uncentered main est} for $k=4$ and $j=1$.}
    \label{Illustration of Notations in ALE main est}
\end{figure}
\par 
More specifically, for each $j\in \{1,2,\ldots,d\}$, let $\{N_j (k)=(z_{k-1,j},z_{k,j}]: k=1,2,\ldots,K\}$ be a sufficiently fine partition of the sample range of $\{x_{i,j}: i=1,2,\ldots,n\}$ into $K$ intervals. Since the estimator is computed for a fixed $K$, we have omitted it as a superscript on the partition, with the understanding that the partition implicitly depends on $K$. In all of our examples later in the paper, we chose $z_{k,j}$ as the $\frac{k}{K}$ quantile of the empirical distribution of $\{x_{i,j}: i=1,2,\ldots,n\}$ with $z_{0,j}$ chosen just below the smallest observation, and $z_{K,j}$ chosen as the largest observation. Figure \ref{Illustration of Notations in ALE main est} illustrates the notation and concepts in computing the ALE main effect estimator $\hat{f}_{j,ALE}(x_j)$ for the first predictor $j=1$ for the case of $d=2$ predictors. For $k=1,2,\ldots,K$, let $n_j(k)$ denote the number of training observations that fall into the $k$th interval $N_j (k)$, so that $\sum_{k=1}^K n_j(k)=n$. For a particular value $x$ of the predictor $x_j$, let $k_j(x)$ denote the index of the interval into which $x$ falls, i.e., $x\in (z_{k_j(x)-1,j} ,z_{k_j (x),j}]$. 
\par 
For general $d$, to estimate the main effect function $f_{j,ALE}(\cdot)$ of a predictor $X_j$, we first compute an estimate of the uncentered effect $g_{j,ALE}(\cdot)$ defined in \eqref{ALE uncentered main def as limit}, which is \begin{equation}\label{ALE uncentered main est} \hat{g}_{j,ALE}(x)=\sum_{k=1}^{k_j(x)} \frac{1}{n_j(k)} \sum_{\{i:x_{i,j} \in N_j(k)\}} [f(z_{k,j},\mathbf{x}_{i,\backslash j})-f(z_{k-1,j},\mathbf{x}_{i,\backslash j})] \end{equation} for each $x \in (z_{0,j},z_{K,j}]$. Analogous to \eqref{ALE centered main def}, the ALE main effect estimator $\hat{f}_{j,ALE}(\cdot)$ is then obtained by subtracting an estimate of $\mathbb{E}[g_{j,ALE}(X_j)]$ from \eqref{ALE uncentered main est}, i.e.,
\begin{equation}\label{ALE centered main est} \begin{split}
\hat{f}_{j,ALE}(x) &= \hat{g}_{j,ALE}(x)-\frac{1}{n}\sum_{i=1}^n \hat{g}_{j,ALE}(x_{i,j})= \hat{g}_{j,ALE}(x)-\frac{1}{n}\sum_{k=1}^K n_j(k) \hat{g}_{j,ALE}(z_{k,j}).	
\end{split}\end{equation}
\par
\begin{figure}[!ht]
    \centering
    \includegraphics[width=\textwidth]{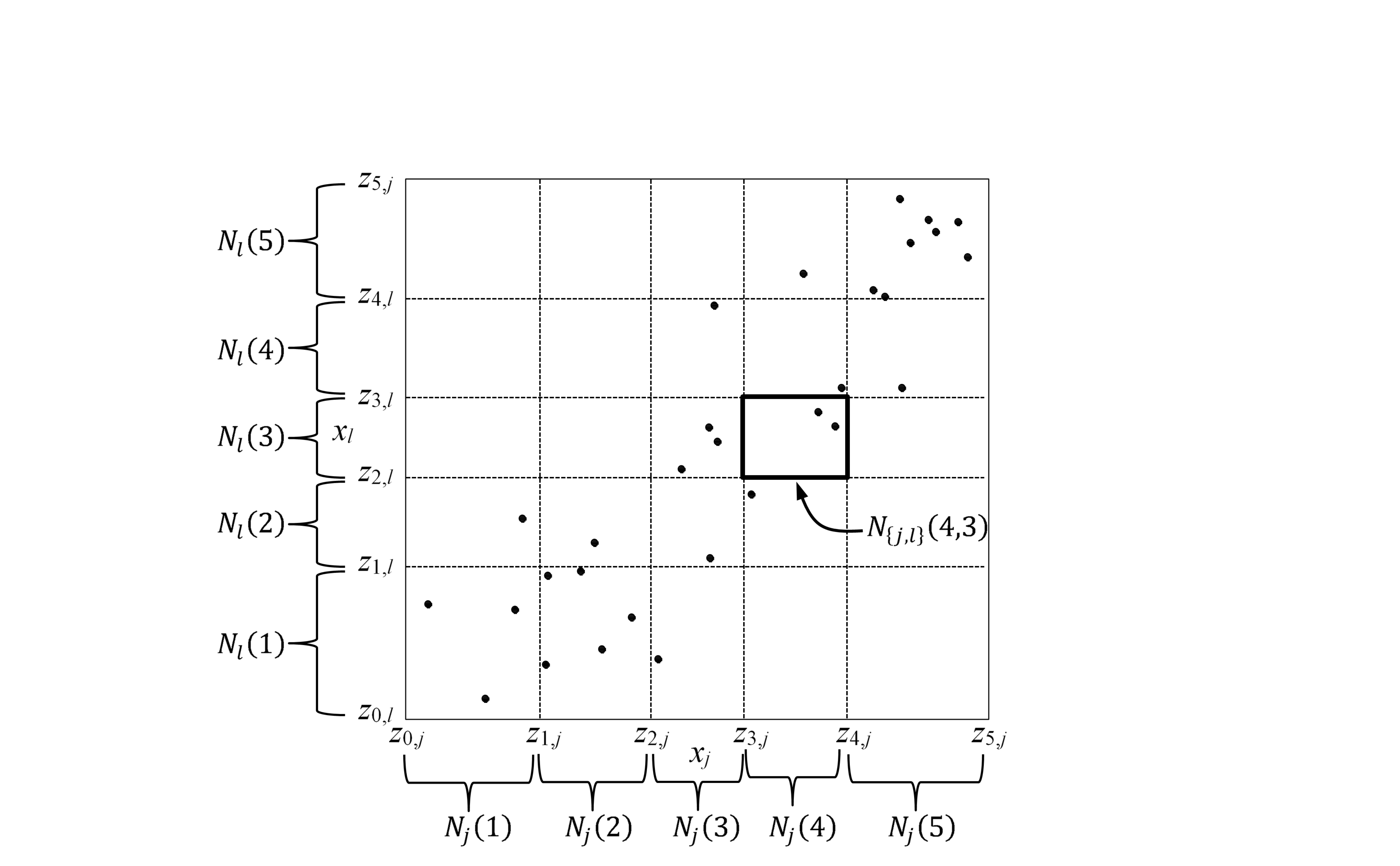}
    \caption{Illustration of the notation used in computing the ALE second-order effect estimator $\hat{f}_{\{j,l\},ALE}(x_j,x_l)$ for $K=5$. The ranges of $\{x_{i,j}: i=1,2,\ldots,n\}$ and $\{x_{i,l}: i=1,2,\ldots,n\}$ are each partitioned into $5$ intervals, and their Cartesian product forms the grid of rectangular cells $\{N_{\{j,l\}}  (k,m)=N_j(k)\times N_l(m): k=1,2,\ldots,5; m=1,2,\ldots,5\}$. The cell with bold borders is the region $N_{\{j,l\}}(4,3)$. The second-order finite differences $\Delta_f^{\{j,l\}}(K, k, m;\mathbf{x}_{i,\backslash \{j,l\}})$ in Eq. \eqref{second order finte diff in est} for $(k,m)=(4,3)$ are calculated across the corners of this cell. In the inner summation of Eq. \eqref{ALE uncentered second est}, these differences are then averaged over the $n_{\{j,l\}}(4,3)=2$ observations in region $N_{\{j,l\}}(4,3)$.}
    \label{Illustration of Notations in ALE second est}
\end{figure}

 
To estimate the ALE second-order effect of a pair of predictors $(X_j,X_l)$, we partition the sample range of $\{(x_{i,j},x_{i,l}): i = 1,2, \ldots, n\}$ into a grid of $K^2$ rectangular cells obtained as the Cartesian product of the individual one-dimensional partitions. Figure \ref{Illustration of Notations in ALE second est} illustrates the notation and concepts. Let $(k,m)$ (with $k$ and $m$ integers between $1$ and $K$) denote the indices into the grid of rectangular cells with $k$ corresponding to $x_j$ and $m$ corresponding to $x_l$. In analogy with $N_j(k)$ and $n_j(k)$ defined in the context of estimating $f_{j,ALE} (\cdot)$, let $N_{\{j,l\}}(k,m)=N_j(k)\times N_l(m)=(z_{k-1,j},z_{k,j}]\times(z_{m-1,l},z_{m,l}]$ denote the cell associated with indices $(k,m)$, and let $n_{\{j,l\}}(k,m)$ denote the number of training observations that fall into cell $N_{\{j,l\}}(k,m)$, so that $\sum_{k=1}^K \sum_{m=1}^K n_{\{j,l\}}(k,m)=n$. 
\par 
To estimate $f_{\{j,l\},ALE}(x_j,x_l)$, we first estimate the uncentered effect $h_{\{j,l\},ALE} (x_j,x_l)$ defined in \eqref{ALE uncentered second def as limit} by 
\begin{equation}\label{ALE uncentered second est} \hat{h}_{\{j,l\},ALE}(x_j,x_l )=\sum_{k=1}^{k_j(x_j)} \sum_{m=1}^{k_l (x_l)}\frac{1}{n_{\{j,l\}}(k,m)} \sum_{\{i: \mathbf{x}_{i,\{j,l\}} \in N_{\{j,l\}}(k,m)\}}\Delta_f^{\{j,l\}}(K, k, m;\mathbf{x}_{i,\backslash\{j,l\}}) \end{equation}
for each $(x_j,x_l) \in (z_{0,j},z_{K,j}]\times (z_{0,l},z_{K,l}]$. In \eqref{ALE uncentered second est},  $\Delta_f^{\{j,l\}}(K, k, m; \mathbf{x}_{i,\backslash\{j,l\}})$
is the second-order finite difference defined in \eqref{second order finte diff}, evaluated at the $i$th observation $\mathbf{x}_{i,\backslash\{j,l\}}$, i.e.
\begin{equation}\label{second order finte diff in est}\begin{split} \Delta_f^{\{j,l\}}(K,k,m; \mathbf{x}_{i, \backslash\{j,l\}})=[f(z_{k,j},z_{m,l},\mathbf{x}_{i,\backslash\{j,l\}})-f(z_{k-1,j},z_{m,l},\mathbf{x}_{i,\backslash\{j,l\}})]\\
		-[f(z_{k,j},z_{m-1,l},\mathbf{x}_{i,\backslash\{j,l\}})-f(z_{k-1,j},z_{m-1,l},\mathbf{x}_{i,\backslash\{j,l\}})]\end{split}\end{equation} Analogous to \eqref{ALE second centering 1 def}, we next compute estimates of the ALE main effects of $X_j$ and $X_l$ for the function $\hat{h}_{\{j,l\},ALE}(x_j,x_l)$ and then subtract these from $\hat{h}_{\{j,l\},ALE}(x_j,x_l)$ to give an estimate of $g_{\{j,l\},ALE}(x_j,x_l)$: \begin{equation}\label{ALE second centering 1 est} \begin{split} &\hat{g}_{\{j,l\},ALE}(x_j,x_l)\\ = &\hat{h}_{\{j,l\},ALE}(x_j,x_l )-\sum_{k=1}^{k_j(x_j)} \frac{1}{n_j(k)}\sum_{\{i: x_{i,j}\in N_j (k)\}}[\hat{h}_{\{j,l\},ALE}(z_{k,j},x_{i,l})- \hat{h}_{\{j,l\},ALE}(z_{k-1,j},x_{i,l})]\\
- &\sum_{m=1}^{k_l(x_l)}\frac{1}{n_l(m)}\sum_{\{i:x_{i,l}\in N_l (m)\}}[\hat{h}_{\{j,l\},ALE}(x_{i,j},z_{m,l})-\hat{h}_{\{j,l\},ALE}(x_{i,j},z_{m-1,l})]\\ = &\hat{h}_{\{j,l\},ALE}(x_j,x_l)-\sum_{k=1}^{k_j(x_j)}\frac{1}{n_j(k)}\sum_{m=1}^K n_{j,l}(k,m)[\hat{h}_{\{j,l\},ALE}(z_{k,j},z_{m,l})-\hat{h}_{\{j,l\},ALE}(z_{k-1,j},z_{m,l})]\\ -&\sum_{m=1}^{k_l(x_l)}\frac{1}{n_l (m)}\sum_{k=1}^K n_{\{j,l\}}(k,m)[\hat{h}_{\{j,l\},ALE}(z_{k,j},z_{m,l})-\hat{h}_{\{j,l\},ALE}(z_{k,j},z_{m-1,l})].\end{split}\end{equation}
Finally, analogous to \eqref{ALE centered second def}, we estimate $f_{\{j,l\},ALE}(x_j,x_l)$ by subtracting an estimate of\\ $\mathbb{E}[\hat{g}_{\{j,l\},ALE} (X_j,X_l)]$ from \eqref{ALE second centering 1 est}, which gives \begin{equation}\label{ALE centered second est}
\begin{split} \hat{f}_{\{j,l\},ALE}(x_j,x_l) &=\hat{g}_{\{j,l\},ALE} (x_j,x_l )-\frac{1}{n}\sum_{i=1}^n \hat{g}_{\{j,l\},ALE}(x_{i,j},x_{i,l})\\ &= \hat{g}_{\{j,l\},ALE}(x_j,x_l)-\frac{1}{n}\sum_{k=1}^K\sum_{m=1}^K n_{\{j,l\}}(k,m)\hat{g}_{\{j,l\},ALE}(z_{k,j},z_{m,l}).
\end{split}\end{equation}
\par 
Theorems \ref{Theorem3} and \ref{Theorem4} in Appendix \ref{statement and proofs of theorems} show that, under mild conditions, \eqref{ALE centered main est} and \eqref{ALE centered second est} are consistent estimators of the ALE main effect \eqref{ALE centered main def} of $X_j$ and ALE second-order effect \eqref{ALE centered second def} of $(X_j, X_l)$, respectively. 
\par 
ALE plots are plots of the ALE effect estimates $\hat{f}_{j,ALE}(x_j)$ and $\hat{f}_{\{j,l\},ALE}(x_j,x_l)$ versus the predictors involved. ALE plots have substantial computational advantages over PD plot, which we discuss in Section \ref{comp advantages}. In addition, ALE plots can produce reliable estimates of the main and interaction effects in situations where PD plots break down, which we illustrate with examples in the next section, as well as an example on real data in Section \ref{illustration with bike sharing}. 
\section{Toy Examples Illustrating when ALE Plots are Reliable but PD Plots Break Down}\label{toy examples ale vs pd}
\begin{example}\label{toy example: noiseless}
\begin{figure}[!ht]
    \centering
    \includegraphics[width=\textwidth]{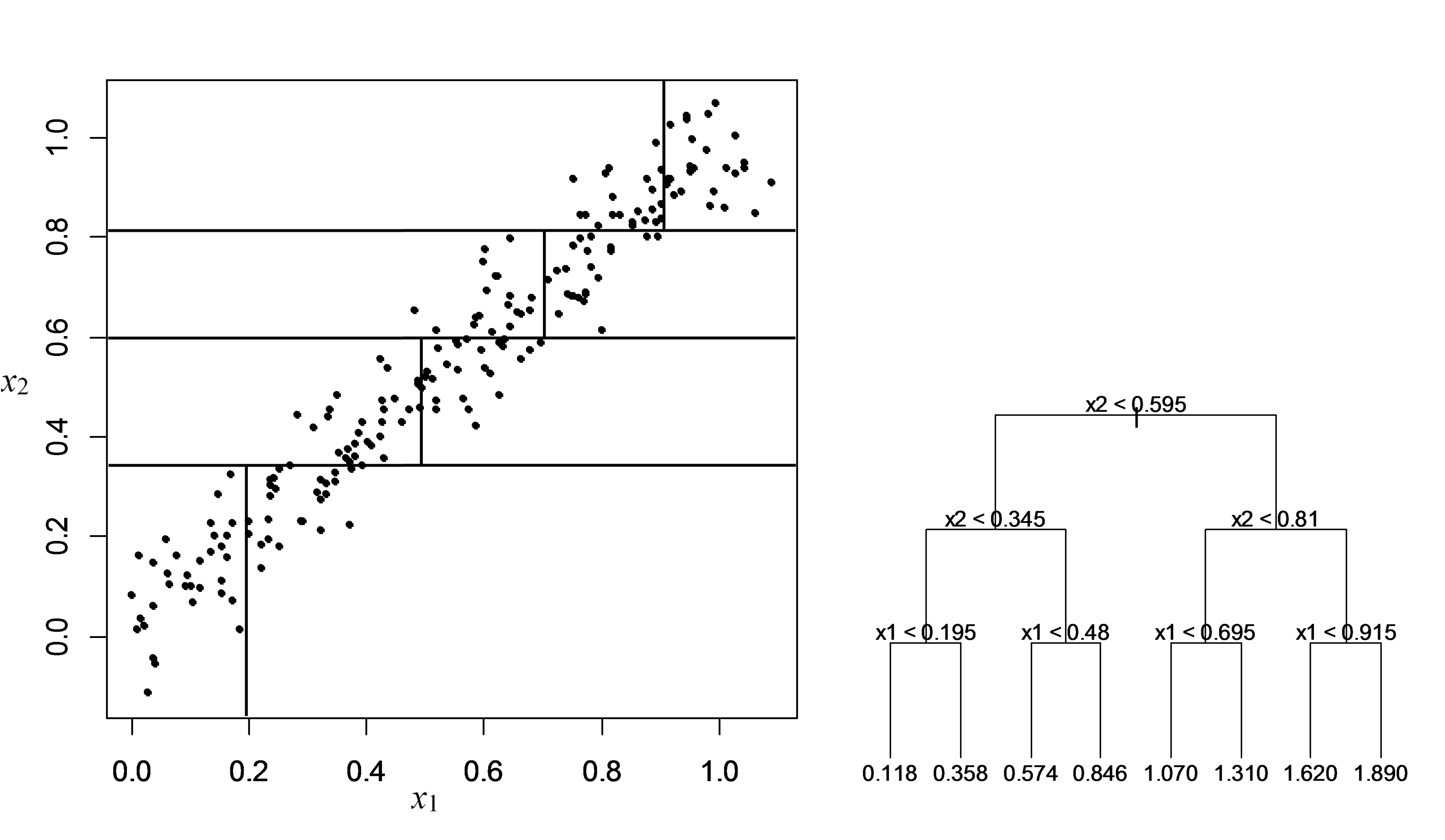}
    \caption{Depiction of the first eight splits in the tree fitted to the Example 1 data. The left panel is a scatterplot of $x_2$ vs.\ $x_1$ showing splits corresponding to the truncated tree in the right panel.}
    \label{1st8split}
\end{figure}
This example was introduced in Section \ref{introduction}. For this example, $d=2$, $n=200$, and $(X_1,X_2)$ follows a uniform distribution along a segment of the line $x_2=x_1$ with independent $N(0,0.05^2)$ variables added to both predictors. Figure \ref{1st8split} shows a scatter plot of $X_2$ vs.\ $X_1$. The true response was generated according to the noiseless model $Y=X_1+X_2^2$ for the $200$ training observations in Figure \ref{1st8split}, to which we fit a tree using the \textbf{tree} package of \texttt{R} (\cite{Ripley}). The tree was overgrown and then pruned back to have $100$ leaf nodes, which was approximately the optimal number of leaf nodes according to a cross-validation error sum of squares criterion. Notice that the optimal size tree is relatively large, because the response here is a deterministic function $X_1+X_2^2$ of the predictors with no response observation error. The first eight splits of the fitted tree $f(\mathbf{x})$ are also depicted in Figure \ref{1st8split}. Figure \ref{ALEandPDtree} shows main effect PD plots, M plots, and ALE plots for the full $100$-node fitted tree $f(\mathbf{x})$, calculated via \eqref{PD estimate}, \eqref{M estimate}, and \eqref{ALE uncentered main est}—\eqref{ALE centered main est}, respectively. For both $j=1$ and $j=2$, $\hat{f}_{j,ALE}(x_j)$ is much more accurate than either $\hat{f}_{j,PD}(x_j)$ or $\hat{f}_{j,M}(x_j)$. By inspection of the fitted tree in Figure \ref{1st8split}, it is clear why $\hat{f}_{j,PD}(x_j)$ performs so poorly in this example. For small $x_1$ values like $x_1\approx 0.2$, the PD plot estimate $\hat{f}_{1,PD}(x_1\approx 0.2)$ is much higher than it should be, because it is based on the extrapolated values of $f(\mathbf{x})$ in the upper-left corner of the scatter plot in Figure \ref{1st8split}, which were substantially overestimated due to the nature of the tree splits and the absence of any data in that region. For similar reasons, $\hat{f}_{2,PD}(x_2)$ for small $x_2$ values is substantially underestimated because of the extrapolation in the lower-right corner of the scatter plot in Figure \ref{1st8split}. In contrast, by avoiding this extrapolation, $\hat{f}_{1,ALE}(x_1)$ and $\hat{f}_{2,ALE}(x_2)$ are estimated quite accurately and are quite close to the true linear (for $x_1$) and quadratic (for $x_2$) effects, as seen in Figures \ref{ALEandPDtree}(a) and  \ref{ALEandPDtree}(b), respectively. 
\begin{figure}
    \centering
    \includegraphics[width=\textwidth]{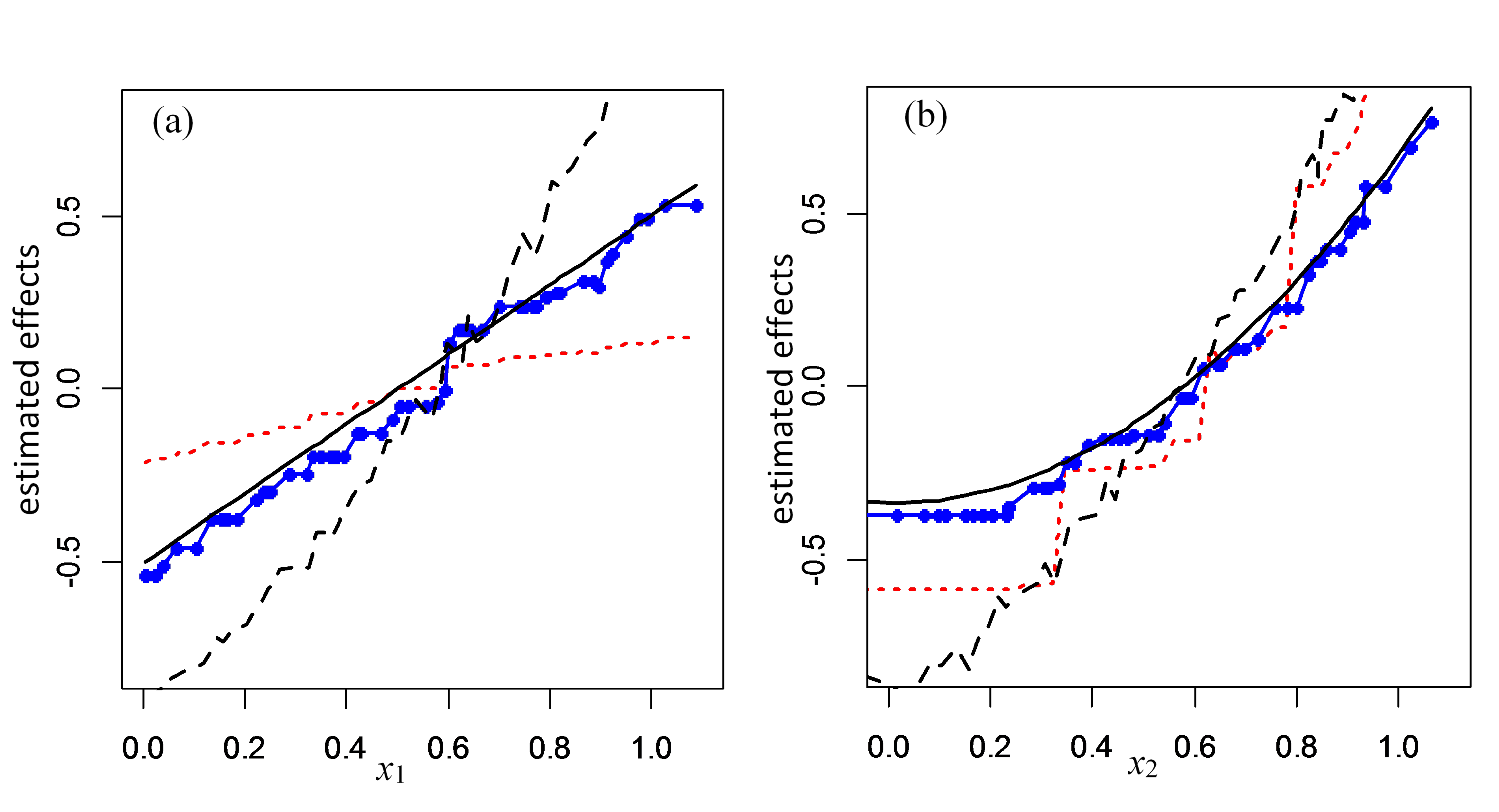}
    \caption{For the tree model fitted to the Example 1 data, plots of $\hat{f}_{j,ALE}(x_j)$ (blue line with bullets), $\hat{f}_{j,PD}(x_j)$ (red dotted line), $\hat{f}_{j,M}(x_j)$ (black dashed line), and the true main effect of $X_j$ (black solid line) for (a) $j=1$, for which the true effect of $X_1$ is linear, and (b) $j=2$, for which the true effect of $X_2$ is quadratic. For both $j=1$ and $j=2$, $\hat{f}_{j,ALE}(x_j)$ is much more accurate than either $\hat{f}_{j,PD}(x_j)$ or $\hat{f}_{j,M}(x_j)$.}
    \label{ALEandPDtree}
\end{figure}
\par 
Also notice that the M plots in Figures  \ref{ALEandPDtree}(a) and  \ref{ALEandPDtree}(b) perform very poorly. As expected, because of the strong correlation between $X_1$ and $X_2$, $\hat{f}_{1,M}(x_1)$ and $\hat{f}_{2,M}(x_2)$ are quite close to each other and are each combinations of the true effects of $X_1$ and $X_2$. In the subsequent examples, we do not further consider M plots.\\
\end{example}
\begin{example}\label{toy example: with noise and 50 reps}
This example is a modification of Example 1 having the same $d=2$, $n=200$, and $(X_1,X_2)$ following a uniform distribution along a segment of the line $x_2=x_1$ with independent $N(0,0.05^2)$ variables added to both predictors. However, the true response is now generated as noisy observations according to the model $Y=X_1+X_2^2+\varepsilon$ with $\varepsilon\sim N(0,0.1^2)$, and we fit a neural network model instead of a tree. For the neural network, we used the \textbf{nnet} package of \texttt{R} (\cite{Venables}) with ten nodes in the single hidden layer, a linear output activation function, and a decay/regularization parameter of $0.0001$, all of which were chosen as approximately optimal via multiple replicates of $10$-fold cross-validation (the cross-validation $r^2$ for this model varied between $0.965$ and $0.975$, depending on the data set generated, which is quite close to the theoretical $r^2$ value of $1-\frac{\mathrm{Var}(\varepsilon)}{\mathrm{Var}(Y)} =0.972$). We repeated the procedure in a Monte Carlo simulation with $50$ replicates, where on each replicate we generated a new training data set of $200$ observations and refit the neural network model with the same tuning parameters mentioned above. The estimated main effect functions $\hat{f}_{j,ALE}(x_j)$ and $\hat{f}_{j,PD}(x_j)$ (for $j = 1, 2$) over all $50$ replicates are shown in Figure \ref{50reps}. For this example too, $\hat{f}_{j,ALE}(x_j)$ is far superior to $\hat{f}_{j,PD}(x_j)$. On every replicate, $\hat{f}_{1,ALE}(x_1)$ and $\hat{f}_{2,ALE}(x_2)$ are quite close to the true linear and quadratic effects, respectively. In sharp contrast, $\hat{f}_{1,PD}(x_1)$ and $\hat{f}_{2,PD}(x_2)$ are so inaccurate on many replicates that they are of little use in understanding the true effects of $X_1$ and $X_2$. 
\begin{figure}
    \centering
    \includegraphics[width=\textwidth]{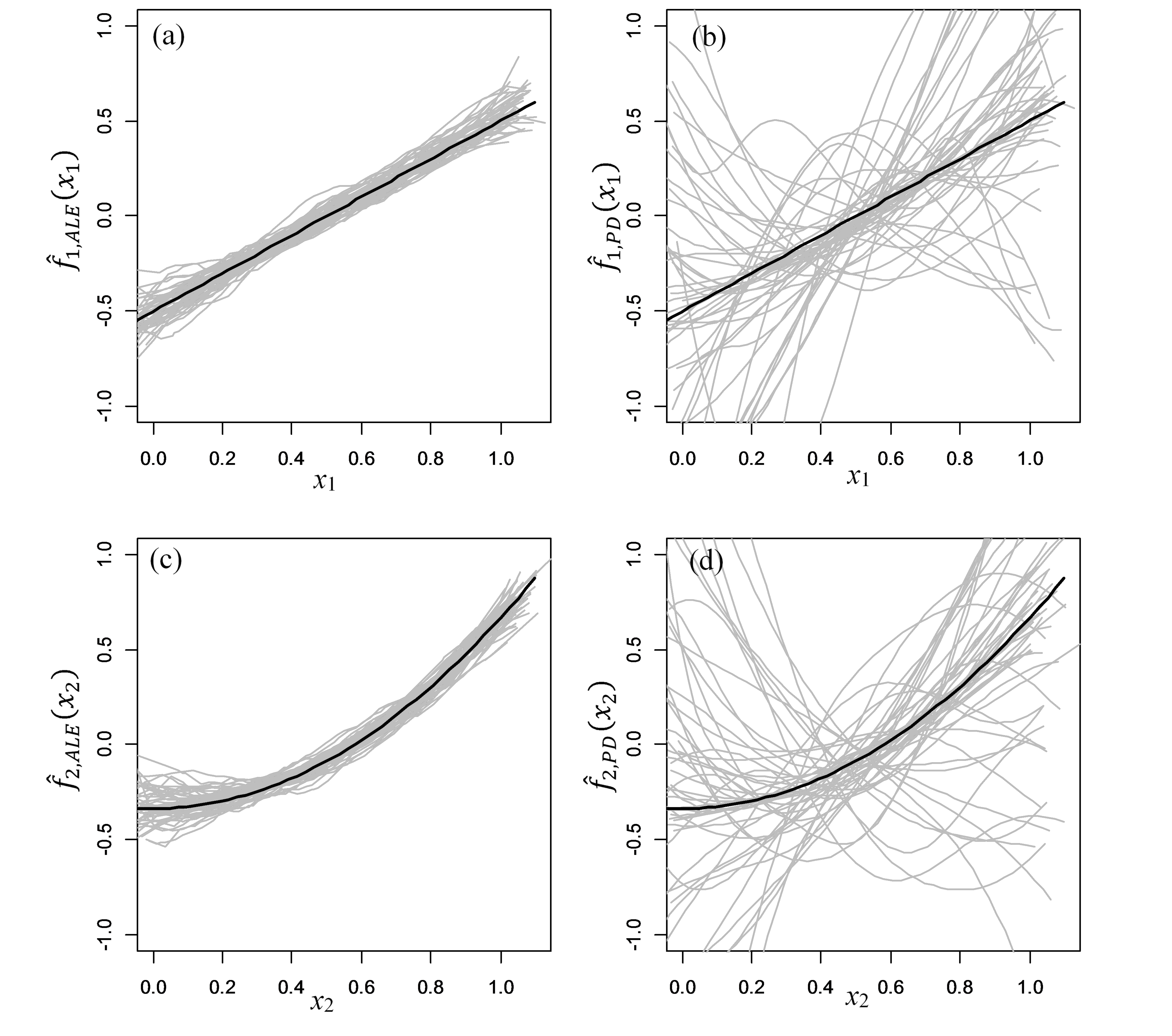}
    \caption{Comparison of (a) $\hat{f}_{1,ALE}(x_1)$, (b) $\hat{f}_{1,PD}(x_1)$, (c) $\hat{f}_{2,ALE}(x_2)$, and (d) $\hat{f}_{2,PD}(x_2)$ for neural network models fitted over $50$ Monte Carlo replicates of the Example 2 data. In each panel, the black curve is the true effect function (linear for $X_1$ and quadratic for $X_2$), and the gray curves are the estimated effect functions over the $50$ Monte Carlo replicates.}
    \label{50reps}
\end{figure}
\end{example}
\section{Discussion}\label{discussion}

\subsection{Illustration with a Bike-Sharing Real Data Example}\label{illustration with bike sharing}
\par 
We now show an example with a real, larger data set. The data are a compilation of the bike-sharing rental counts from the Capital Bikeshare system (Washington D.C., USA) over the two-year period 2011-2012, aggregated on an hourly basis, together with hourly weather and seasonal information over the same time period. The data file can be found at https://archive.ics.uci.edu/\\ml/datasets/Bike+Sharing+Dataset (\cite{BikeSharing}). There are $n=17393$ cases/rows in the training data set corresponding to $17393$ hours of data. The response is the total number of bike rental counts in each hour. We use the following $d=11$ predictors: year ($X_1$, categorical with $2$ categories: $0 = 2011$, $1 = 2012$), month ($X_2$, treated as numerical: $1 =$ January, $2 =$ February, $\ldots$, $12 =$ December), hour ($X_3$, treated as numerical: $\{0,1,\ldots,23\}$), holiday ($X_4$, categorical: $0 =$ non-holiday, $1 =$ holiday), weekday ($X_5$, treated as numerical: $\{0, 1, \ldots, 6\}$ representing day of a week with $0 = $ Sunday), workingday ($X_6$, categorical: $1=$ neither weekend nor holiday, $0 =$ otherwise), weather situation ($X_7$, treated as numerical: $\{1,2,3,4\}$, smaller values correspond to nicer weather situations), temp ($X_8$, numerical: temperature in Celsius), atemp ($X_9$, numerical: feeling temperature in Celsius), hum ($X_{10}$, numerical: humidity), windspeed ($X_{11}$, numerical: wind speed). We do not use date and season in the data file as predictors since they are dependent on the other predictors. Notice that the set of predictors are highly correlated. For example, temperature and feeling temperature are highly correlated, and so are month and temperature. 
\par
\begin{figure}[!ht]
    \centering
    \includegraphics[trim={1cm 0cm 0.7cm 1cm},clip,width = \textwidth]{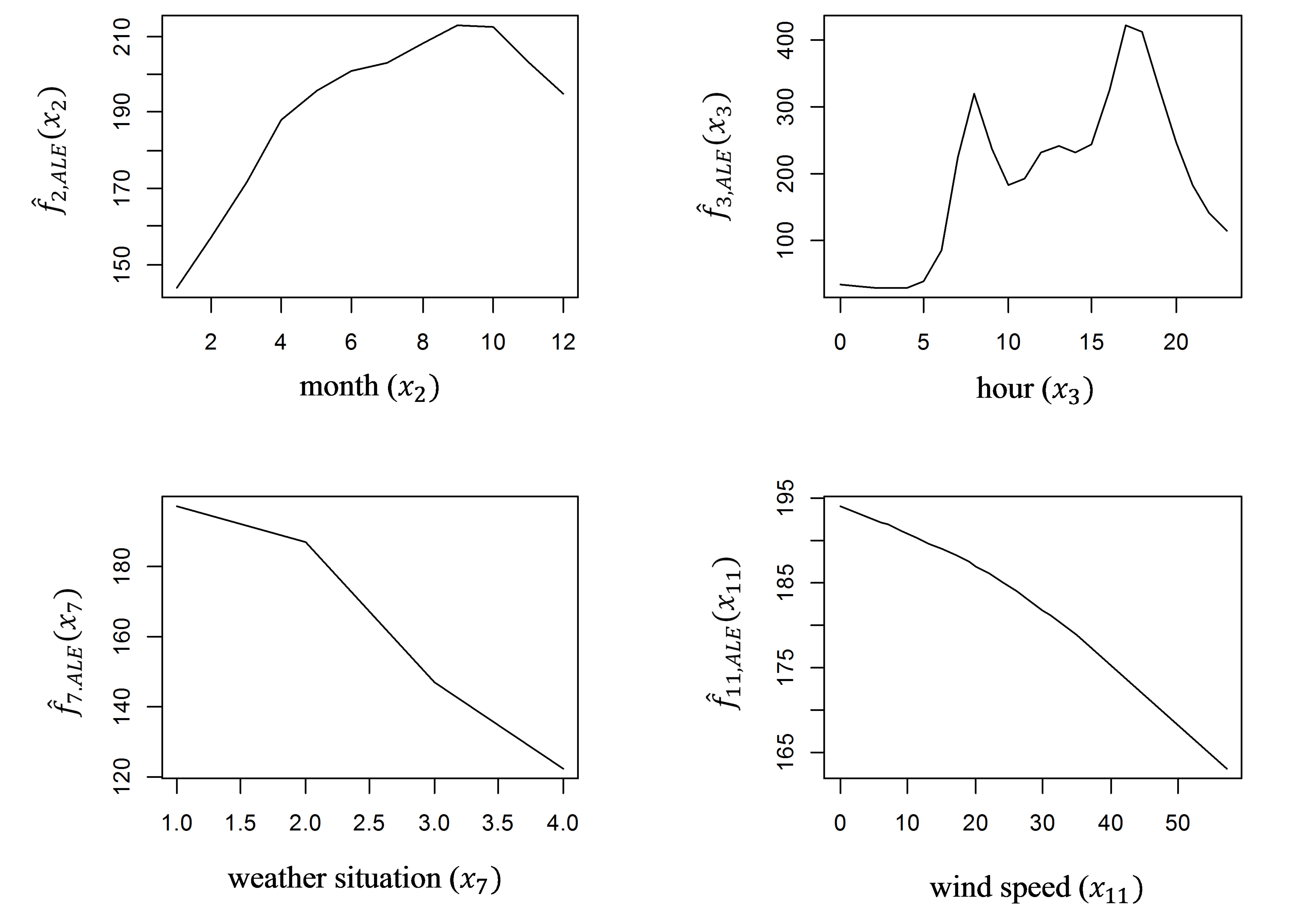}
    \caption{For the bike-sharing example with $f(\mathbf{x})$ a neural network model for predicting hourly bike rental counts, ALE main-effect plots for month ($X_2$, top left), hour-of-day ($X_3$, top right), weather situation ($X_7$, bottom left), and wind speed ($X_{11}$, bottom right) predictors. The zero-order effects have been included, i.e., the plots are of $f_{j,ALE}(x_j) + \mathbb{E}[f(\mathbf{X})]$. }
    \label{bs-main}
\end{figure}

\begin{figure}
    \centering
    \includegraphics[trim={1cm 0cm 0cm 0cm},clip, width = \textwidth]{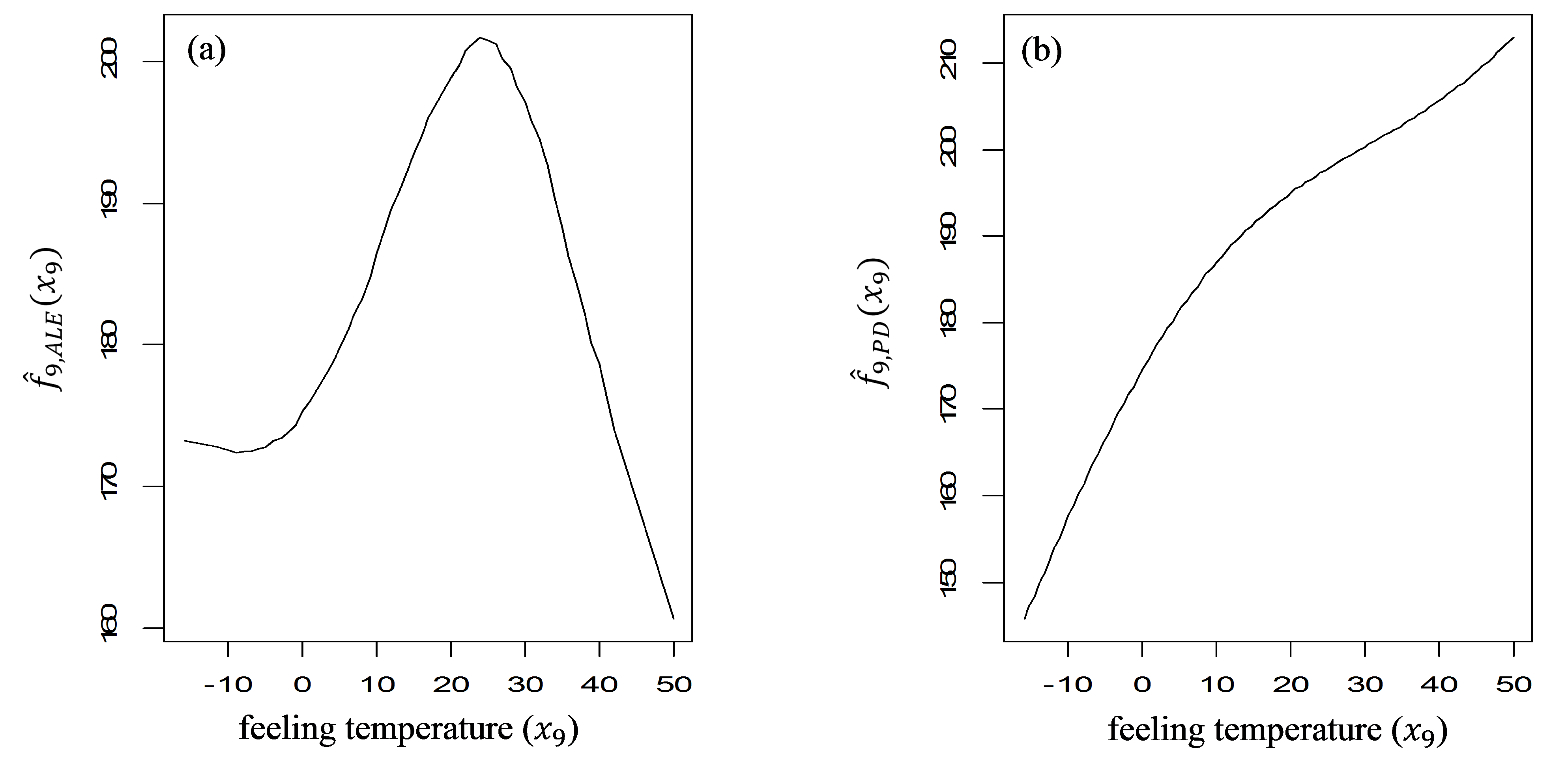}
    \caption{For the bike-sharing data example with neural network predicted counts for $f(\mathbf{x})$, ALE main-effect plot (left panel) and PD main-effect plot (right panel) for feeling temperature ($X_9$). Both plots include the zero-order effect $\mathbb{E}[f(\mathbf{X})]$. The two plots differ substantially, and the ALE plot seems to agree more with intuition.}
    \label{bs-vsPD}
\end{figure}

\begin{figure}
    \centering
    \includegraphics[trim={1cm 0cm 0cm 0cm},clip,width = \textwidth]{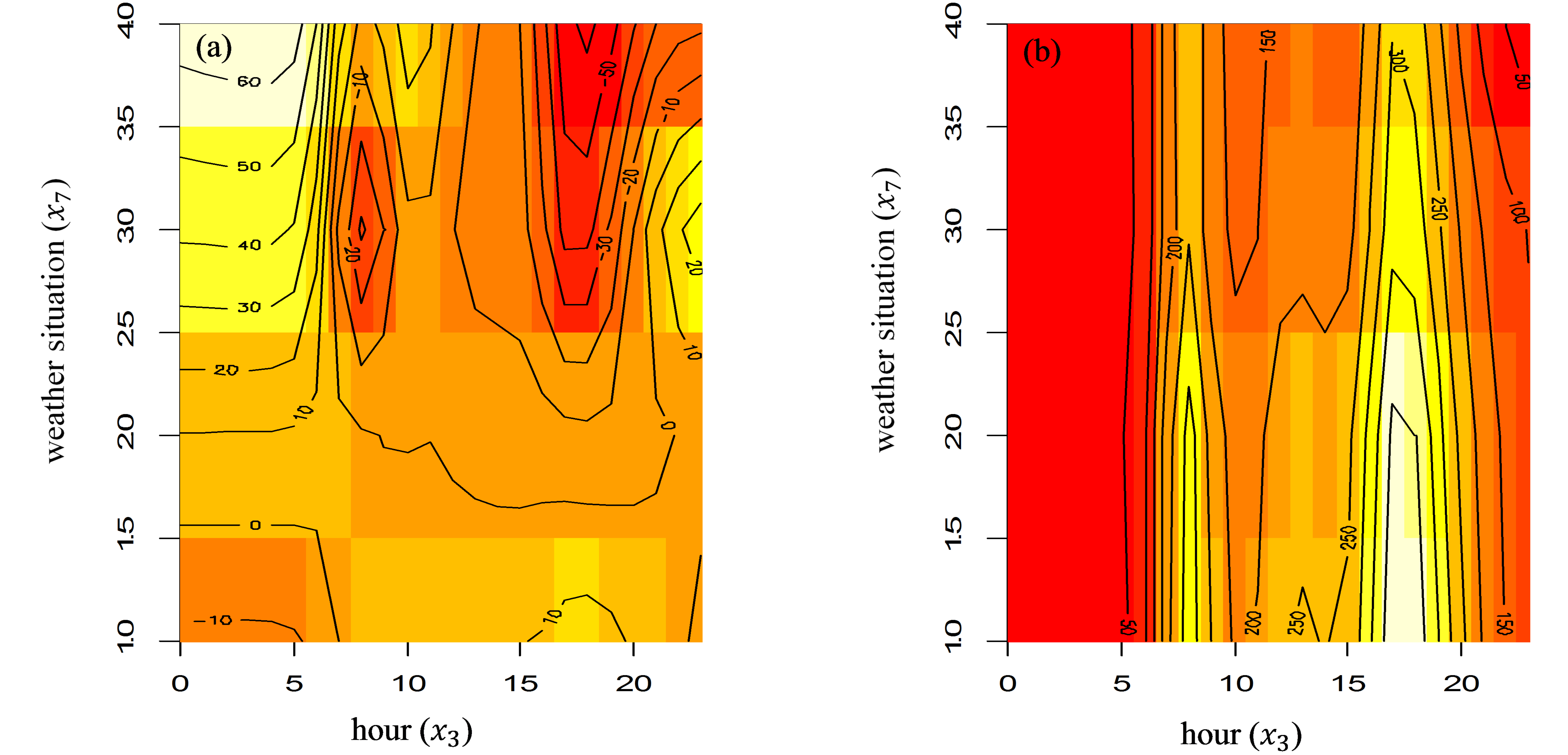}
    \caption{ALE second-order interaction plots for the predictors hour ($X_3$) and weather situation ($X_7$) without (left panel) and with (right panel) the main effects of $X_3$ and $X_7$ included. The left panel plots $\hat{f}_{\{3,7\},ALE}(x_3, x_7)$, and the right panel plots $\mathbb{E}[f(\mathbf{X})] + \hat{f}_{3,ALE}(x_3) + \hat{f}_{7,ALE}(x_7) + \hat{f}_{\{3,7\},ALE}(x_3, x_7)$. The numbers on the contours are the function values.}
    \label{bs-interaction}
\end{figure}

 
We fit a neural network model using the \texttt{R} \textbf{\texttt{nnet}} package (\cite{Venables}) with 10 nodes in the single hidden layer (\texttt{size} = 10), a logistic output activation function (\texttt{linout = FALSE}), and a regularization parameter of $0.05$ (\texttt{decay} $=0.05$). These parameters are approximately optimal according to multiple replicates of 3-fold cross validation (CV $r^2\approx 0.90$). For the ALE and PD plots, the function $f(\mathbf{x})$ is the predicted hourly count of rental bikes from the fitted neural network model. Figure \ref{bs-main}, Figure \ref{bs-vsPD}(a), and Figure \ref{bs-interaction} show ALE main and interaction effects plots for various predictors. We used $K=100$ for both the main-effect plots and the interaction plots. 
\par
The ALE main-effect plots are shown for month ($X_2$), hour ($X_3$), weather situation ($X_7$), and wind speed ($X_{11}$) in Figure \ref{bs-main}, and for  feeling temperature ($X_9$) in Figure \ref{bs-vsPD}(a). All of the ALE main-effect plots provide clear interpretations of the (main) effects of the predictors. For the effect of month ($X_2$), the number of rentals is lowest in January and gradually increases month-by-month until it peaks in September-October (months 9-10), after which it declines in the winter months. For the effect of hour of day ($X_3$), the number of rentals increases until it first peaks at the morning rush hour around 8:00 am (hour $8$), after which it decreases to moderate levels over the late morning and early afternoon hours, and then peaks again at the evening rush hour around 5:00-6:00 pm (hours $17-18$). For the effect of weather situation ($X_7$), the number of rentals monotonically decreases as the weather situation worsens. Recall that a larger value of $X_7$ corresponds to worse weather conditions. For the effect of wind speed ($X_{11}$), the number of rentals also monotonically decreases as the wind speed increases. For the effect of atemp ($X_9$, in Figure \ref{bs-vsPD}(a)), the number of rentals steadily increases as atemp (i.e., feeling temperature) increases up until about $26$ degrees Celsius ($79$ degrees Fahrenheit), after which it steadily decreases. This makes perfect sense, since a feeling temperature of $26$ degrees Celsius might be considered as nearly optimal for comfortably biking around a city (note that feeling temperature takes into factors such as humidity and breeze, so the actual temperature would be somewhat lower), and feeling temperatures that are either substantially higher or lower than this will make bike rental less appealing for many people.
In comparison, Figure \ref{bs-vsPD}(b) shows the PD main effect plot for feeling temperature, which is substantially different from the ALE main effect plot in Figure \ref{bs-vsPD}(a), even though the two are for the exact same fitted neural network model. The difference is due to the high correlation between feeling temperature and some of the other predictors, and the resulting extrapolation that makes PD plots unreliable. In this case, the PD plot indicates that the number of bike rentals monotonically increases as feeling temperature increases, even at feeling temperatures over $40$ degrees Celsius ($104$ Fahrenheit). The ALE plot for feeling temperature in Figure \ref{bs-vsPD}(a), which indicates that bike rentals will decrease as feeling temperature increases beyond the comfortable range, is in much better agreement with common sense. In addition to providing better interpretability, the ALE plots are much faster to compute than the PD plots (see Section \ref{comp advantages}).
\par 
Figure \ref{bs-interaction} shows two versions of the ALE second-order interaction effect plot for the hour and weather situation predictors ($\{X_3, X_7\}$), without and with the main effects of $X_3$ and $X_7$ included. The latter (Figure \ref{bs-interaction}(b)) plots $\mathbb{E}[f(\mathbf{X})] + \hat{f}_{3,ALE}(x_3) + \hat{f}_{7,ALE}(x_7) + \hat{f}_{\{3,7\},ALE}(x_3, x_7)$, whereas the former (Figure \ref{bs-interaction}(a)) plots only $\hat{f}_{\{3,7\},ALE}(x_3, x_7)$. Generally speaking, the latter provides a clearer picture of the joint effects of two predictors, whereas the former allows the overall magnitude of the interaction effect to be more easily assessed. Our \textbf{\texttt{ALEPlot}} \texttt{R} package allows either to be plotted.
\par
The interaction ALE plot in Figure \ref{bs-interaction} reveals an interesting relationship and is an important supplement to the main effects ALE plots for $X_3$ and $X_7$ in Figure \ref{bs-main}. Consider the decrease in bike rental counts that is due to larger weather situation values (i.e., less pleasant weather). If there were no interactions, this decrease would be the same regardless of the hour or the levels of the other predictors. But from Figure \ref{bs-interaction}(a), there is clearly a strong interaction between $X_3$ and $X_7$, since the contour values vary over a range of about $110$ units (from $-50$ to $+60$), which is almost as large as the range for the main effect $\hat{f}_{7,ALE}(x_7)$ in Figure \ref{bs-main}. The effect of weather situation, which is a decrease in rentals as weather situation increases, is clearly amplified during the rush hour peaks, and in general at hours when the overall rental counts are expected to be higher. One must be careful interpreting interaction plots without the main effects included. From Figure \ref{bs-interaction}(a), at some hours (e.g., around hour $0$, which is midnight) $\hat{f}_{\{3,7\},ALE}(x_3, x_7)$ \textit{increases} as weather situation increases. However, when the main effects of $X_3$ and $X_7$ are included as in Figure \ref{bs-interaction}(b), it is clear that increasing weather situation decreases bike rentals at any hour. 
\par
It makes sense that the effects of weather situation are amplified by the effects of hour (on which the overall bike rental counts depend heavily), and this example illustrates how visualizations like this can aid the model building process by suggesting modifications of the model that one might consider. For example, Figure \ref{bs-interaction}(b) indicates that the effects of weather situation and/or hour on rental counts might be better modeled as multiplicative. 

\subsection{The Wrong and Right Ways to Interpret ALE Plots (and PD Plots)}\label{wrong and right ways to interpret}
\par 
This section provides a word of caution on how not to interpret ALE plots when the predictors are highly correlated, which also applies to interpreting PD plots. Reconsider Example \ref{toy example: noiseless}, in which $X_1$ and $X_2$ are highly correlated (see Figure \ref{1st8split}), and the ALE and PD plots are as in Figure \ref{ALEandPDtree}. The \textit{wrong} way to interpret the ALE plot is that it implies that if we fix (say) $x_1$ and then vary $x_2$ over its entire range, the response (of which $f(\mathbf{x})$ is a predictive model) is expected to vary as in Figure \ref{ALEandPDtree}(b). And this interpretation is wrong even if $f(\mathbf{x})$ is truly additive in the predictors. Indeed, varying $x_2$ over its entire range with $x_1$ held fixed would take $\mathbf{x}$ far outside the envelope of the training data to which $f$ was fit, as can be seen in Figure \ref{1st8split}. $f(\mathbf{x})$ is obviously unreliable for this level of extrapolation, which was the main motivation for ALE plots, and we have highly uncertain knowledge of the hypothetical values of the response far outside the data envelope.
\par
\begin{figure}
    \centering
    \includegraphics[trim={1cm 0.5cm 0cm 0cm},clip,width = \textwidth]{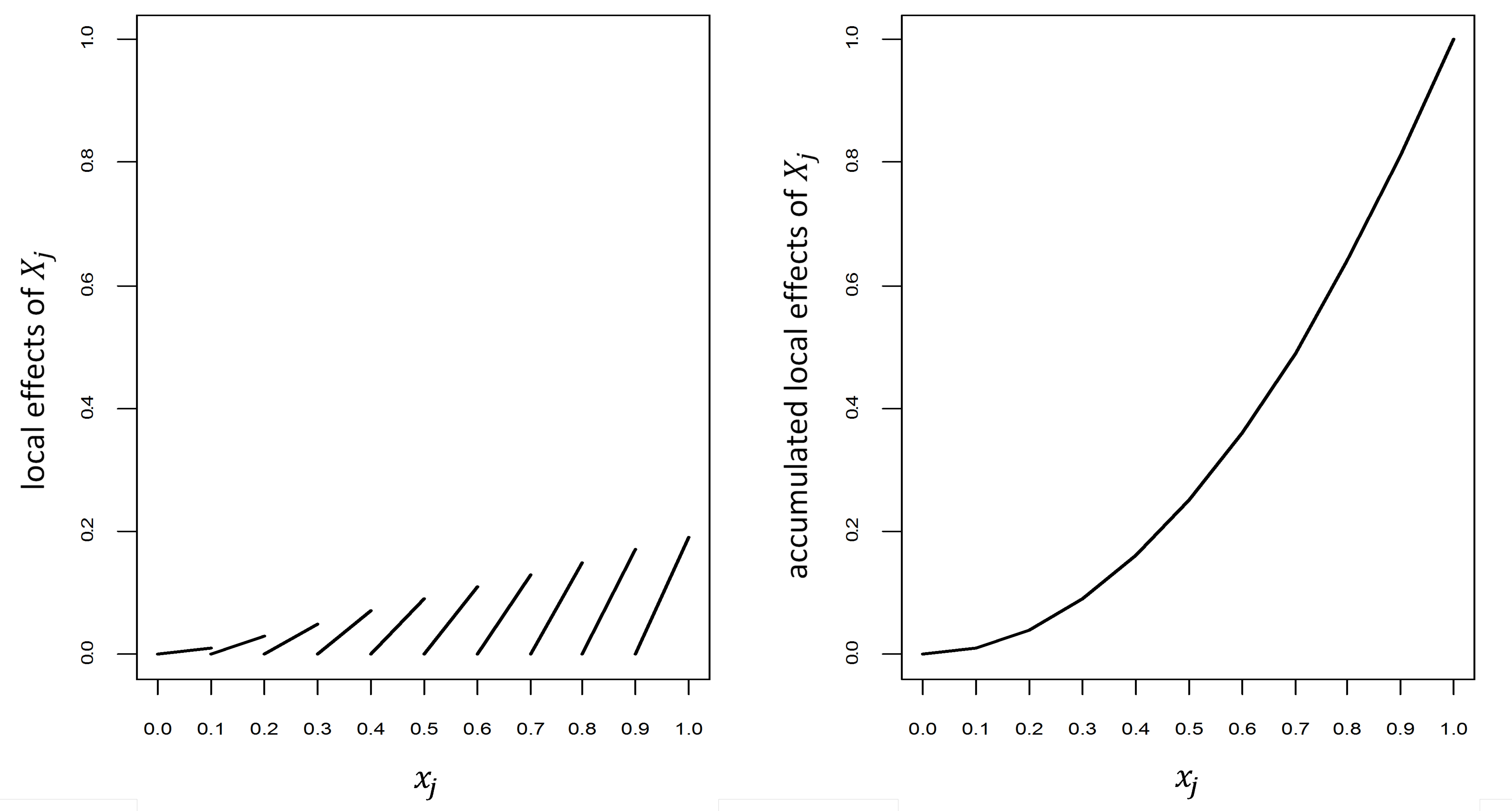}
    \caption{Illustration of the right way to interpret ALE plots for an example in which $f(\mathbf{x}) = \sum_{j=1}^{d}f_j(x_j)$ is additive with quadratic $f_j(x_j)=x_j^2$, and $K=10$ equally-spaced bins over the support $[0,1]$ were used. The left panel shows the local effects of $X_j$ within each bin (i.e., the summand in Eq. \eqref{ALE uncentered main def as limit}). The local effects are local and require no extrapolation outside the data envelope. The right plot is of $g_{j,ALE}(x_j)$ and can be viewed as piecing together (or accumulating) the local effects in a manner that facilitates easier visualization of the underlying global effect.}
    \label{interpret fig}
\end{figure}

However, ALE plots are still very useful if we interpret them correctly, and the correct interpretation is illustrated in Figure \ref{interpret fig}. In this toy example, suppose $f(\mathbf{x}) = \sum_{j=1}^{d} f_j(x_j)$ is additive, the effect of $X_j$ is the quadratic function $f_j(x_j)=x_j^2$, and $X_j$ is highly correlated with the other predictors. The left panel of Figure \ref{interpret fig} shows the local effects of $X_j$ within each bin, for $K=10$ equally-spaced bins over the support $[0,1]$. Here, the local effect within a bin is defined as the summand in the Eq. \eqref{ALE uncentered main def as limit} definition of $g_{j,ALE}(x_j)$, which represents the average change in $f(\mathbf{X})$ as $X_j$ changes from the left endpoint to the right endpoint of the bin. 
\par
The local effects are exactly that -- local -- and require no extrapolation beyond the envelope of the data, since the changes in $f(\mathbf{X})$ are averaged across the conditional distribution of $\mathbf{X}_{\backslash j}$, given that $X_j$ falls in that bin. Consequently, if the bin widths are not too small and $f$ is not too noisy, a local effect plot like the one in the left panel of Figure \ref{interpret fig} could be interpreted to reveal the effect of $X_j$ on $f$.
\par
However, the effect of $X_j$ is much easier to visualize if we  \textit{accumulate} the local effects via Eq. \eqref{ALE uncentered main def as limit} and plot $g_{j,ALE}(x_j)$ instead, as in the right panel of Figure \ref{interpret fig}. Aside from vertical centering, this is exactly the ALE plot, and it is best viewed as a way of piecing together the local effects in a manner that provides easier visualization of the underlying global effect of a predictor. The additive recovery property discussed in the next subsection provides even stronger justification for this manner of piecing together the local effects. Namely, if $f(\mathbf{x}) = \sum_{j=1}^{d} f_j(x_j)$ is additive, the ALE plot manner of piecing together the local effects produces the correct global effect function $f_{j,ALE}(x_j) = f_j(x_j)$. Of course, one must still keep in mind that the global effect $f_j(x_j)$ may only hold when the set of predictors $\mathbf{x}$ jointly fall within the data envelope.
\subsection{Paired Differencing and Additive Recovery}\label{paired diff and add recovery}
The ALE functions have an attractive additive recovery property mentioned in Remark \ref{Remark: explanations additive recovery}. Suppose $f(\mathbf{x})=\sum_{j=1}^d f_j(x_j)$ is an additive function of the individual predictors. Then it is straightforward to show that the ALE main effects are $f_{j,ALE}(x_j)=f_j(x_j)$ for $(j=1,2,\ldots,d)$, up to an additive constant. That is, the ALE effects recover the correct additive functions. More generally, the following result states that higher-order ALE effects $f_{J,ALE}(x_J)$ have a similar additive recovery property.
\par 
\textbf{Additive recovery for ALE plots.} Suppose $f$ is of the form $f(\mathbf{x})=\sum_{J\subseteq\{1,2,\ldots,d\},|J|\leq k} f_J (\mathbf{x}_J)$ for some $1\leq k\leq d$. That is, $f$ has interactions of order $k$, but no higher-order interactions than that. Then for every subset $J$ with $|J|=k$, $f_{J,ALE}(\mathbf{x}_J)=f_J(\mathbf{x}_J)+\sum_{u\subset J} h_u(\mathbf{x}_u)$ for some functions $h_u(\mathbf{x}_u)$ that are of strictly lower order than $k$. In other words, for every $J$ with $|J|=k$, the ALE effect $f_{J,ALE}(\mathbf{x}_J)$ returns the correct $k$-order interaction $f_J(\mathbf{x}_J)$, since the presence of strictly-lower-order functions do not alter the interpretation of a $k$-order interaction. 
\par 
The proof of this additive recovery property for ALE plots follows directly from the decomposition theorem in Appendix \ref{ale decomposition theorem and properties of L}. It also follows that if the functions $\{f_J(\mathbf{x}_J)\}$ in the expression for $f(\mathbf{x})$ are adjusted so that each has no lower-order ALE effects, then $f_{J,ALE}(\mathbf{x}_J)=f_J (\mathbf{x}_J)$ for each $J\subseteq\{1,2,\ldots,d\}$. Although PD plots have a similar additive recovery property (see below), M plots have no such property. For example, if $f(\mathbf{x})=\sum_{j=1}^d f_j(x_j)$, and the predictors are dependent, then each $f_{j,M}(x_j)$ may be a combination of the main effects of many predictors. As discussed previously, this can be viewed as the omitted variable bias in regression, whereby a regression of $Y$ on (say) $X_1$, omitting a correlated nuisance variable $X_2$ on which $Y$ also depends, will bias the effect of $X_1$ on $Y$.
\begin{figure}[!ht]
    \centering
    \includegraphics[width=\textwidth]{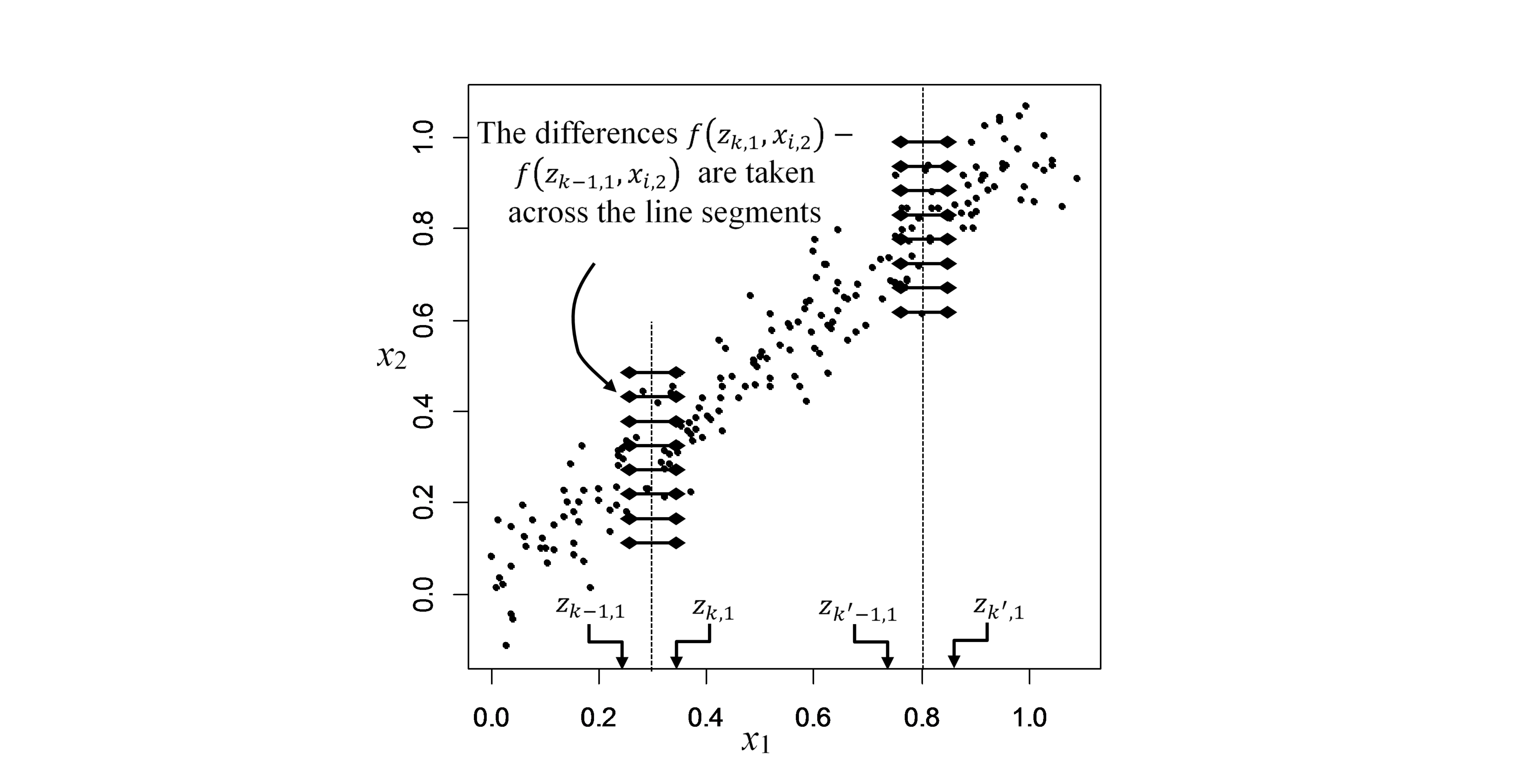}
    \caption{Illustration of how, when estimating $f_{1,ALE}(x_1)$, the differences $f(z_{k,1},x_{i,2})-f(z_{k-1,1},x_{i,2})$ and $f(z_{k',1},x_{i,2} )-f(z_{k'-1,1},x_{i,2})$ in \eqref{ALE uncentered main est} are paired differences that block out the nuisance variable $X_2$. Here, $k=k_1(0.3)$ and $k'=k_1(0.8)$. }
    \label{paireddiff}
\end{figure}
\par 
The mechanism by which ALE plots avoid this omitted nuisance variable bias is illustrated in Figure \ref{paireddiff}, for the same example depicted in Figures \ref{PD and M difference}, \ref{Illustration of ALE Comp}, and \ref{1st8split}. First note that the M plot functions in this example are severely biased, because $f_{1,M}(x_1)$ averages the function $f(x_1,X_2)$ itself (as opposed to its derivative) with respect to the conditional distribution of $X_2|X_1=x_1$ (see \eqref{M def} and \eqref{M estimate}). For example, for $f(\mathbf{x})=x_1+x_2^2$ considered in Figure \ref{ALEandPDtree}, the M plot averaged function is $f_{1,M} (x_1)=\mathbb{E}[f(X_1,X_2)|X_1=x_1]=x_1+\mathbb{E}[X_2^2 |X_1=x_1 ]\neq x_1$, which is biased by the functional dependence of $f(\mathbf{x})$ on the correlated nuisance variable $X_2$. In contrast to averaging the function $f$ itself, the ALE effect $f_{1,ALE}(x_1)$ estimated via \eqref{ALE uncentered main est}—\eqref{ALE centered main est} averages only the local effect of $f$ represented by the paired differences $f(z_{k,1},x_{i,2} )-f(z_{k-1,1},x_{i,2})$ in \eqref{ALE uncentered main est}. As illustrated in Figure \ref{paireddiff}, this paired differencing is what blocks out the effect of the correlated nuisance variable $X_2$. Continuing the $f(\mathbf{x})=x_1+x_2^2$ example, the paired differences $f(z_{k,1},x_{i,2} )-f(z_{k-1,1},x_{i,2})=(z_{k,1}+x_{i,2}^2 )-(z_{k-1,1}+x_{i,2}^2 )=z_{k,1}-z_{k-1,1}$ for the ALE plot completely block out the effect of $X_2$, so that the accumulated local effect $\sum_{k=1}^{k_1(x_1 )} [z_{k,1}-z_{k-1,1}] = x_1+\text{constant}$ is correct.\\
\par  
\textbf{Multiplicative recovery for ALE plots for independent subsets of predictors.}  Suppose the model is of the form $f(\mathbf{x})=f_J(\mathbf{x}_J ) f_{\backslash J}(\mathbf{x}_{\backslash J})$ for some $J\subset \{1,2,\ldots,d\}$ with $\mathbf{X}_J$ independent of $\mathbf{X}_{\backslash J}$. In this case it is straightforward to show that the ALE $|J|$-order interaction effect of $\mathbf{X}_J$ is $f_{J,ALE}(\mathbf{x}_J)=f_J (\mathbf{x}_J )\mathbb{E}[f_{\backslash J}(\mathbf{X}_{\backslash J})]+\sum_{u\subset J} h_u(\mathbf{x}_u)$ for some lower-order functions $h_u(\mathbf{x}_u)$. That is, the ALE $|J|$-order interaction effect $f_{J,ALE}(\mathbf{x}_J)$ recovers the correct function $f_J (\mathbf{x}_J)$, except for a multiplicative constant $\mathbb{E}[f_{\backslash J}(\mathbf{X}_{\backslash J})]$ and the additive presence of strictly lower-order functions. \\
\par  \textbf{Comparison to PD plots.} PD plots also have the same additive and multiplicative recovery properties just discussed. Moreover, for $f(\mathbf{x})=f_J(\mathbf{x}_J )f_{\backslash J}(\mathbf{x}_{\backslash J})$, PD plots have multiplicative recovery (up to a multiplicative constant) even when $\mathbf{X}_J$ and $\mathbf{X}_{\backslash J}$ are dependent (\cite{Hastie}). Although it is probably desirable to have multiplicative recovery when $\mathbf{X}_J$ and $\mathbf{X}_{\backslash J}$ are independent, it is unclear whether multiplicative recovery is even desirable if $\mathbf{X}_J$ and $\mathbf{X}_{\backslash J}$ are dependent. 
\par 
For example, suppose $f(\mathbf{x})=x_1 x_2$ with $X_1$ $(J=1)$ and $X_2$ $(\backslash J=2)$ standard normal random variables with correlation coefficient $\rho$. It is straightforward to show that $f_{\{1,2\},ALE}(x_1,x_2)=x_1 x_2-\frac{1}{2}\rho (x_1^2+x_2^2)$, $f_{1,ALE} (x_1 )=\frac{1}{2}\rho(x_1^2-1)$, and $f_{2,ALE}(x_2)=\frac{1}{2}\rho(x_2^2-1)$, compared to $f_{\{1,2\},PD}(x_1,x_2 )=x_1 x_2$, $f_{1,PD}(x_1 )=0$, and $f_{2,PD}(x_2)=0$. Because of the strong interaction, it is essential to look at the second-order interaction effects in order to understand the functional dependence of $f(\cdot)$ on the predictors. Both $f_{\{1,2\},ALE}(x_1,x_2)$ and $f_{\{1,2\},PD}(x_1,x_2)$ correctly recover the interaction, up to lower-order functions of the individual predictors. Regarding the main effects, however, the picture is more ambiguous. First, it is important to note that with a strong interaction and dependent predictors, it is unclear whether the main effects are even meaningful. And it is equally unclear whether the PD main effect $f_{1,PD}(x_1)=0$ is any more or less meaningful than the ALE main effect $f_{1,ALE}(x_1)=\frac{1}{2}\rho(x_1^2-1)$. If $X_1$ and $X_2$ were independent in this example, then it would probably be desirable to view the main effects of $X_1$ and $X_2$ as zero, but in this case $f_{1,PD}(x_1 )=f_{1,ALE}(x_1 )=0$ would actually be in agreement.  The situation is murkier with dependent predictors in this example. The local effect $\frac{\partial f(\mathbf{x})}{\partial x_1} =x_2$ of $X_1$ depends strongly on the value of $x_2$, in that it is amplified for larger $|x_2|$ and changes sign if $x_2$ changes sign. Thus, if $\rho$ is large and positive, the local effect of $X_1$ is positive for $x_1>0$ and negative for $x_1<0$, which is the local effect of a quadratic relationship. In this case one might argue that the quadratic $f_{1,ALE}(x_1)=\frac{1}{2}\rho(x_1^2-1)$ is more revealing than $f_{1,PD}(x_1)=0$. However, the debate is largely academic, because when strong interactions are present the lower-order effects should not be interpreted in a vacuum. 
\subsection{Computational Advantages of ALE Plots over PD Plots}\label{comp advantages}
\par 
For general supervised learning models $f(\mathbf{x})$, ALE plots have an enormous computational advantage over PD plots. Suppose we want to compute $\hat{f}_{J,ALE}(\mathbf{x}_J)$ for one subset $J\subseteq\{1,2,\ldots,d\}$ over a grid in the $\mathbf{x}_J$-space with $K$ discrete locations for each variable. Computation of $\hat{f}_{J,ALE}(\mathbf{x}_J)$ over this grid requires a total of $2^{|J|}\times n$ evaluations of the supervised learning model $f(\mathbf{x})$ (see \eqref{ALE uncentered main est}—\eqref{ALE centered second est} or \eqref{Lu estimate} for $|J|>2$). In comparison, computation of $\hat{f}_{J,PD}(\mathbf{x}_J)$ over this grid requires a total of $K^{|J|}\times n$ evaluations of $f(\mathbf{x})$. For example, for $K=50$, PD main effects and second-order interaction effects require, respectively, $25$ and $625$ times more evaluations of $f(\mathbf{x})$ than the corresponding ALE effects. Moreover, as we discuss in Appendix \ref{implementation details}, the evaluations of $f(\mathbf{x})$ can be easily vectorized (in \texttt{R}, for example, by appropriately calling the \textbf{\texttt{predict}} function that is built into most supervised learning packages in \texttt{R}). 
\par 
Also notice that the number of evaluations of $f(\mathbf{x})$ for ALE plots does not depend on $K$, which is convenient. As $n$ increases, the observations become denser, in which case we may want the fineness of the discretization to increase as well. If we choose $K^{|J|} \propto n$ (which results in the same average number of observations per cell as $n$ increases), then the number of evaluations of $f(\mathbf{x})$ is $\mathcal{O}(n)$ for ALE plots versus $\mathcal{O}(n^2 )$ for PD plots. 
\par
For the bike sharing example in Section \ref{illustration with bike sharing}, we implemented the ALE and PD plots using our \texttt{R} package \textbf{\texttt{ALEPlot}} on a Windows\texttrademark{} laptop with Intel(R) Core(TM) i7-7600U CPU @ 2.80 GHz processor. The ALE main-effect plots and the ALE second-order interaction plots took less than 1 second each. In comparison, with the same $K=100$, the PD main-effect plots took about 5 seconds each, and the PD interaction plots (not shown in Section \ref{illustration with bike sharing}) took about 8 minutes each. The PD plot computational expense is proportional to $K$ for main effects and to $K^2$ for second-order interaction effects, whereas the ALE plot computational expense is largely independent of $K$. The ALE interaction plots were orders of magnitude faster to compute than the PD interaction plots (less than 1 second vs.\ 8 minutes).

\subsection{Relation to Functional ANOVA with Dependent Inputs}\label{relation to functional anova}
\par 
In the context of the closely-related problem of functional ANOVA with dependent input (i.e., predictor) variables, the extrapolation issue that motivated ALE plots has been previously considered. \cite{Hooker} proposed a functional ANOVA decomposition of $f(\mathbf{x})$ into component functions $\{f_{J,ANOVA}(\mathbf{x}_J):J\subseteq\{1,2,\ldots,d\}\}$ by adopting the \cite{Stone} approach of using weighted integrals in the function approximation optimality criterion and in the component function orthogonality constraints. \cite{Hooker} used $p_{\{1,2,\ldots,d\}}(\mathbf{x})$ as a weighting function, which indirectly avoids extrapolation of $f(\mathbf{x})$ in regions in which there are no training observations, because any such extrapolations are assigned little or no weight. The resulting ANOVA component functions are \emph{hierarchically orthogonal} under the correlation inner product, in the sense that $f_{J,ANOVA}(\mathbf{X}_J)$ and $f_{u,ANOVA}(\mathbf{X}_u)$ are uncorrelated when $u\subset J$. However, $f_{J,ANOVA}(\mathbf{X}_J)$ and $f_{u,ANOVA}(\mathbf{X}_u)$ are not uncorrelated for general $u\neq J$. 
\par 
In comparison, we show in Appendix \ref{ale decomposition theorem and properties of L} that the ALE decomposition\\ $f(\mathbf{x})=\sum_{J\subseteq \{1,2,\ldots,d\}} f_{J,ALE}(\mathbf{x}_J)$ mentioned in Remark \ref{Remark: Appendices B and C} has the following orthogonality-like property. For each $J\subseteq \{1,2,\ldots,d\}$, let $\mathcal{H}_J(\cdot)$ denote the operator that maps a function $f$ to its ALE effect $f_{J,ALE}$, i.e., such that $f_{J,ALE}=\mathcal{H}_J(f)$ (see Appendix \ref{def of higher-order effects} for details). The collection of operators $\{\mathcal{H}_J:J\subseteq\{1,2,\ldots,d\}\}$ behaves like a collection of operators that project onto orthogonal subspaces of an inner product space. Namely, if ``$\circ$" denotes the composite function operator, then $\mathcal{H}_J\circ \mathcal{H}_J(f)=\mathcal{H}_J(f)$, and $\mathcal{H}_{J'}\circ \mathcal{H}_J(f)=0$ for each $J'\subseteq \{1,2,\ldots,d\}$ with $J' \neq J$. 
\par
In other words, the ALE $|J'|$-order effect of the predictors $\mathbf{X}_{J'}$ for the function $f_{J,ALE}$ is identically zero when $J \neq J'$; and the ALE $|J|$-order effect of the predictors $\mathbf{X}_J$ for the function $f_{J,ALE}$ is the same function $f_{J,ALE}$. For example, for any pair of predictors $\{X_j,X_l\}$, the ALE main effect of any predictor (including $X_j$ or $X_l$) for the function $f_{\{j,l\},ALE}(x_j,x_l)$ is identically zero. Thus each ALE second-order interaction effect function has ALE main effects that are all identically zero. Likewise, each ALE main effect function has ALE second-order interaction effects that are all identically zero. And for the function $f_{\{j,l\},ALE}(x_j,x_l)$, the ALE second-order interaction effect of any other pair of predictors is identically zero. Similarly, the ALE first- and second-order interaction effects for any ALE third-order effect function are all identically zero, and vice-versa.
\par 
For the purpose of visualizing the effects of the predictors on black box supervised learning models, the correlation orthogonality in other functional ANOVA decompositions may be less relevant and less useful than the ALE pseudo-orthogonality. As discussed in \cite{Roosen}, if the predictors are dependent, it may even be preferable to artificially impose a product $p_{\{1,2,\ldots,d\}}(\mathbf{x})$ in the functional ANOVA decomposition to avoid conflating direct and indirect effects of a predictor, and this will typically result in ANOVA component functions that are no longer uncorrelated. For example, suppose $f(\mathbf{x})=x_1+x_2$, and $X_1$ and $X_2$ are correlated. Any functional ANOVA decomposition that gives uncorrelated main effects will not give the correct main effects $f_1(x_1)=x_1$ and $f_2(x_2)=x_2$ that are needed to understand the true functional dependence of $f(\mathbf{x})$ on $x_1$ and $x_2$. In contrast, the ALE and PD main effect functions are the correct functions $f_1(x_1)=x_1$ and $f_2(x_2)=x_2$, up to an additive constant (see Section \ref{paired diff and add recovery}). Functional ANOVA can be coerced into producing the correct main effects $f_{1,ANOVA}(x_1 )=x_1$ and $f_{2,ANOVA}(x_2)=x_2$ by artificially imposing a product distribution $p_{\{1,2\}}(\mathbf{x})=p_1(x_1)p_2(x_2)$, but then the ANOVA component functions are no longer uncorrelated. Moreover, artificially imposing a product $p_{\{1,2,\ldots,d\}}(\mathbf{x})$ in functional ANOVA still leaves the extrapolation problem that plagues PD plots and that motivated ALE plots and the work of \cite{Hooker}. 
\par 
In addition, practical implementation is far more cumbersome for functional ANOVA decompositions than for ALE (or PD) plots, for multiple reasons. First, $p_{\{1,2,\ldots,d\}}(\mathbf{x})$ must be estimated in the functional ANOVA approach of \cite{Hooker}, which is problematic in high-dimensions. In contrast, the ALE effect estimators \eqref{ALE uncentered main est}—\eqref{ALE centered second est} involve summations over the training data but require no estimate of $p_{\{1,2,\ldots,d\}}(\mathbf{x})$. Second, each ALE plot effect function can be computed one-at-a-time using straightforward and computationally efficient calculations that involve only finite differencing, averaging, and summing. In contrast, the functional ANOVA component functions must be computed simultaneously, which requires the solution of a complex system of equations. Follow-up work in \cite{Li}, \cite{Chastaing}, and \cite{Rahman} improved the solution techniques, sometimes restricting the component ANOVA functions to be expansions in basis functions such as polynomials and splines, but these are more restrictive (perhaps negating the benefits of fitting a black box supervised learning model in the first place), as well as more cumbersome and computationally expensive than ALE plots. 
\section{Conclusions}\label{conclusion}
\par 
For visualizing the effects of the predictor variables in black box supervised learning models, PD plots are the most widely used method. The ALE plots that we have proposed in this paper are an alternative that has two important advantages over PD plots. First, by design, ALE plots avoid the extrapolation problem that can render PD plots unreliable when the predictors are highly correlated (see Figures \ref{ALEandPDtree}, \ref{50reps}, and \ref{bs-vsPD}). Second, ALE plots are substantially less computationally expensive than PD plots, requiring only $2^{|J|}\times n$ evaluations of the supervised learning model $f(\mathbf{x})$ to compute each $\hat{f}_{J,ALE}(\mathbf{x}_J)$, compared to $K^{|J|}\times n$ evaluations to compute each $\hat{f}_{J,PD}(\mathbf{x}_J)$. In light of this, we suggest that ALE plots should be adopted as a standard visualization component in supervised learning software. We have also provided, as supplementary material, an \texttt{R} package \textbf{\texttt{ALEPlot}} to implement the ALE plots.
\bibliography{citations}

\begin{thebibliography}{15}
\expandafter\ifx\csname natexlab\endcsname\relax\def\natexlab#1{#1}\fi
\expandafter\ifx\csname url\endcsname\relax
  \def\url#1{\texttt{#1}}\fi
\expandafter\ifx\csname urlprefix\endcsname\relax\def\urlprefix{URL: }\fi

\bibitem[{Chambers and Hastie(1992)}]{Chambers}
Chambers, J.~M. and Hastie, T.~J. (1992) \textit{Statistical Models in S}.
\newblock Pacific Grove, CA: Wadsworth \& Brooks/Cole.

\bibitem[{Chastaing and Prieur(2012)}]{Chastaing}
Chastaing, G., G.~F. and Prieur, C. (2012) Generalized hoeffding-sobol
  decomposition for dependent variables - application to sensitivity analysis.
\newblock \textit{Electronic Journal of Statistics}, \textbf{6}, 2420–2448.

\bibitem[{Cleveland(1993)}]{Cleveland}
Cleveland, W.~S. (1993) \textit{Visualizing Data}.
\newblock Summit, NJ: Hobart Press.

\bibitem[{Cook(1995)}]{Cook}
Cook, D.~R. (1995) Graphics for studying the net effects of regression
  predictors.
\newblock \textit{Statistica Sinica}, \textbf{5}, 689--708.

\bibitem[{Fanaee-T and Gama(2013)}]{BikeSharing}
Fanaee-T, H. and Gama, J. (2013) Event labeling combining ensemble detectors
  and background knowledge.
\newblock \textit{Progress in Artificial Intelligence}, 1--15.

\bibitem[{Friedman(2001)}]{Friedman}
Friedman, J.~H. (2001) Greedy function approximation: A gradient boosting
  machine.
\newblock \textit{Annals of Statistics}, \textbf{29}, 1189--1232.

\bibitem[{Goldstein and Pitkin(2015)}]{Goldstein}
Goldstein, A., K. A. B.~J. and Pitkin, E. (2015) Peeking inside the black box:
  Visualizing statistical learning with plots of individual conditional
  expectation.
\newblock \textit{Journal of Computational \& Graphical Statistics},
  \textbf{24}, 44--65.

\bibitem[{Hastie and Friedman(2009)}]{Hastie}
Hastie, T., T.~R. and Friedman, J. (2009) \textit{The elements of statistical
  learning}.
\newblock New York: Springer.

\bibitem[{Hooker(2007)}]{Hooker}
Hooker, G. (2007) Generalized functional anova diagnostics for high dimensional
  functions of dependent variables.
\newblock \textit{Journal of Computational and Graphical Statistics},
  \textbf{16}, 709--732.

\bibitem[{Li and Rabitz(2012)}]{Li}
Li, G. and Rabitz, H. (2012) General formulation of hdmr component functions
  with independent and correlated variables.
\newblock \textit{Journal of Mathematical Chemistry}, \textbf{50}, 99--130.

\bibitem[{Rahman(2014)}]{Rahman}
Rahman, S. (2014) A generalized anova dimensional decomposition for dependent
  probability measures.
\newblock \textit{SIAM/ASA Journal of Uncertainty Quantification}, \textbf{2},
  670--697.

\bibitem[{Ripley(2015)}]{Ripley}
Ripley, B.~D. (2015) tree: Classification and regression trees. r package
  version 1.0-36.
\newblock \urlprefix\url{http://CRAN.R-project.org/package=tree}.

\bibitem[{Roosen(1995)}]{Roosen}
Roosen, C.~B. (1995) \textit{Visualization and Exploration of High-Dimensional
  Functions Using the Functional Anova Decomposition}.
\newblock Ph.D. thesis, Stanford University.

\bibitem[{Stone(1994)}]{Stone}
Stone, C.~J. (1994) The use of polynomial splines and their tensor products in
  multivariate function estimation.
\newblock \textit{Annals of Statistics}, \textbf{22}, 118--171.

\bibitem[{Venables and Ripley(2002)}]{Venables}
Venables, W.~N. and Ripley, B.~D. (2002) \textit{Modern Applied Statistics with
  S. Fourth Edition}.
\newblock New York: Springer.

\end{thebibliography}
\bibliographystyle{rss.bst}
\clearpage
\begin{appendices}
\numberwithin{equation}{section}
\renewcommand{\theequation}{\Alph{section}.\arabic{equation}}
\section{Statements and Proofs of Theorems A.1 and 1 - 4}\label{statement and proofs of theorems}
\begin{thmA.1} [\textbf{sufficient conditions for the limit in \eqref{ALE uncentered main def as limit} to exist, independent of the sequence of partitions}] \label{TheoremA.1} Define the functions $h(u,z)\equiv \mathbb{E}[f(u,\mathbf{X}_{\backslash j})|X_j=z]$ and $h'(u,z)\equiv\frac{\partial h(u,z)}{\partial u}$ where the derivative exists, and suppose that $p$ and $f$ are such that: 
\begin{enumerate}[(i)]
    \item $f$ is bounded on $\mathcal{S}$.
    \item \label{assumption 2}$h(\cdot,\cdot)$ is differentiable in its first argument everywhere on the set $\{(z,z): z\in \mathcal{S}_j\backslash \mathcal{J}\}$, where the set of points $\mathcal{J}=\{u_1,u_2,\ldots,u_M\}\subset \mathcal{S}_j$ at which $h(\cdot,\cdot)$ is nondifferentiable (and possibly discontinuous) in its first argument is a finite collection. $\mathcal{J}$ may be empty.
    \item \label{assumption 3} The differentiability of $h(\cdot,\cdot)$ with respect to its first argument is uniform on the set $\{(z,z): z\in \mathcal{S}_j\backslash \mathcal{J}\}$, in the sense that $\forall\varepsilon>0$, $\exists\delta>0$ such that
    \begin{equation*}
    \left|h'(z,z)-\frac{h(u,z')-h(v,z')}{u-v}\right|<\varepsilon 
    \end{equation*}
	whenever $|u-v|<\delta$ and $(v,u]\cap \mathcal{J}=\emptyset$ with $v<z\leq u$ and $v<z'\leq u$.
    \item  \label{assumption 4}	$h(\cdot,\cdot)$ is continuous in its second argument everywhere on the set $\{(z,z): z\in \mathcal{S}_j\}$, and the continuity is uniform in the sense that $\forall \varepsilon > 0, \exists \delta >0$ such that
		$|h(z,z)-h(z,z')|<\varepsilon$
	whenever $|z-z'|<\delta$.
    \item  \label{assumption 5} For each $l=1,2,\ldots,M$, the potential discontinuity of $h(u,u_l)$ in its first argument at $u = u_l$ is such that $\lim_{u\to u_l^{-}} h(u,u_l)$ exists and $\lim_{u\to u_l^{+}}h(u,u_l)=h(u_l,u_l)$ (i.e., all discontinuities are of the jump type and are right continuous).
\end{enumerate}
Under these conditions, the limit in \eqref{ALE uncentered main def as limit} exists independent of the sequence of partitions $\{\mathcal{P}_{j}^K: K = 1,2, \ldots\}$, and the limit is
\begin{equation} \label{1b}
  	g_{j,ALE}(x_j)=\int_{x_{\min,j}}^{x_j} \widetilde{h}(z) dz + \sum_{l=1, u_l\leq {x_j}}^M J_l,  
\end{equation}
where $J_l\equiv \lim_{u\to u_l^{+}}h(u,u_l)-\lim_{u\to u_l^{-}}h(u,u_l)$ is the jump at $u_l$, and $\widetilde{h}(z)$ is defined as $h'(z,z)$ on $\mathcal{S}_j\backslash \mathcal{J}$ and $0$ on $\mathcal{J}$.
\end{thmA.1}
\begin{proof}
For a partition sequence $\{\mathcal{P}_{j}^K\}$ satisfying $\lim_{K\to\infty} \delta_{j,K} = 0$ in Definition \ref{def as limit}, let $\mathcal{K}_K=\{k\in \{1,2,\ldots,K\}:(z_{k-1,j}^K,z_{k,j}^K]\cap J\neq \emptyset\}$, i.e., $\mathcal{K}_K$ is the set of indices of the partition intervals in $\mathcal{P}_{j}^K$ that contain one or more of the discontinuity points in $\mathcal{J}$. The condition $\lim_{K\to\infty} \delta_{j,K} = 0$, together with conditions (\ref{assumption 2}) and (\ref{assumption 3}), imply that $\forall \varepsilon >0, \exists  K_1$ such that $\forall K>K_1$ and $k\in\{1,2,\ldots,K\}\backslash \mathcal{K}_K$,
\begin{align} \label{2b}
\begin{split}
    &\left|\widetilde{h}(z_{k,j}^K)(z_{k,j}^K-z_{k-1,j}^K)-\mathbb{E}[f(z_{k,j}^K,\mathbf{X}_{\backslash j})-f(z_{k-1,j}^K,\mathbf{X}_{\backslash j})|X_j\in (z_{k-1,j}^K,z_{k,j}^K]]\right| \\
    &=\left|h'(z_{k,j}^K,z_{k,j}^K )(z_{k,j}^K-z_{k-1,j}^K)-\frac{\int_{z\in(z_{k-1,j}^K,z_{k,j}^K]} [h(z_{k,j}^K,z)-h(z_{k-1,j}^K,z)]dp_j(z)}{p_j((z_{k-1,j}^K,z_{k,j}^K])}\right|   \\
     &=\frac{(z_{k,j}^K-z_{k-1,j}^K)}{p_j((z_{k-1,j}^K,z_{k,j}^K])} \left|\int_{z\in(z_{k-1,j}^K,z_{k,j}^K]}\left[\frac{h(z_{k,j}^K,z)-h(z_{k-1,j}^K,z)}{(z_{k,j}^K-z_{k-1,j}^K)}-h'(z_{k,j}^K,z_{k,j}^K)\right]dp_j(z)\right|   \\
     &\leq\frac{(z_{k,j}^K-z_{k-1,j}^K)}{p_j ((z_{k-1,j}^K,z_{k,j}^K])} \int_{z\in(z_{k-1,j}^K,z_{k,j}^K]}\left|\frac{h(z_{k,j}^K,z)-h(z_{k-1,j}^K,z)}{(z_{k,j}^K-z_{k-1,j}^K)}-h'(z_{k,j}^K,z_{k,j}^K)\right|dp_j(z)      	\\
     &<\frac{(z_{k,j}^K-z_{k-1,j}^K)}{p_j ((z_{k-1,j}^K,z_{k,j}^K])} \int_{z\in(z_{k-1,j}^K,z_{k,j}^K]}\frac{\varepsilon}{4(x_{\max,j}-x_{\min,j})}dp_j(z)   \\
     &=\frac{(z_{k,j}^K-z_{k-1,j}^K)\varepsilon}{4(x_{\max,j}-x_{\min,j})},
     \end{split}
\end{align}
so that
\begin{align} \label{3b}
\begin{split}
     & \left|\sum_{k=1,k\notin \mathcal{K}_K}^{ k_j^K(x_j)} \widetilde{h}(z_{k,j}^K )(z_{k,j}^K-z_{k-1,j}^K)-\sum_{k=1, k\notin \mathcal{K}_K}^{ k_j^K(x_j)} \mathbb{E}[f(z_{k,j}^K,\mathbf{X}_{\backslash j})-f(z_{k-1,j}^K,\mathbf{X}_{\backslash j} )|X_j\in(z_{k-1,j}^K,z_{k,j}^K]]\right|  \\
     &\leq \sum_{k=1, k\notin \mathcal{K}_K}^{ k_j^K(x_j)}\left|\widetilde{h}(z_{k,j}^K )(z_{k,j}^K-z_{k-1,j}^K )-\mathbb{E}[f(z_{k,j}^K,\mathbf{X}_{\backslash j})-f(z_{k-1,j}^K,\mathbf{X}_{\backslash j})|X_j\in(z_{k-1,j}^K,z_{k,j}^K]]\right|    \\
      &<\sum_{k=1, k\notin 
  \mathcal{K}_K}^{ k_j^K(x_j)}\frac{(z_{k,j}^K-z_{k-1,j}^K)\varepsilon}{4(x_{\max,j}-x_{\min,j})}\leq \frac{\varepsilon}{4}.
  \end{split}
\end{align}
Regarding the partition intervals $(z_{k-1,j}^K,z_{k,j}^K]$ that contain at least one of the nondifferentiable points in $\mathcal{J}$ (i.e., $k\in \mathcal{K}_K$), since $|\mathcal{J}| = M<\infty$, for sufficiently large $K$ each such interval contains only one discontinuity. Let $l=l(k,K)$ denote the index of the point $u_l\in(z_{k-1,j}^K,z_{k,j}^K]$ at which there is a discontinuity.  Then using conditions (\ref{assumption 4}) and (\ref{assumption 5}), $\forall \varepsilon > 0, \exists K_2$ such that $\forall K>K_2$ and $k\in \mathcal{K}_K$, 
\begin{align}\label{4b}
\begin{split}
    &\left|J_l-\mathbb{E}[f(z_{k,j}^K,\mathbf{X}_{\backslash j} )-f(z_{k-1,j}^K,\mathbf{X}_{\backslash j} )|X_j\in(z_{k-1,j}^K,z_{k,j}^K]]\right|  \\
   &=\left|\lim_{u\to u_l^{+}}h(u,u_l)-\lim_{u\to u_l^{-}}h(u,u_l)-\frac{\int_{z\in(z_{k-1,j}^K,z_{k,j}^K]}[h(z_{k,j}^K,z)-h(z_{k-1,j}^K,z)]dp_j(z)}{p_j ((z_{k-1,j}^K,z_{k,j}^K])} \right|   \\
   &\leq\left|\lim_{u\to u_l^{+}}h(u,u_l)-\lim_{u\to u_l^{-}}h(u,u_l)-\frac{\int_{z\in(z_{k-1,j}^K,z_{k,j}^K]}[h(z_{k,j}^K,u_l )-h(z_{k-1,j}^K,u_l)]dp_j(z)}{p_j ((z_{k-1,j}^K,z_{k,j}^K])}\right|\\   
   &+\left|\frac{\int_{z\in(z_{k-1,j}^K,z_{k,j}^K]}{[h(z_{k,j}^K,u_l )-h(z_{k-1,j}^K,u_l )]-[h(z_{k,j}^K,z)-h(z_{k-1,j}^K,z)]}dp_j(z)}{p_j ((z_{k-1,j}^K,z_{k,j}^K]) } \right|\\
    & =\left|\lim_{u\to u_l^{+}}h(u,u_l)-\lim_{u\to u_l^{-}} h(u,u_l )-[h(z_{k,j}^K,u_l )-h(z_{k-1,j}^K,u_l)]\right|  \\ 
   &+ \frac{\left|\int_{z\in(z_{k-1,j}^K,z_{k,j}^K]}{[h(z_{k,j}^K,u_l )-h(z_{k,j}^K,z)]-[h(z_{k-1,j}^K,u_l )-h(z_{k-1,j}^K,z)]}dp_j (z)\right|}{p_j((z_{k-1,j}^K,z_{k,j}^K])} \\
    &\leq\left|\lim_{u\to u_l^{+}}h(u,u_l )-h(z_{k,j}^K,u_l)\right|+\left|h(z_{k-1,j}^K,u_l )-\lim_{u\to u_l^{-}}h(u,u_l)\right|\\ 
   &\quad\quad+\left|\frac{\int_{z\in(z_{k-1,j}^K,z_{k,j}^K]}\left|h(z_{k,j}^K,u_l)-h(z_{k,j}^K,z)\right|dp_j(z)}{p_j ((z_{k-1,j}^K,z_{k,j}^K]) }\right|\\
   &\quad\quad+\left|\frac{\int_{z\in(z_{k-1,j}^K,z_{k,j}^K]}\left|[h(z_{k-1,j}^K,u_l )-h(z_{k-1,j}^K,z)]\right|dp_j(z)}{p_j ((z_{k-1,j}^K,z_{k,j}^K])}\right|\\
   &< \frac{\varepsilon}{16M} +\frac{\varepsilon}{16M}  +\frac{\varepsilon}{16M}\frac{\int_{z\in (z_{k-1,j}^K,z_{k,j}^K]}dp_j(z)}{p_j ((z_{k-1,j}^K,z_{k,j}^K]) }+\frac{\varepsilon}{16M}\frac{\int_{z\in (z_{k-1,j}^K,z_{k,j}^K]}dp_j (z)}{p_j((z_{k-1,j}^K,z_{k,j}^K]) }\\
   &= \frac{\varepsilon}{4M}.
   \end{split}
\end{align}
Consequently, $\forall K>K_2$,
\begin{align}
    \begin{split}\label{5b}
   &\left|\sum_{k=1, k\in \mathcal{K}_K}^{ k_j^K(x_j)}\mathbb{E}[f(z_{k,j}^K,\mathbf{X}_{\backslash j} )-f(z_{k-1,j}^K,\mathbf{X}_{\backslash j} )|X_j\in(z_{k-1,j}^K,z_{k,j}^K]] -\sum_{l=1  ,u_l\leq x_j}^M J_l\right|\\   
	& \leq \sum_{k=1,k\in \mathcal{K}_K}^{ k_j^K(x_j)}\left|\mathbb{E}[f(z_{k,j}^K,\mathbf{X}_{\backslash j} )-f(z_{k-1,j}^K,\mathbf{X}_{\backslash j} )|X_j\in(z_{k-1,j}^K,z_{k,j}^K]]-J_{l(k,K)}\right|\\   
	& <\sum_{k=1, k\in \mathcal{K}_K}^{ k_j^K(x_j)}\frac{\varepsilon}{4M}\leq \frac{\varepsilon}{4}, 		
    \end{split}
\end{align}
Now condition (\ref{assumption 3}) implies that $\widetilde{h}(z)$ is uniformly continuous on $\mathcal{S}_j\backslash\mathcal{J}$. To see this, let $\varepsilon$, $\delta$, $u$, $v$, and $z^{'}$ be as in condition (\ref{assumption 3}). Applying condition (\ref{assumption 3}) twice, first with $z=v$ and second with $z=u$, implies that 
\begin{equation*}
\left|\widetilde{h}(u)-\widetilde{h}(v)\right|\leq\left|\widetilde{h}(u)-\frac{h(u,z^{'} )-h(v,z^{'} )}{u-v}\right|+\left|\widetilde{h}(v)-\frac{h(u,z^{'} )-h(v,z^{'})}{u-v}\right|<2\varepsilon.
\end{equation*}
In other words, $\forall \varepsilon>0$, $\exists\delta>0$ such that $|\widetilde{h}(u)-\widetilde{h}(v)|<2\varepsilon$ whenever $|u-v|<
\delta$ and $(v,u]\cap J=\emptyset$, which is uniform continuity on $\mathcal{S}_j\backslash \mathcal{J}$. This in turn implies that $\widetilde{h}(z)$ is both bounded and Riemann integrable on $\mathcal{S}_j$ (the latter, because a bounded function on a closed interval is Riemann integrable if and only if it is continuous almost everywhere, per Lebesgue measure). \\
By definition of Riemann integrability, $\exists K_3$ such that $\forall K>K_3$, 
\begin{equation}\label{6b}
    \left|\int_{x_{\min,j}}^{x_j} \widetilde{h}(z)dz-\sum_{k=1}^{ k_j^K(x_j)}\widetilde{h}(z_{k,j}^K)(z_{k,j}^K-z_{k-1,j}^K)\right|<\frac{\varepsilon}{4}. 
\end{equation}
In addition, since $|\mathcal{K}_K|\leq |\mathcal{J}|=M<\infty$ and since $\widetilde{h}(z)$ is bounded on $\mathcal{S}_j$, $\exists K_4$ such that $\forall K>K_4$,
\begin{align}\label{7b}
    \begin{split}
 & \left|\sum_{k=1}^{ k_j^K(x_j)} \widetilde{h}(z_{k,j}^K )(z_{k,j}^K-z_{k-1,j}^K)-\sum_{k=1,k\notin \mathcal{K}_K}^{ k_j^K(x_j)}\widetilde{h}(z_{k,j}^K )(z_{k,j}^K-z_{k-1,j}^K)\right| \\
	&=\left|\sum_{k=1, k\in \mathcal{K}_K}^{ k_j^K(x_j)}\widetilde{h}(z_{k,j}^K)(z_{k,j}^K-z_{k-1,j}^K)\right|<\frac{\varepsilon}{4}.
    \end{split}
\end{align}
Thus, combining \eqref{3b}, \eqref{5b}, \eqref{6b}, and \eqref{7b}, $\forall K>\max\{K_1,K_2,K_3,K_4\}$
{\allowdisplaybreaks
\begin{align*}
    &\left|\sum_{k=1}^{ k_j^K(x_j)}\mathbb{E}[f(z_{k,j}^K,\mathbf{X}_{\backslash j})-f(z_{k-1,j}^K,\mathbf{X}_{\backslash j} )|X_j\in(z_{k-1,j}^K,z_{k,j}^K]] -\int_{x_{\min,j}}^{x_j}\widetilde{h}(z)dz+\sum_{l=1,u_l\leq x_j}^M J_l\right|\\
    &\leq\left|\sum_{k=1, k\in\mathcal{K}_K}^{k_j^K(x_j)}\mathbb{E}[f(z_{k,j}^K,\mathbf{X}_{\backslash j} )-f(z_{k-1,j}^K,\mathbf{X}_{\backslash j} )|X_j\in (z_{k-1,j}^K,z_{k,j}^K]] -\sum_{l=1  ,u_l\leq x_j}^M J_l\right|\\
    & \quad +\left|\sum_{k=1,k\notin \mathcal{K}_K}^{k_j^K(x_j)}\widetilde{h}(z_{k,j}^K )(z_{k,j}^K-z_{k-1,j}^K)-\sum_{k=1,k\notin \mathcal{K}_K}^{k_j^K(x_j)}\mathbb{E}[f(z_{k,j}^K,\mathbf{X}_{\backslash j})-f(z_{k-1,j}^K,\mathbf{X}_{\backslash j} )|X_j\in(z_{k-1,j}^K,z_{k,j}^K]] \right|  \\
    &\quad +\left|\int_{x_{\min,j}}^{x_j}\widetilde{h} (z)dz-\sum_{k=1}^{k_j^K(x_j)}\widetilde{h}(z_{k,j}^K )(z_{k,j}^K-z_{k-1,j}^K)\right|\\
    &\quad +\left|\sum_{k=1}^{k_j^K(x_j)}\widetilde{h}(z_{k,j}^K)(z_{k,j}^K-z_{k-1,j}^K)-\sum_{k=1,k\notin K_K }^{k_j^K(x_j)}\widetilde{h}(z_{k,j}^K )(z_{k,j}^K-z_{k-1,j}^K)\right|  \\
    &<\frac{\varepsilon}{4}+\frac{\varepsilon}{4}+\frac{\varepsilon}{4}+\frac{\varepsilon}{4}=\varepsilon.  
\end{align*}}
Since $\varepsilon$ was arbitrary, this shows that the limit in \eqref{ALE uncentered main def as limit} exists and is equal to \eqref{1b} .

\end{proof}
\begin{thm} [\textbf{Uncentered ALE Main Effect for differentiable $f(\cdot)$}]\label{Theorem1} Let $f^j(x_j,\mathbf{x}_{\backslash j}) \equiv \frac{\partial{f(x_j,\mathbf{x}_{\backslash j})}} {\partial{x_j}}$ denote the partial derivative of $f(\mathbf{x})$ with respect to $x_j$ when the derivative exists. In Definition \ref{def as limit}, assume that the limit in \eqref{ALE uncentered main def as limit} exists independent of the sequence of partitions (e.g., if the conditions of Theorem A.1 are satisfied). If in addition, 
\begin{enumerate} [(i)]
    \item $f(x_j,\mathbf{x}_{\backslash j})$ is differentiable in $x_j$ on $\mathcal{S}$,
    \item $f^j(x_j,\mathbf{x}_{\backslash j})$ is continuous in $(x_j,\mathbf{x}_{\backslash j})$ on $\mathcal{S}$, and
    \item  $\mathbb{E}[f^j(X_j,\mathbf{X}_{\backslash j})|X_j=z_j]$ is continuous in $z_j$ on $\mathcal{S}_j$,
\end{enumerate}then for each $x_j \in \mathcal{S}_j$. 
\begin{equation*}
    	g_{j,ALE}(x_j)=\int_{x_{\min,j}}^{x_j} \mathbb{E}[f^j(X_j,\mathbf{X}_{\backslash j} )|X_j=z_j]dz_j.	
\end{equation*}

\end{thm}
\begin{proof}
Let $\{\mathcal{P}_{j}^K: K=1,2,\ldots\}$ denote any sequence of partitions in Definition \ref{def as limit}. Since $\mathbb{E}[f^j(X_j,\mathbf{X}_{\backslash j})|X_j=z_j]$ is continuous in $z_j$, it is Riemann integrable on $\mathcal{S}_j$. By definition of Riemann-integrability, for any $\varepsilon>0$ there exists a $K_1(\varepsilon)$ such that for all $K>K_1(\varepsilon)$,  
\begin{align} \label{pf thm differentiable eq 1}
\begin{split}
  	\Bigg|\int_{x_{\min,j}}^{x_j} \mathbb{E}[f^j(X_j,\mathbf{X}_{\backslash j} )\big|X_j=z_j]dz_j&-\sum_{k=1}^{ k_j^K(x_j)} \mathbb{E}[f^j(X_j,\mathbf{X}_{\backslash j} )\big|X_j=z_{k,j}^K](z_{k,j}^K-z_{k-1,j}^K)\Bigg| \\
  	&\quad\quad<\frac{\varepsilon}{4}.  
 \end{split}
\end{align}
Notice that in \eqref{pf thm differentiable eq 1}, the upper limit $x_j$ in the integral does not necessarily coincide exactly with the upper limit $z_{k,j}^K$ in the Riemann sum when $k= k_j^K(x_j)$. However, $x_j$ and $z_{k,j}^K$ for $k = k_j^K(x_j)$ must become arbitrarily close as $K\to\infty$, so that \eqref{pf thm differentiable eq 1} still holds. This follows because the continuity of $f^j (x_j,\mathbf{x}_{\backslash j})$ on the compact $\mathcal{S}$ implies that $\mathbb{E}[f^j(X_j,\mathbf{X}_{\backslash j})|X_j=z_j]$ is bounded, so that for $k = k_j^K(x_j)$, we have $\int_{x_j}^{z_{k,j}^K}\mathbb{E}[f^j (X_j,\mathbf{X}_{\backslash j})|X_j=z_j]dz_j\to 0$ as $K\to\infty$.
\par

By Definition \ref{def as limit}, there also exists a $K_2(\varepsilon)$ such that for all $K>K_2(\varepsilon)$
\begin{equation}\label{pf thm differentiable eq 2}
    	\left|g_{j,ALE}(x_j)-\sum_{k=1}^{ k_j^K(x_j)}\mathbb{E}[f(z_{k,j}^K,\mathbf{X}_{\backslash j} )-f(z_{k-1,j}^K,\mathbf{X}_{\backslash j} )|X_j\in (z_{k-1,j}^K,z_{k,j}^K]] \right|<\frac{\varepsilon}{4}.
\end{equation}
If we can show that the summations in \eqref{pf thm differentiable eq 1} and \eqref{pf thm differentiable eq 2} are within $\frac{\varepsilon}{2}$ of each other for sufficiently large $K$, then combining this with \eqref{pf thm differentiable eq 1} and \eqref{pf thm differentiable eq 2} will imply that
\begin{equation*}
    \left|g_{j,ALE}(x_j)-\int_{x_{\min,j}}^{x_j}\mathbb{E}[f^j(X_j,\mathbf{X}_{\backslash j} )|X_j=z_j]dz_j\right|<\varepsilon
\end{equation*}
for sufficiently large $K$, and the proof will be complete. Towards this end, write the summand in \eqref{pf thm differentiable eq 2} as
\begin{align} \label{pf thm differentiable eq 3}
\begin{split}
     &\mathbb{E}[f(z_{k,j}^K,\mathbf{X}_{\backslash j} )-f(z_{k-1,j}^K,\mathbf{X}_{\backslash j})|X_j \in (z_{k-1,j}^K,z_{k,j}^K]]\\
     &=\frac{1}{p_j ((z_{k-1,j}^K,z_{k,j}^K])} \int_{z_j\in (z_{k-1,j}^K,z_{k,j}^K]} \mathbb{E}[f(z_{k,j}^K,\mathbf{X}_{\backslash j} )-f(z_{k-1,j}^K,\mathbf{X}_{\backslash j} )|X_j=z_j]dp_j(z_j)\\
     & =\frac{1}{p_j ((z_{k-1,j}^K,z_{k,j}^K])} \int_{z_j\in (z_{k-1,j}^K,z_{k,j}^K]}\int_{\mathbf{x}_{\backslash j}}[f(z_{k,j}^K,\mathbf{x}_{\backslash j} )-f(z_{k-1,j}^K,\mathbf{x}_{\backslash j} )]dp_{\backslash j|j}(\mathbf{x}_{\backslash j} |z_j)dp_j (z_j) \\
     &=\frac{(z_{k,j}^K-z_{k-1,j}^K)}{p_j ((z_{k-1,j}^K,z_{k,j}^K])} \int_{z_j\in (z_{k-1,j}^K,z_{k,j}^K]}\int_{\mathbf{x}_{\backslash j}}   \frac{[f(z_{k,j}^K,\mathbf{x}_{\backslash j} )-f(z_{k-1,j}^K,\mathbf{x}_{\backslash j} )]}{(z_{k,j}^K-z_{k-1,j}^K )} dp_{\backslash j|j}(\mathbf{x}_{\backslash j}|z_j)dp_j(z_j) \\
     &=\frac{(z_{k,j}^K-z_{k-1,j}^K )}{p_j((z_{k-1,j}^K,z_{k,j}^K])} \int_{z_j\in (z_{k-1,j}^K,z_{k,j}^K]}\int_{\mathbf{x}_{\backslash j}}f^j (z(k,K,\mathbf{x}_{\backslash j} ),\mathbf{x}_{\backslash j} )dp_{\backslash j|j} (\mathbf{x}_{\backslash j} |z_j)dp_j (z_j),	
 \end{split}
\end{align}
for some $z(k,K,\mathbf{x}_{\backslash j})\in (z_{k-1,j}^K,z_{k,j}^K]$, where the last equality in \eqref{pf thm differentiable eq 3} follows by the mean value theorem applied to $f(x_j,\mathbf{x}_{\backslash j})$ over the interval $x_j\in(z_{k-1,j}^K,z_{k,j}^K]$. In addition, by the continuity of $f^j(x_j,\mathbf{x}_{\backslash j})$ on $\mathcal{S}$ (which implies uniform continuity since $\mathcal{S}$ is compact), there exists a $K_3(\varepsilon)$ such that for all $K>K_3 (\varepsilon)$ and for all $k$, $\mathbf{x}_{\backslash j}$, and $z_j\in (z_{k-1,j}^K,z_{k,j}^K]$
\begin{equation}\label{pf thm differentiable eq 4}
    \left|f^j(z(k,K,\mathbf{x}_{\backslash j}),\mathbf{x}_{\backslash j} )-f^j(z_j,\mathbf{x}_{\backslash j})\right|< \frac{\varepsilon}{4(x_{\max,j}-x_{\min,j})}.
\end{equation}
And by the assumed continuity of $\mathbb{E}[f^j(X_j,\mathbf{X}_{\backslash j} )|X_j=z_j]$ on $\mathcal{S}_j$ (which again implies uniform continuity on the compact $\mathcal{S}_j$), there exists a $K_4(\varepsilon)$ such that for all $K>K_4(\varepsilon)$ and for all $k$ and $z_j\in (z_{k-1,j}^K,z_{k,j}^K]$ 
\begin{equation}\label{pf thm differentiable eq 5}
    \left|\mathbb{E}[f^j(X_j,\mathbf{X}_{\backslash j})|X_j=z_j]-\mathbb{E}[f^j(X_j,\mathbf{X}_{\backslash j})|X_j=z_{k,j}^K ]\right|< \frac{\varepsilon}{4(x_{\max,j}-x_{\min,j})}.	
\end{equation}
Using \eqref{pf thm differentiable eq 4} and \eqref{pf thm differentiable eq 5}, for all $K>\max\{K_3(\varepsilon),K_4(\varepsilon)\}$ and for all $k$ and $z_j\in(z_{k-1,j}^K,z_{k,j}^K]$, 
\begin{align} \label{pf thm differentiable eq 6}
    \begin{split}
      & \left|\int_{\mathbf{x}_{\backslash j}}f^j(z(k,K,\mathbf{x}_{\backslash j} ),\mathbf{x}_{\backslash j} )dp_{\backslash j|j} (\mathbf{x}_{\backslash j}|z_j) - \mathbb{E}[f^j (X_j,\mathbf{X}_{\backslash j} )|X_j=z_{k,j}^K ]\right|\\
      &\leq\left|\int_{\mathbf{x}_{\backslash j}}f^j (z(k,K,\mathbf{x}_{\backslash j}),\mathbf{x}_{\backslash j} )dp_{\backslash j|j} (\mathbf{x}_{\backslash j}|z_j) -\int_{\mathbf{x}_{\backslash j}}f^j(z_j,\mathbf{x}_{\backslash j})dp_{\backslash j|j}(\mathbf{x}_{\backslash j}|z_j)\right|\\
      & +\left|\int_{\mathbf{x}_{\backslash j}}f^j (z_j,\mathbf{x}_{\backslash j} )dp_{\backslash j|j} (\mathbf{x}_{\backslash j}|z_j) -\mathbb{E}[f^j(X_j,\mathbf{X}_{\backslash j} )|X_j=z_{k,j}^K ]\right|\\
       &\leq \int_{\mathbf{x}_{\backslash j}}|f^j (z(k,K,\mathbf{x}_{\backslash j} ),\mathbf{x}_{\backslash j} )-f^j(z_j,\mathbf{x}_{\backslash j} )|dp_{\backslash j|j} (\mathbf{x}_{\backslash j}|z_j) \\
       & +\left|\mathbb{E}[f^j(X_j,\mathbf{X}_{\backslash j} )|X_j=z_j]-\mathbb{E}[f^j (X_j,\mathbf{X}_{\backslash j})|X_j=z_{k,j}^K]\right|\\
	  &\leq\int_{\mathbf{x}_{\backslash j}}\frac{\varepsilon}{4(x_{\max,j}-x_{\min,j} )} dp_{\backslash j|j} (\mathbf{x}_{\backslash j} |z_j) +\frac{\varepsilon}{4(x_{\max,j}-x_{\min,j})} =\frac{\varepsilon}{2(x_{\max,j}-x_{\min,j})}.
    \end{split}
\end{align}
Thus, using \eqref{pf thm differentiable eq 3} and \eqref{pf thm differentiable eq 6}, for all $K>\max\{K_3(\varepsilon),K_4(\varepsilon)\}$ the difference between the summations in \eqref{pf thm differentiable eq 1} and \eqref{pf thm differentiable eq 2} satisfy
{\allowdisplaybreaks
\begin{align*}
    &\Bigg|\sum_{k=1}^{ k_j^K(x_j)}\mathbb{E}[f(z_{k,j}^K,\mathbf{X}_{\backslash j})-f(z_{k-1,j}^K,\mathbf{X}_{\backslash j} )|X_j\in(z_{k-1,j}^K,z_{k,j}^K]]\\
    &- \sum_{k=1}^{ k_j^K(x_j)}\mathbb{E}[f^j(X_j,\mathbf{X}_{\backslash j})|X_j=z_{k,j}^K](z_{k,j}^K-z_{k-1,j}^K )\Bigg|\\
    &\leq \sum_{k=1}^{ k_j^K(x_j)}\Big|\mathbb{E}[f(z_{k,j}^K,\mathbf{X}_{\backslash j} )-f(z_{k-1,j}^K,\mathbf{X}_{\backslash j} )|X_j\in(z_{k-1,j}^K,z_{k,j}^K]\\ 
	&-\mathbb{E}[f^j(X_j,\mathbf{X}_{\backslash j} )|X_j=z_{k,j}^K ](z_{k,j}^K-z_{k-1,j}^K)\Big|\\
	&=\sum_{k=1}^{ k_j^K(x_j)}\Bigg|\frac{(z_{k,j}^K-z_{k-1,j}^K )}{p_j((z_{k-1,j}^K,z_{k,j}^K])} \int_{z_j\in(z_{k-1,j}^K,z_{k,j}^K]}\int_{\mathbf{x}_{\backslash j}}f^j(z(k,K,\mathbf{x}_{\backslash j} ),\mathbf{x}_{\backslash j})dp_{\backslash j|j}(\mathbf{x}_{\backslash j}|z_j)dp_j(z_j)\\ 
    & - \mathbb{E}[f^j(X_j,\mathbf{X}_{\backslash j} )|X_j=z_{k,j}^K](z_{k,j}^K-z_{k-1,j}^K)\Bigg|\quad\quad(\text{using \eqref{pf thm differentiable eq 3}})\\
	&=\sum_{k=1}^{ k_j^K(x_j)}(z_{k,j}^K-z_{k-1,j}^K )\\
	&\times\left|\dfrac{\int_{z_j\in (z_{k-1,j}^K,z_{k,j}^K]}\{\int_{\mathbf{x}_{\backslash j}}f^j (z(k,K,\mathbf{x}_{\backslash j}) ),\mathbf{x}_{\backslash j} )dp_{\backslash j|j}(\mathbf{x}_{\backslash j}|z_j) -\mathbb{E}[f^j(X_j,\mathbf{X}_{\backslash j} )|X_j=z_{k,j}^K ]\}dp_j(z_j)}{p_j ((z_{k-1,j}^K,z_{k,j}^K])}\right| \\
	&\leq \sum_{k=1}^{ k_j^K(x_j))}(z_{k,j}^K-z_{k-1,j}^K ) \times\\
	&\dfrac{\int_{z_j\in (z_{k-1,j}^K,z_{k,j}^K]}\left|\int_{\mathbf{x}_{\backslash j}}f^j(z(k,K,\mathbf{x}_{\backslash j} ),\mathbf{x}_{\backslash j})dp_{\backslash j|j}(\mathbf{x}_{\backslash j}|z_j)-\mathbb{E}[f^j (X_j,\mathbf{X}_{\backslash j})|X_j=z_{k,j}^K ]\right|dp_j(z_j)}{p_j((z_{k-1,j}^K,z_{k,j}^K])}\\
	&\leq \sum_{k=1}^{ k_j^K(x_j)}(z_{k,j}^K-z_{k-1,j}^K)  \frac{\int_{z_j\in (z_{k-1,j}^K,z_{k,j}^K]}\frac{\varepsilon}{2(x_{\max,j}-x_{\min,j})}dp_j (z_j)}{p_j ((z_{k-1,j}^K,z_{k,j}^K])} \quad\quad(\text{using \eqref{pf thm differentiable eq 6}})\\
	&=\sum_{k=1}^{ k_j^K(x_j)}(z_{k,j}^K-z_{k-1,j}^K)  \frac{\varepsilon p_j ((z_{k-1,j}^K,z_{k,j}^K])}{2(x_{\max,j}-x_{\min,j} ) p_j((z_{k-1,j}^K,z_{k,j}^K]) }\\
	&=\frac{\varepsilon}{2(x_{\max,j}-x_{\min,j})}  \sum_{k=1}^{ k_j^K(x_j)}(z_{k,j}^K-z_{k-1,j}^K)
\end{align*}}
\begin{flalign}\label{pf thm differentiable eq 7}
&\quad\leq \frac{\varepsilon(x_{\max,j}-x_{\min,j} )}{2(x_{\max,j}-x_{\min,j})} =\frac{\varepsilon}{2}.&	
\end{flalign}
Finally, combining \eqref{pf thm differentiable eq 1},\eqref{pf thm differentiable eq 2}, and \eqref{pf thm differentiable eq 7} , for all $K>\max\{K_1(\varepsilon),K_2(\varepsilon),K_3(\varepsilon),K_4(\varepsilon)\}$,
{\allowdisplaybreaks
\begin{align*}
     &\left|g_{j,ALE}(x_j)-\int_{x_{\min,j}}^{x_j} \mathbb{E}[f^j(X_j,\mathbf{X}_{\backslash j} )|X_j=z_j]dz_j\right| \\
     &\leq \left|g_{j,ALE}(x_j)-\sum_{k=1}^{ k_j^K(x_j)}\mathbb{E}[f(z_{k,j}^K,\mathbf{X}_{\backslash j} )-f(z_{k-1,j}^K,\mathbf{X}_{\backslash j} )|X_j\in(z_{k-1,j}^K,z_{k,j}^K]] \right|\\
	&+\left|\int_{x_{\min,j}}^{x_j} \mathbb{E}[f^j(X_j,\mathbf{X}_{\backslash j})|X_j=z_j]dz_j-\sum_{k=1}^{ k_j^K(x_j)}\mathbb{E}[f^j(X_j,\mathbf{X}_{\backslash j})|X_j=z_{k,j}^K](z_{k,j}^K-z_{k-1,j}^K)\right| \\  
	&+\Bigg|\sum_{k=1}^{ k_j^K(x_j)}\mathbb{E}[f(z_{k,j}^K,\mathbf{X}_{\backslash j} )-f(z_{k-1,j}^K,\mathbf{X}_{\backslash j} )|X_j\in(z_{k-1,j}^K,z_{k,j}^K]]\\
	       & \quad\quad-\sum_{k=1}^{ k_j^K(x_j)}\mathbb{E}[f^j(X_j,\mathbf{X}_{\backslash j} )|X_j=z_{k,j}^K ](z_{k,j}^K-z_{k-1,j}^K)\Bigg|\\
	&\leq \frac{\varepsilon}{4}+\frac{\varepsilon}{4}+\frac{\varepsilon}{2}=\varepsilon,  
\end{align*}}
which completes the proof.
\end{proof}
\begin{thm}[\textbf{Uncentered ALE Second-Order Effect for differentiable $f(\cdot)$}] \label{Theorem2}Let\\ $f^{\{j,l\}}(x_j,x_l,\mathbf{x}_{\backslash\{j,l\}})\equiv \frac{\partial^2 f(x_j,x_l,\mathbf{x}_{\backslash\{j,l\}})}{\partial x_j \partial x_l}$ denote the second-order partial derivative of $f(\mathbf{x})$ with respect to $x_j$ and $x_l$ when the derivative exists. In Definition \ref{def 2nd as limit}, assume that the limit in \eqref{ALE uncentered second def as limit} exists independent of the sequences of partitions. If in addition, 
\begin{enumerate} [(i)]
    \item $f(x_j, x_l, \mathbf{x}_{\backslash\{j,l\}})$ is differentiable in $(x_j, x_l)$ on $\mathcal{S}$,
    \item $f^{\{j,l\}}(x_j, x_l, \mathbf{x}_{\backslash \{j,l\}})$ is continuous in $(x_j, x_l, \mathbf{x}_{\backslash\{j,l\}})$ on $\mathcal{S}$, and
    \item  $\mathbb{E}[f^{\{j,l\}}(X_j,X_l,\mathbf{X}_{\backslash \{j,l\}})|X_j=z_j, X_l = z_l]$ is continuous in $(z_j, z_l)$ on $\mathcal{S}_j\times\mathcal{S}_l$,
\end{enumerate} then, for each $(x_j, x_l) \in \mathcal{S}_j\times \mathcal{S}_l$, 
\begin{equation*}
    h_{\{j,l\},ALE}(x_j,x_l)\equiv \int_{x_{\min,l}}^{x_l} \int_{x_{\min,j}}^{x_j} \mathbb{E}[f^{\{j,l\}}(X_j,X_l,\mathbf{X}_{\backslash\{j,l\}})|X_j=z_j,X_l=z_l]dz_j dz_l.
\end{equation*}
\end{thm}
\begin{proof}
This proof follows the same general course as the proof of Theorem \ref{Theorem1}. Let $\{\mathcal{P}_{j}^K: K=1,2,\ldots\}$ and $\{\mathcal{P}_{l}^K: K=1,2,\ldots\}$ denote any two sequences of partitions of $\mathcal{S}_j$ and $\mathcal{S}_l$ in Definition \ref{def 2nd as limit}. Since $\mathbb{E}[f^{\{j,l\}}(X_j,X_l,\mathbf{X}_{\backslash \{j,l\}})|X_j=z_j, X_l = z_l]$ is continuous in $(z_j, z_l)$, it is Riemann integrable on $\mathcal{S}_j\times\mathcal{S}_l$. By definition of Riemann-integrability of real-valued functions defined on rectangular domains, for any $\varepsilon>0$ there exists a $K_1(\varepsilon)$ such that for all $K>K_1(\varepsilon)$,  
\begin{align} \label{pf thm differentiable 2 eq 1}
\begin{split}
  	&\Bigg|\int_{x_{\min,l}}^{x_l}\int_{x_{\min,j}}^{x_j} \mathbb{E}[f^{\{j,l\}}(X_j,X_l,\mathbf{X}_{\backslash \{j,l\}} )|X_j=z_j,X_l = z_l]dz_jdz_l\\
  	&-\sum_{k=1}^{ k_j^K(x_j)}\sum_{m=1}^{ k_l^K(x_l)} \mathbb{E}[f^{\{j,l\}}(X_j,X_l,\mathbf{X}_{\backslash \{j,l\}} )|X_j=z_{k,j}^K,X_l=z_{m,l}^K](z_{k,j}^K-z_{k-1,j}^K)(z_{m,l}^K-z_{m-1,l}^K)\Bigg|\\
  	&<\frac{\varepsilon}{4}.  
 \end{split}
\end{align}
 By similar argument as in Theorem \ref{Theorem1}, \eqref{pf thm differentiable 2 eq 1} holds even though the upper limits $x_j$ and $x_l$ in the double integral do not necessarily coincide exactly with the upper limits $z_{k,j}^K$ for $k= k_j^K(x_j)$ and $z_{m,l}^K$ for $m= k_l^K(x_l)$ in the Riemann sum. This follows because the continuity of $f^{\{j,l\}} (x_j,x_l,\mathbf{x}_{\backslash \{j,l\}})$ on the compact $\mathcal{S}$ implies that $\mathbb{E}[f^{\{j,l\}}(X_j,X_l,\mathbf{X}_{\backslash \{j,l\}})|X_j=z_j, X_l = z_l]$ is bounded, and because the two corresponding upper limits for each predictor become arbitrarily close as $K\to\infty$. Thus, for $k = k_j^K(x_j)$ and $m = k_l^K(x_l)$, it follows that \[\int_{x_l}^{z_{m,l}^K}\int_{x_j}^{z_{k,j}^K}\mathbb{E}[f^{\{j,l\}} (X_j,X_l,\mathbf{X}_{\backslash \{j,l\}})|X_j=z_j,X_l = z_l]dz_jdz_l\to 0\] as $K\to\infty$. \par
By Definition \ref{def 2nd as limit}, there exists a $K_2(\varepsilon)$ such that for all $K>K_2(\varepsilon)$
\begin{equation}\label{pf thm differentiable 2 eq 2}
    	\Bigg|h_{\{j,l\},ALE}(x_j,x_l)-\sum_{k=1}^{ k_j^K(x_j)}\sum_{m=1}^{ k_l^K(x_l)}\mathbb{E}[\Delta_f^{\{j,l\}}(K,k,m;\mathbf{X}_{\backslash \{j,l\}})|X_j\in (z_{k-1,j}^K,z_{k,j}^K], X_l\in (z_{m-1,l}^K,z_{m,l}^K]] \Bigg|< \frac{\varepsilon}{4}.
\end{equation}
The remainder of the proof shows that the summations in \eqref{pf thm differentiable 2 eq 1} and \eqref{pf thm differentiable 2 eq 2} are within $\frac{\varepsilon}{2}$ of each other for sufficiently large $K$, which, when combined with \eqref{pf thm differentiable 2 eq 1} and \eqref{pf thm differentiable 2 eq 2}, proves the desired result that $|h_{\{j,l\},ALE}(x_j,x_l)- \int_{x_{\min,l}}^{x_l} \int_{x_{\min,j}}^{x_j} \mathbb{E}[f^{\{j,l\}}(X_j,X_l,\mathbf{X}_{\backslash\{j,l\}})|X_j=z_j,X_l=z_l]dz_j dz_l|<\varepsilon$ for sufficiently large $K$.\\
The summand in \eqref{pf thm differentiable 2 eq 2} can be written as
\begin{align} \label{pf thm differentiable 2 eq 3}
\begin{split}
     &\mathbb{E}[\Delta_f^{\{j,l\}}(K,k,m;\mathbf{X}_{\backslash \{j,l\}})|X_j\in (z_{k-1,j}^K,z_{k,j}^K], X_l\in (z_{m-1,l}^K,z_{m,l}^K]] \\
     &=\frac{\int_{(z_j,z_l)\in (z_{k-1,j}^K,z_{k,j}^K] \times (z_{m-1,l}^K,z_{m,l}^K]} \mathbb{E}[\Delta_f^{\{j,l\}}(K,k,m;\mathbf{X}_{\backslash \{j,l\}})|X_j=z_j, X_l=z_l]dp_{\{j,l\}}(z_j,z_l)}{p_{\{j,l\}}((z_{k-1,j}^K,z_{k,j}^K]\times(z_{m-1,l}^K,z_{m,l}^K])} \\
     &=\frac{\int_{(z_j,z_l)\in (z_{k-1,j}^K,z_{k,j}^K] \times (z_{m-1,l}^K,z_{m,l}^K]} \int_{\mathbf{x}_{\backslash \{j,l\}}}\Delta_f^{\{j,l\}}(K,k,m;\mathbf{x}_{\backslash \{j,l\}})dp_{\backslash \{j,l\}|\{j,l\}}(\mathbf{x}_{\backslash \{j,l\}} |z_j, z_l) dp_{\{j,l\}}(z_j,z_l)} {p_{\{j,l\}}((z_{k-1,j}^K,z_{k,j}^K]\times(z_{m-1,l}^K,z_{m,l}^K])} \\
     &=\frac{(z_{k,j}^K-z_{k-1,j}^K)(z_{m,l}^K-z_{m-1,l}^K)}{p_{\{j,l\}}((z_{k-1,j}^K,z_{k,j}^K]\times(z_{m-1,l}^K,z_{m,l}^K])}\times\\
     &\int_{(z_j,z_l)\in (z_{k-1,j}^K,z_{k,j}^K] \times (z_{m-1,l}^K,z_{m,l}^K]}
     \int_{\mathbf{x}_{\backslash \{j,l\}}}\frac{\Delta_f^{\{j,l\}}(K,k,m;\mathbf{x}_{\backslash \{j,l\}})}{(z_{k,j}^K-z_{k-1,j}^K)(z_{m,l}^K-z_{m-1,l}^K)}dp_{\backslash \{j,l\}|\{j,l\}}(\mathbf{x}_{\backslash \{j,l\}} |z_j, z_l) dp_{\{j,l\}}(z_j,z_l)\\
     &=\frac{(z_{k,j}^K-z_{k-1,j}^K)(z_{m,l}^K-z_{m-1,l}^K)}{p_{\{j,l\}}((z_{k-1,j}^K,z_{k,j}^K]\times(z_{m-1,l}^K,z_{m,l}^K])}\times\\
     &\int_{(z_j,z_l)\in (z_{k-1,j}^K,z_{k,j}^K] \times (z_{m-1,l}^K,z_{m,l}^K]}
     \int_{\mathbf{x}_{\backslash \{j,l\}}}f^{\{j,l\}}(\mathbf{z}(k,m,K,\mathbf{x}_{\backslash\{j,l\}}),\mathbf{x}_{\backslash\{j,l\}})dp_{\backslash \{j,l\}|\{j,l\}}(\mathbf{x}_{\backslash \{j,l\}} |z_j, z_l) dp_{\{j,l\}}(z_j,z_l),	
 \end{split}
\end{align}
for some $\mathbf{z}(k,m, K,\mathbf{x}_{\backslash \{j,l\}}) \equiv (z_1(k,K, \mathbf{x}_{j}), z_2(k,m, K,\mathbf{x}_{\backslash \{j,l\}})) \in (z_{k-1,j}^K,z_{k,j}^K]\times (z_{m-1,l}^K,z_{m,l}^K]$, where the last equality in \eqref{pf thm differentiable 2 eq 3} follows by applying the mean value theorem twice to $f(x_j, x_l, \mathbf{x}_{\backslash \{j,l\}})$ over the rectangle $(x_j,x_l) \in(z_{k-1,j}^K,z_{k,j}^K]\times(z_{m-1,l}^K,z_{m,l}^K]$. Specifically, the first application of the mean value theorem gives
\begin{equation} \label{1st mvt eq}
 \frac{f(z_{k,j}^K,\mathbf{x}_{\backslash j})-f(z_{k-1,j}^K,\mathbf{x}_{\backslash j})}{z_{k,j}^K-z_{k-1,j}^K} = f^j(z_1(k,K,\mathbf{x}_{\backslash j}),\mathbf{x}_{\backslash j})   
\end{equation} for some $z_1(k,K,\mathbf{x}_{\backslash j})\in(z_{k-1,j}^K,z_{k,j}^K]$. A second application of the mean value theorem to \eqref{1st mvt eq} gives
{\allowdisplaybreaks
\begin{align*}
     &\frac{\Delta_f^{\{j,l\}}(K,k,m;\mathbf{x}_{\backslash \{j,l\}})}{(z_{k,j}^K-z_{k-1,j}^K)(z_{m,l}^K-z_{m-1,l}^K)} \\
     &= \frac{\frac{f(z_{k,j}^K,z_{m,l}^K,\mathbf{x}_{\backslash \{j,l\}})-f(z_{k-1,j}^K,z_{m,l}^K,\mathbf{x}_{\backslash \{j,l\}})}{z_{k,j}^K-z_{k-1,j}^K} -\frac{f(z_{k,j}^K,z_{m-1,l}^K,\mathbf{x}_{\backslash \{j,l\}})-f(z_{k-1,j}^K,z_{m-1,l}^K,\mathbf{x}_{\backslash \{j,l\}})}{z_{k,j}^K-z_{k-1,j}^K} }{z_{m,l}^K-z_{m-1,l}^K} \\
     &= f^{\{j,l\}}(z_1(k,K,\mathbf{x}_{\backslash j}), z_2(k,m, K,\mathbf{x}_{\backslash \{j,l\}}),\mathbf{x}_{\backslash \{j,l\}})
\end{align*}}
for some $z_2(k,m, K,\mathbf{x}_{\backslash \{j,l\}})\in (z_{m-1,l}^K,z_{m,l}^K]$.\\
Further, by the continuity of $f^{\{j,l\}}(x_j,x_l,\mathbf{x}_{\backslash \{j,l\}})$ (which implies uniform continuity on the compact $\mathcal{S}$), there exists a $K_3(\varepsilon)$ such that for all $K>K_3(\varepsilon)$ and for all $k$, $m$, $\mathbf{x}_{\backslash \{j,l\}}$, $z_j \in (z_{k-1,j}^K,z_{k,j}^K]$, and $z_l\in (z_{m-1,l}^K,z_{m,l}^K]$,
\begin{equation}\label{pf thm differentiable 2 eq 4}
    \Big|f^{\{j,l\}}(\mathbf{z}(k,m,K,\mathbf{x}_{\backslash \{j,l\}}),\mathbf{x}_{\backslash \{j,l\}} )-f^{\{j,l\}}(z_j,z_l,\mathbf{x}_{\backslash \{j,l\}})\Big|< \frac{\varepsilon}{4(x_{\max,j}-x_{\min,j})(x_{\max,l}-x_{\min,l})}.
\end{equation}
Likewise, by the assumed continuity of $\mathbb{E}[f^{\{j,l\}}(X_j,X_l,\mathbf{X}_{\backslash \{j,l\}} )|X_j=z_j, X_l = z_l]$ on the compact set $\mathcal{S}_j\times\mathcal{S}_l$ (which implies uniform continuity on this set), there exists a $K_4(\varepsilon)$ such that for all $K>K_4(\varepsilon)$ and for all $k$, $m$, $z_j\in (z_{k-1,j}^K,z_{k,j}^K]$, and $z_l\in (z_{m-1,l}^K,z_{m,l}^K]$, 
\begin{equation}\label{pf thm differentiable 2 eq 5}
\begin{split}
     \Big|\mathbb{E}[f^{\{j,l\}}(X_j,X_l,\mathbf{X}_{\backslash \{j,l\}})|X_j=z_j,X_l = z_l]&-\mathbb{E}[f^{\{j,l\}}(X_j,X_l,\mathbf{X}_{\backslash \{j,l\}})|X_j=z_{k,j}^K, X_l = z_{m,l}^K ]\Big|\\
     &< \frac{\varepsilon}{4(x_{\max,j}-x_{\min,j})(x_{\max,l}-x_{\min,l})}.
\end{split}
\end{equation}
Using \eqref{pf thm differentiable 2 eq 4} and \eqref{pf thm differentiable 2 eq 5}, for all $K>\max\{K_3(\varepsilon), K_4(\varepsilon)\}$ and for all $k$, $m$, $z_j\in(z_{k-1,j}^K,z_{k,j}^K]$, and $z_l\in (z_{m-1,l}^K,z_{m,l}^K]$, 
\begin{align} \label{pf thm differentiable 2 eq 6}
    \begin{split}
      & \Bigg| \int_{\mathbf{x}_{\backslash \{j,l\}}}f^{\{j,l\}}(\mathbf{z}(k,m,K,\mathbf{x}_{\backslash\{j,l\}}),\mathbf{x}_{\backslash\{j,l\}})dp_{\backslash \{j,l\}|\{j,l\}}(\mathbf{x}_{\backslash \{j,l\}} |z_j, z_l) \\
      &- \mathbb{E}[f^{\{j,l\}}(X_j,X_l,\mathbf{X}_{\backslash \{j,l\}})|X_j=z_{k,j}^K, X_l = z_{m,l}^K ]\Bigg|\\
      &\leq\Bigg| \int_{\mathbf{x}_{\backslash \{j,l\}}}f^{\{j,l\}}(\mathbf{z}(k,m,K,\mathbf{x}_{\backslash\{j,l\}}),\mathbf{x}_{\backslash\{j,l\}})dp_{\backslash \{j,l\}|\{j,l\}}(\mathbf{x}_{\backslash \{j,l\}} |z_j, z_l)  \\
      &- \int_{\mathbf{x}_{\backslash \{j,l\}}}f^{\{j,l\}}(z_j,z_l,\mathbf{x}_{\backslash\{j,l\}})dp_{\backslash \{j,l\}|\{j,l\}}(\mathbf{x}_{\backslash \{j,l\}} |z_j, z_l) \bigg|\\
      & +\Bigg|\int_{\mathbf{x}_{\backslash \{j,l\}}}f^{\{j,l\}}(z_j,z_l,\mathbf{x}_{\backslash\{j,l\}})dp_{\backslash \{j,l\}|\{j,l\}}(\mathbf{x}_{\backslash \{j,l\}} |z_j, z_l) \\
      &- \mathbb{E}[f^{\{j,l\}}(X_j,X_l,\mathbf{X}_{\backslash \{j,l\}})|X_j=z_{k,j}^K, X_l = z_{m,l}^K ]\Bigg|\\
       &\leq \int_{\mathbf{x}_{\backslash \{j,l\}}}\Big|f^{\{j,l\}}(\mathbf{z}(k,m,K,\mathbf{x}_{\backslash\{j,l\}}),\mathbf{x}_{\backslash\{j,l\}})-f^{\{j,l\}}(z_j,z_l,\mathbf{x}_{\backslash\{j,l\}})\Big|dp_{\backslash \{j,l\}|\{j,l\}}(\mathbf{x}_{\backslash \{j,l\}} |z_j, z_l) \\
       & +\Big|\mathbb{E}[f^{\{j,l\}}(X_j,X_l,\mathbf{X}_{\backslash\{j,l\}} )|X_j=z_j,X_l = z_l]-\mathbb{E}[f^{\{j,l\}} (X_j,X_l,\mathbf{X}_{\backslash \{j,l\}})|X_j=z_{k,j}^K, X_l = z_{m,l}^K]\Big|\\
	  &\leq\int_{\mathbf{x}_{\backslash \{j,l\}}}\frac{\varepsilon}{4(x_{\max,j}-x_{\min,j})(x_{\max,l}-x_{\min,l})} dp_{\backslash \{j,l\}|\{j,l\}} (\mathbf{x}_{\backslash \{j,l\}} |z_j,z_l)\\ &+\frac{\varepsilon}{4(x_{\max,j}-x_{\min,j})(x_{\max,l}-x_{\min,l})} =\frac{\varepsilon}{2(x_{\max,j}-x_{\min,j})(x_{\max,l}-x_{\min,l})}.
    \end{split}
\end{align}
Thus, using \eqref{pf thm differentiable 2 eq 3} and \eqref{pf thm differentiable 2 eq 6}, for all $K>\max\{K_3(\varepsilon),K_4(\varepsilon)\}$ the difference between the summations in \eqref{pf thm differentiable 2 eq 1} and \eqref{pf thm differentiable 2 eq 2} satisfies
{\allowdisplaybreaks
\begin{align*}
    &\Bigg|\sum_{k=1}^{ k_j^K(x_j)}\sum_{m=1}^{ k_l^K(x_l)}\mathbb{E}[\Delta_f^{\{j,l\}}(K,k,m;\mathbf{X}_{\backslash \{j,l\}})|X_j\in (z_{k-1,j}^K,z_{k,j}^K], X_l\in (z_{m-1,l}^K,z_{m,l}^K]]\\
    &- \sum_{k=1}^{ k_j^K(x_j)}\sum_{m=1}^{ k_l^K(x_l)} \mathbb{E}[f^{\{j,l\}}(X_j,X_l,\mathbf{X}_{\backslash \{j,l\}} )|X_j=z_{k,j}^K,X_l=z_{m,l}^K](z_{k,j}^K-z_{k-1,j}^K)(z_{m,l}^K-z_{m-1,l}^K)\Bigg|\\
    &\leq\sum_{k=1}^{ k_j^K(x_j)}\sum_{m=1}^{ k_l^K(x_l)}\Big|\mathbb{E}[\Delta_f^{\{j,l\}}(K,k,m;\mathbf{X}_{\backslash \{j,l\}})|X_j\in (z_{k-1,j}^K,z_{k,j}^K], X_l\in (z_{m-1,l}^K,z_{m,l}^K]]\\
    &- \mathbb{E}[f^{\{j,l\}}(X_j,X_l,\mathbf{X}_{\backslash \{j,l\}} )|X_j=z_{k,j}^K,X_l=z_{m,l}^K](z_{k,j}^K-z_{k-1,j}^K)(z_{m,l}^K-z_{m-1,l}^K)\Big|\\
	&=\sum_{k=1}^{ k_j^K(x_j)}\sum_{m=1}^{ k_l^K(x_l)}\Bigg|\frac{(z_{k,j}^K-z_{k-1,j}^K )(z_{m,l}^K-z_{m-1,l}^K )}{p_{\{j,l\}}((z_{k-1,j}^K,z_{k,j}^K])\times(z_{m-1,l}^K,z_{m,l}^K])} \\
	&\times  \int_{(z_j,z_l)\in (z_{k-1,j}^K,z_{k,j}^K] \times (z_{m-1,l}^K,z_{m,l}^K]}\int_{\mathbf{x}_{\backslash \{j,l\}}}f^{\{j,l\}}(\mathbf{z}(k,m,K,\mathbf{x}_{\backslash \{j,l\}} ),\mathbf{x}_{\backslash \{j,l\}})dp_{\backslash \{j,l\}|\{j,l\}}(\mathbf{x}_{\backslash \{j,l\}}|z_j,z_l)dp_{\{j,l\}}(z_j,z_l)\\ 
    & - \mathbb{E}[f^{\{j,l\}}(X_j,X_l,\mathbf{X}_{\backslash \{j,l\}} )|X_j=z_{k,j}^K,X_l=z_{m,l}^K](z_{k,j}^K-z_{k-1,j}^K)(z_{m,l}^K-z_{m-1,l}^K)\Bigg|\quad\quad(\text{using \eqref{pf thm differentiable 2 eq 3}})\\
	&=\sum_{k=1}^{ k_j^K(x_j)}\sum_{m=1}^{ k_l^K(x_l)}\frac{(z_{k,j}^K-z_{k-1,j}^K )(z_{m,l}^K-z_{m-1,l}^K)}{p_{\{j,l\}} ((z_{k-1,j}^K,z_{k,j}^K]\times(z_{m-1,l}^K,z_{m,l}^K])}\\
	&\times\Bigg|\int_{(z_j,z_l)\in (z_{k-1,j}^K,z_{k,j}^K] \times (z_{m-1,l}^K,z_{m,l}^K]}\{\int_{\mathbf{x}_{\backslash \{j,l\}}}f^{\{j,l\}} (\mathbf{z}(k,m,K,\mathbf{x}_{\backslash \{j,l\}}) ),\mathbf{x}_{\backslash \{j,l\}} )dp_{\backslash \{j,l\}|\{j,l\}}(\mathbf{x}_{\backslash \{j,l\}}|z_j,z_l)\\ &-\mathbb{E}[f^{\{j,l\}}(X_j,X_l,\mathbf{X}_{\backslash \{j,l\}} )|X_j=z_{k,j}^K,X_l = z_{m,l}^K ]\}dp_{\{j,l\}}(z_j,z_l)\Bigg| \\
	&\leq \sum_{k=1}^{ k_j^K(x_j)}\sum_{m=1}^{ k_l^K(x_l)}\frac{(z_{k,j}^K-z_{k-1,j}^K )(z_{m,l}^K-z_{m-1,l}^K )}{p_{\{j,l\}}((z_{k-1,j}^K,z_{k,j}^K]\times(z_{m-1,l}^K,z_{m,l}^K])}\\
	&\times \int_{(z_j,z_l)\in (z_{k-1,j}^K,z_{k,j}^K] \times (z_{m-1,l}^K,z_{m,l}^K]}\Bigg|\int_{\mathbf{x}_{\backslash j}}f^{\{j,l\}}(\mathbf{z}(k,m, K,\mathbf{x}_{\backslash \{j,l\}} ),\mathbf{x}_{\backslash \{j,l\}})dp_{\backslash \{j,l\}|\{j,l\}}(\mathbf{x}_{\backslash \{j,l\}}|z_j,z_l)\\
	&\quad\quad-\mathbb{E}[f^{\{j,l\}} (X_j,X_l,\mathbf{X}_{\backslash \{j,l\}})|X_j=z_{k,j}^K,X_l=z_{m,l}^K ]\Bigg|dp_{\{j,l\}}(z_j,z_l)\\
	&\leq \sum_{k=1}^{ k_j^K(x_j)}\sum_{m=1}^{ k_l^K(x_l)}\frac{(z_{k,j}^K-z_{k-1,j}^K)(z_{m,l}^K-z_{m-1,l}^K)}{p_{\{j,l\}} ((z_{k-1,j}^K,z_{k,j}^K]\times(z_{m-1,l}^K,z_{m,l}^K])}\\ 
	&\times\int_{(z_j,z_l)\in (z_{k-1,j}^K,z_{k,j}^K] \times (z_{m-1,l}^K,z_{m,l}^K]}\frac{\varepsilon}{2(x_{\max,j}-x_{\min,j})(x_{\max,l}-x_{\min,l})}dp_{\{j,l\}}(z_j,z_l)\quad\quad(\text{using \eqref{pf thm differentiable 2 eq 6}})\\
	&=\sum_{k=1}^{ k_j^K(x_j)}\sum_{m=1}^{ k_l^K(x_l)}(z_{k,j}^K-z_{k-1,j}^K)(z_{m,l}^K-z_{m-1,l}^K)  \\ &\times \frac{\varepsilon p_{\{j,l\}} ((z_{k-1,j}^K,z_{k,j}^K]\times(z_{m-1,l}^K,z_{m,l}^K])}{2(x_{\max,j}-x_{\min,j} )(x_{\max,l}-x_{\min,l}) p_{\{j,l\}}((z_{k-1,j}^K,z_{k,j}^K]\times(z_{m-1,l}^K,z_{m,l}^K]) }\\
	&=\frac{\varepsilon}{2(x_{\max,j}-x_{\min,j})(x_{\max,l}-x_{\min,l})}  \sum_{k=1}^{ k_j^K(x_j)}\sum_{m=1}^{ k_l^K(x_l)}(z_{k,j}^K-z_{k-1,j}^K)(z_{m,l}^K-z_{m-1,l}^K)
\end{align*}}
\begin{flalign}\label{pf thm differentiable 2 eq 7}
&\leq \frac{\varepsilon(x_{\max,j}-x_{\min,j} )(x_{\max,l}-x_{\min,l})}{2(x_{\max,j}-x_{\min,j})(x_{\max,l}-x_{\min,l})} =\frac{\varepsilon}{2}.&
\end{flalign}
Finally, combining \eqref{pf thm differentiable 2 eq 1},\eqref{pf thm differentiable 2 eq 2}, and \eqref{pf thm differentiable 2 eq 7} , for all $K>\max\{K_1(\varepsilon),K_2(\varepsilon),K_3(\varepsilon),K_4(\varepsilon)\}$,
{\allowdisplaybreaks
\begin{align*}
     &\left|h_{\{j,l\},ALE}(x_j,x_l)-\int_{x_{\min,l}}^{x_l}\int_{x_{\min,j}}^{x_j} \mathbb{E}[f^{\{j,l\}}(X_j,X_l,\mathbf{X}_{\backslash \{j,l\}} )|X_j=z_j, X_l = z_l]dz_jdz_l\right| \\
     &\leq \left|h_{\{j,l\},ALE}(x_j,x_l)-\sum_{k=1}^{ k_j^K(x_j)}\sum_{m=1}^{ k_l^K(x_l)}\mathbb{E}[\Delta_f^{\{j,l\}}(K,k,m;\mathbf{X}_{\backslash \{j,l\}})|X_j\in(z_{k-1,j}^K,z_{k,j}^K],X_l\in(z_{m-1,l}^K,z_{m,l}^K]] \right|\\
	&\qquad+\Bigg|\int_{x_{\min,l}}^{x_l}\int_{x_{\min,j}}^{x_j} \mathbb{E}[f^{\{j,l\}}(X_j,X_l,X_{\backslash \{j,l\}})|X_j=z_j,X_l = z_l]dz_jdz_l\\
	&\qquad\qquad-\sum_{k=1}^{ k_j^K(x_j)}\sum_{m=1}^{ k_l^K(x_l)}\mathbb{E}[f^{\{j,l\}}(X_j,X_l,\mathbf{X}_{\backslash \{j,l\}})|X_j=z_{k,j}^K,X_l=z_{m,l}^K](z_{k,j}^K-z_{k-1,j}^K)(z_{m,l}^K-z_{m-1,l}^K)\Bigg| \\  
	&\qquad +\Bigg|\sum_{k=1}^{ k_j^K(x_j)}\sum_{m=1}^{ k_l^K(x_l)}\mathbb{E}[\Delta_f^{\{j,l\}}(K,k,m;\mathbf{X}_{\backslash \{j,l\}} )|X_j\in(z_{k-1,j}^K,z_{k,j}^K],X_l\in(z_{m-1,l}^K,z_{m,l}^K]]\\
	&\qquad\qquad-\sum_{k=1}^{ k_j^K(x_j)}\sum_{m=1}^{ k_l^K(x_l)}\mathbb{E}[f^{\{j,l\}}(X_j,X_l,\mathbf{X}_{\backslash \{j,l\}} )|X_j=z_{k,j}^K,X_l=z_{m,l}^K ](z_{k,j}^K-z_{k-1,j}^K)(z_{m,l}^K-z_{m-1,l}^K)\Bigg|\\
	&\leq \frac{\varepsilon}{4}+\frac{\varepsilon}{4}+\frac{\varepsilon}{2}=\varepsilon,  
\end{align*}}
which completes the proof.
\end{proof}
\begin{thm}[\textbf{consistency of the ALE main effect estimator}]\label{Theorem3} Consider a sequence of partitions $\{\mathcal{P}_{j}^K:K=1,2,\ldots\}$ as in Definition \ref{def as limit}, such that $\delta_{j,K} \to 0$ as $K\to\infty$. Denote the estimator \eqref{ALE uncentered main est} of the uncentered ALE main effect of $X_j$ using partition $\mathcal{P}_{j}^K$ and sample size $n$ by
\begin{equation} \label{consistency thm eq 1st-1}
    \hat{g}_{j,ALE,K,n}(x) \equiv \sum_{k=1}^{ k_j^K(x)}\frac{\sum_{i=1}^{n}\mathbb{I}(X_{i,j}\in(z_{k-1,j}^K,z_{k,j}^K])[f(z_{k,j}^K,\mathbf{X}_{i,\backslash j})-f(z_{k-1,j}^K,\mathbf{X}_{i,\backslash j} )]} {\sum_{i=1}^{n}\mathbb{I}(X_{i,j}\in(z_{k-1,j}^K,z_{k,j}^K])},
\end{equation}
where $\mathbb{I}(\cdot)$ denotes the indicator function of an event. Assume that $\{\mathbf{X}_i:i=1,2,\ldots\}$ is an i.i.d. sequence of random vectors drawn from $p(\cdot)$, and let $P$ denote the probability measure on the underlying probability space of this random sequence. Let 
\begin{equation}\label{consistency thm eq 1st-2}
    g_{j,ALE,K}(x) \equiv \sum_{k=1}^{k_j^K(x)} \mathbb{E}[f(z_{k,j}^K,\mathbf{X}_{\backslash j})-f(z_{k-1,j}^K,\mathbf{X}_{\backslash j} )|X_j\in (z_{k-1,j}^K,z_{k,j}^K]]	 
\end{equation}
denote the finite-$K$ version of $g_{j,ALE}(x_j)$ in \eqref{ALE uncentered main def as limit} using the same partition $\mathcal{P}_{j}^K$ as the estimator. 

Then $\hat{g}_{j,ALE,K,n}(x)$ is a strongly consistent estimator of $g_{j,ALE}(x)$ in the following sense. For each $x\in \mathcal{S}_j$, each $\varepsilon > 0$, and each $K$, 
\begin{equation*}
    \lim_{n \to \infty} \hat{g}_{j,ALE,K,n}(x) = g_{j,ALE,K}(x)  \quad \mathrm{a.s.-}P,
\end{equation*}
with $|g_{j,ALE,K}(x)-g_{j,ALE}(x)| < \varepsilon$ if $K$ is sufficiently large. 

Moreover, there exists a sequence of sample sizes $\{n_K: K=1,2,\ldots\}$ such that, as $K \to \infty$, $\hat{g}_{j,ALE,K,n_K}(x) \to g_{j,ALE}(x)$ both in probability and in mean. 
\end{thm}
\begin{proof}
Consider any fixed $K$ and $k\in\{1,2,\ldots,K\}$. Applying the strong law of large numbers to both the numerator and denominator in the summand of \eqref{consistency thm eq 1st-1}, as $n\to\infty$, the summand converges a.s.-$P$: 
{\allowdisplaybreaks
\begin{align*}
         &\frac{n^{-1}\sum_{i=1}^n\mathbb{I}(X_{i,j}\in(z_{k-1,j}^K,z_{k,j}^K])[f(z_{k,j}^K,\mathbf{X}_{i,\backslash j} )-f(z_{k-1,j}^K,\mathbf{X}_{i,\backslash j} )] }{n^{-1}\sum_{i=1}^n\mathbb{I}(X_{i,j}\in(z_{k-1,j}^K,z_{k,j}^K]) }\\
         &\overset{\mathrm{a.s.}}{\to} \frac{\mathbb{E}[\mathbb{I}(X_j\in(z_{k-1,j}^K,z_{k,j}^K])[f(z_{k,j}^K,\mathbf{X}_{\backslash j} )-f(z_{k-1,j}^K,\mathbf{X}_{\backslash j} )]]}{\mathbb{E}[\mathbb{I}(X_j\in(z_{k-1,j}^K,z_{k,j}^K])]} \\
         &=\frac{\mathbb{E}[\mathbb{E}[\mathbb{I}(X_j\in(z_{k-1,j}^K,z_{k,j}^K])\{f(z_{k,j}^K,\mathbf{X}_{\backslash j} )-f(z_{k-1,j}^K,\mathbf{X}_{\backslash j} )\}|X_j ]]}{p_j((z_{k-1,j}^K,z_{k,j}^K]))}\\
         &=\frac{\mathbb{E}[\mathbb{I}(X_j\in(z_{k-1,j}^K,z_{k,j}^K])\mathbb{E}[\{f(z_{k,j}^K,\mathbf{X}_{\backslash j} )-f(z_{k-1,j}^K,\mathbf{X}_{\backslash j} )\}|X_j ]]}{p_j ((z_{k-1,j}^K,z_{k,j}^K])}\\
	    &=\frac{\int_{z\in(z_{k-1,j}^K,z_{k,j}^K]}\mathbb{E}[\{f(z_{k,j}^K,\mathbf{X}_{\backslash j} )-f(z_{k-1,j}^K,\mathbf{X}_{\backslash j} )\}|X_j=z]dp_j(z)}{p_j((z_{k-1,j}^K,z_{k,j}^K])}\\
	   &=\frac{\mathbb{E}[f(z_{k,j}^K,\mathbf{X}_{\backslash j} )-f(z_{k-1,j}^K,\mathbf{X}_{\backslash j} )|X_j\in(z_{k-1,j}^K,z_{k,j}^K]]p_j ((z_{k-1,j}^K,z_{k,j}^K])}{p_j ((z_{k-1,j}^K,z_{k,j}^K])}\\
	   & =\mathbb{E}[f(z_{k,j}^K,\mathbf{X}_{\backslash j} )-f(z_{k-1,j}^K,\mathbf{X}_{\backslash j} )|X_j\in(z_{k-1,j}^K,z_{k,j}^K]].
\end{align*}}
Consequently, since the finite sum of almost-surely convergent random variables also converges almost surely, as $n\to\infty$,
\begin{equation}\label{consistency thm eq 1st-3}
    \begin{split}
    &\hat{g}_{j,ALE,K,n}(x) = \sum_{k=1}^{ k_j^K(x)}\frac{\sum_{i=1}^{n}\mathbb{I}( X_{i,j}\in(z_{k-1,j}^K,z_{k,j}^K])[f(z_{k,j}^K,\mathbf{X}_{i,\backslash j} )-f(z_{k-1,j}^K,\mathbf{X}_{i,\backslash j} )]}{\sum_{i=1}^{n}\mathbb{I}( X_{i,j}\in(z_{k-1,j}^K,z_{k,j}^K])}\\
    &\overset{\mathrm{a.s.}}{\to} \sum_{k=1}^{ k_j^K(x)}\mathbb{E}[f(z_{k,j}^K,\mathbf{X}_{\backslash j} )-f(z_{k-1,j}^K,\mathbf{X}_{\backslash j} )|X_j\in(z_{k-1,j}^K,z_{k,j}^K]]\\
    &=g_{j,ALE,K}(x).
    \end{split}
\end{equation}
By definition of $g_{j,ALE}(x)$ in \eqref{ALE uncentered main def as limit}, for any $\varepsilon > 0$ there exists some sufficiently large $K_1=K_1(\varepsilon)$ such that
\begin{equation}\label{consistency thm eq 1st-4}
    |g_{j,ALE,K}(x)-g_{j,ALE}(x)| < \frac{\varepsilon}{2} \quad\forall K > K_1.
\end{equation}
Eqs. \eqref{consistency thm eq 1st-3} and \eqref{consistency thm eq 1st-4} together imply the strong consistency claim.

Because the almost-sure convergence in \eqref{consistency thm eq 1st-3} implies convergence in probability as well, by definition of convergence in probability we have that for each $\varepsilon > 0$,
\begin{equation}\label{consistency thm eq 1st-5}
    \lim_{n \to \infty}P\{|\hat{g}_{j,ALE,K,n}(x) - g_{j,ALE,K}(x)| > \frac{\varepsilon}{2} \}=0,
\end{equation}
so that for each $K$, each $\varepsilon > 0$, and each $\delta > 0$ there exists an integer $n_K = n_K(\varepsilon, \delta)$ such that for all $n \geq n_K$,

\begin{equation}\label{consistency thm eq 1st-6}
    P\{|\hat{g}_{j,ALE,K,n}(x) - g_{j,ALE,K}(x)| > \frac{\varepsilon}{2} \}<\delta.
\end{equation}

Thus, since $|\hat{g}_{j,ALE,K,n_K}(x) - g_{j,ALE}(x)| \leq |\hat{g}_{j,ALE,K,n_K}(x) - g_{j,ALE,K}(x)| + |g_{j,ALE,K}(x) - g_{j,ALE}(x)|$, for $K > K_1$ in \eqref{consistency thm eq 1st-4} we have 
\begin{equation}\label{consistency thm eq 1st-7}
    \begin{split}
    &\quad P\{|\hat{g}_{j,ALE,K,n_K}(x) - g_{j,ALE}(x)| > \varepsilon \}\\
    &\leq P\{|\hat{g}_{j,ALE,K,n_K}(x) - g_{j,ALE,K}(x)| + |g_{j,ALE,K}(x) - g_{j,ALE}(x)| > \varepsilon \}\\
    &\leq P\{|\hat{g}_{j,ALE,K,n_K}(x) - g_{j,ALE,K}(x)| + \frac{\varepsilon}{2} > \varepsilon \}\\
    &= P\{|\hat{g}_{j,ALE,K,n_K}(x) - g_{j,ALE,K}(x)| > \frac{\varepsilon}{2} \}\qquad(\text{using } \eqref{consistency thm eq 1st-6})\\
    &<\delta.
    \end{split}
\end{equation}
Since $\delta$ and $\varepsilon$ were arbitrary, \eqref{consistency thm eq 1st-7} implies that $\hat{g}_{j,ALE,K,n_K}(x)$ converges in probability to $g_{j,ALE}(x)$ as $K \to \infty$, as claimed.

Finally, convergence in mean follows similarly to convergence in probability, if, for fixed $K$, we can replace the convergence in probability \eqref{consistency thm eq 1st-5} by convergence in mean
\begin{equation}\label{consistency thm eq 1st-8}
    \lim_{n \to \infty}\mathbb{E}[|\hat{g}_{j,ALE,K,n}(x) - g_{j,ALE,K}(x)|]=0.
\end{equation}
Because convergence in probability \eqref{consistency thm eq 1st-5}, along with uniform integrability of $\hat{g}_{j,ALE,K,n}(x)$, implies convergence in mean \eqref{consistency thm eq 1st-8}, it suffices to show uniform (over all $n$) integrability of $\hat{g}_{j,ALE,K,n}(x)$, which follows if we can show that $\hat{g}_{j,ALE,K,n}(x)$ is bounded for all $n$ and for fixed $K$. The boundedness follows because $|f|$ is bounded (say by $M < \infty$), so that for every $n$, 
\begin{equation}\label{consistency thm eq 1st-9}
    \begin{split}
    &|\hat{g}_{j,ALE,K,n}(x)| \leq \sum_{k=1}^{ k_j^K(x)}\frac{\sum_{i=1}^{n}\mathbb{I}( X_{i,j}\in(z_{k-1,j}^K,z_{k,j}^K])|f(z_{k,j}^K,\mathbf{X}_{i,\backslash j} )-f(z_{k-1,j}^K,\mathbf{X}_{i,\backslash j} )|}{\sum_{i=1}^{n}\mathbb{I}( X_{i,j}\in(z_{k-1,j}^K,z_{k,j}^K])}\\
    &\leq 2M\sum_{k=1}^{ k_j^K(x)}\frac{\sum_{i=1}^{n}\mathbb{I}( X_{i,j}\in(z_{k-1,j}^K,z_{k,j}^K])}{\sum_{i=1}^{n}\mathbb{I}( X_{i,j}\in(z_{k-1,j}^K,z_{k,j}^K])}\\
    &\leq 2MK,
    \end{split}
\end{equation}
which proves \eqref{consistency thm eq 1st-8}. Eq. \eqref{consistency thm eq 1st-8} in turn implies that for each fixed $K$ and each $\varepsilon > 0$, there exists an integer $n_K = n_K(\varepsilon)$ such that for all $n \geq n_K$, 
\begin{equation}\label{consistency thm eq 1st-10}
    \mathbb{E}[|\hat{g}_{j,ALE,K,n}(x) - g_{j,ALE,K}(x)|] < \frac{\varepsilon}{2}.
\end{equation}
Thus, for $K > K_1$ in \eqref{consistency thm eq 1st-4} we have
\begin{equation}\label{consistency thm eq 1st-11}
    \begin{split}
    &\mathbb{E}[|\hat{g}_{j,ALE,K,n_K}(x) - g_{j,ALE}(x)|] \\
    &\leq \mathbb{E}[|\hat{g}_{j,ALE,K,n_K}(x) - g_{j,ALE,K}(x)|] + \mathbb{E}[|g_{j,ALE,K}(x) - g_{j,ALE}(x)|]\\
    &< \frac{\varepsilon}{2} + \frac{\varepsilon}{2} = \varepsilon.
    \end{split}
\end{equation}
Since $\varepsilon$ was arbitrary, this proves the claimed convergence in mean
\begin{equation*}
    \lim_{K \to \infty}\mathbb{E}[|\hat{g}_{j,ALE,K,n_K}(x) - g_{j,ALE}(x)|]=0.
\end{equation*}
\end{proof}

\begin{thm}[\textbf{consistency of the ALE second-order effect estimator}]\label{Theorem4} For $\{j,l\}\subseteq \{1, 2, \ldots, d\}$, consider two sequences of partitions $\{\mathcal{P}_{j}^K:K=1,2,\ldots\}$ and $\{\mathcal{P}_{l}^K:K=1,2,\ldots\}$ in Definition 2 such that $\lim_{K\to\infty} \delta_{j,K} = \lim_{K\to\infty}\delta_{l,K}= 0$. 
Denote the estimator \eqref{ALE uncentered second est} of the uncentered ALE second-order effect of $(X_j,X_l)$ using partitions $\mathcal{P}_{j}^K$ and $\mathcal{P}_{l}^K$ and sample size $n$ by
\begin{align} \label{consistency thm eq 2nd-1}
\begin{split}
    &\quad \hat{h}_{\{j,l\},ALE,K,n}(x_j, x_l)=\\
    &\sum_{k=1}^{ k_j^K(x_j)}\sum_{m = 1}^{k_l^K(x_l)}\frac{\sum_{i=1}^{n}\mathbb{I}(X_{i,j}\in(z_{k-1,j}^K,z_{k,j}^K],X_{i,l}\in(z_{m-1,l}^K,z_{m,l}^K])\Delta_f^{\{j,l\}}(K, k, m; \mathbf{X}_{i,\backslash {\{j,l\}}})} {\sum_{i=1}^{n}\mathbb{I}(X_{i,j}\in(z_{k-1,j}^K,z_{k,j}^K],X_{i,l}\in(z_{m-1,l}^K,z_{m,l}^K])}.
\end{split}
\end{align}
Let $\{\mathbf{X}_i:i=1,2,\ldots,n\}$ and $P(\cdot)$ be as in Theorem \ref{Theorem3}, and let
\begin{equation}\label{consistency thm eq 2nd-2}
    h_{\{j,l\},ALE,K}(x_j, x_l)\equiv \sum_{k=1}^{k_j^K(x_j)}\sum_{m=1}^{k_l^K(x_l)} \mathbb{E}[\Delta_f^{\{j,l\}}(K, k, m;\mathbf{X}_{\backslash\{j,l\}})|X_j\in (z_{k-1,j}^K,z_{k,j}^K], X_l\in (z_{m-1,l}^K,z_{m,l}^K]],
\end{equation}
denote the finite-$K$ version of $h_{\{j,l\},ALE}(x_j, x_l)$ in \eqref{ALE uncentered second def as limit} using the same partitions as the estimator. Assume $p_{\{j,l\}}(\cdot)$ is strictly positive everywhere in $\mathcal{S}_j\times\mathcal{S}_l$.

Then $\hat{h}_{\{j,l\},ALE,n}(x_j, x_l)$ is a strongly consistent estimator of $h_{\{j,l\},ALE}(x_j, x_l)$ in the following sense. For each $(x_j, x_l) \in \mathcal{S}_j\times\mathcal{S}_l$, each $\varepsilon >0$, and each $K$, 
\begin{equation*}
    \lim_{n \to \infty} \hat{h}_{\{j,l\},ALE,K,n}(x_j, x_l) = h_{\{j,l\},ALE,K}(x_j,x_l) \quad \mathrm{a.s.-}P
\end{equation*} with $|h_{\{j,l\},ALE,K}(x_j,x_l) - h_{\{j,l\},ALE}(x_j,x_l)|<\varepsilon$ if $K$ is sufficiently large. 

In addition, there exists a sequence of sample sizes $\{n_K: K=1,2,\ldots\}$ such that, as $K\to\infty$,
\begin{equation*}
    \hat{h}_{\{j,l\},ALE,n_K}(x_j, x_l)\to h_{\{j,l\},ALE}(x_j, x_l)
\end{equation*} both in probability and in mean for each $(x_j, x_l)\in \mathcal{S}_j\times\mathcal{S}_l$. 
\end{thm}
\begin{proof}
The proof is similar to that of Theorem \ref{Theorem3}. Consider any fixed $K$ and $k, m \in\{1,2,\ldots,K\}$. Applying the strong law of large numbers to both the numerator and denominator in the summand of \eqref{consistency thm eq 2nd-1}, as $n\to\infty$, the summand converges a.s.-$P$: 
{\allowdisplaybreaks
\begin{align*}
         &\frac{n^{-1}\sum_{i=1}^n\mathbb{I}(X_{i,j}\in(z_{k-1,j}^K,z_{k,j}^K], X_{i,l}\in(z_{m-1,l}^K,z_{m,l}^K])\Delta_f^{\{j,l\}}(K, k, m; \mathbf{X}_{i,\backslash {\{j,l\}}}) }{n^{-1}\sum_{i=1}^n\mathbb{I}(X_{i,j}\in(z_{k-1,j}^K,z_{k,j}^K],X_{i,l}\in(z_{m-1,l}^K,z_{m,l}^K]) }\\
         &\overset{\mathrm{a.s.}}\to \frac{\mathbb{E}[\mathbb{I}(X_j\in(z_{k-1,j}^K,z_{k,j}^K],X_l\in(z_{m-1,l}^K,z_{m,l}^K])\Delta_f^{\{j,l\}}(K,k,m;\mathbf{X}_{\backslash {\{j,l\}}})]}{\mathbb{E}[\mathbb{I}(X_j\in(z_{k-1,j}^K,z_{k,j}^K],X_l\in(z_{m-1,l}^K,z_{m,l}^K])]} \\
         &=\frac{\mathbb{E}[\mathbb{E}[\mathbb{I}(X_j\in(z_{k-1,j}^K,z_{k,j}^K],X_l\in(z_{m-1,l}^K,z_{m,l}^K])\Delta_f^{\{j,l\}}(K,k,m;\mathbf{X}_{\backslash {\{j,l\}}})|X_j,X_l]]}{p_{\{j,l\}}((z_{k-1,j}^K,z_{k,j}^K]\times(z_{m-1,l}^K,z_{m,l}^K])}\\
         &=\frac{\mathbb{E}[\mathbb{I}(X_j\in(z_{k-1,j}^K,z_{k,j}^K],X_l\in(z_{m-1,l}^K,z_{m,l}^K])\mathbb{E}[\Delta_f^{\{j,l\}}(K,k,m;\mathbf{X}_{\backslash {\{j,l\}}})|X_j,X_l]]}{p_{\{j,l\}}((z_{k-1,j}^K,z_{k,j}^K]\times(z_{m-1,l}^K,z_{m,l}^K])}\\
	    &=\frac{\int_{(z_j,z_l)\in (z_{k-1,j}^K,z_{k,j}^K]\times (z_{m-1,l}^K,z_{m,l}^K]}\mathbb{E}[\Delta_f^{\{j,l\}}(K,k,m;\mathbf{X}_{\backslash {\{j,l\}}})|X_j=z_j, X_l = z_l]dp_{\{j,l\}}(z_j,z_l)}{p_{\{j,l\}}((z_{k-1,j}^K,z_{k,j}^K]\times(z_{m-1,l}^K,z_{m,l}^K])}\\
	   &=\frac{\mathbb{E}[\Delta_f^{\{j,l\}}(K,k,m;\mathbf{X}_{\backslash {\{j,l\}}})|X_j\in(z_{k-1,j}^K,z_{k,j}^K],X_l\in(z_{m-1,l}^K,z_{m,l}^K]]p_{\{j,l\}} ((z_{k-1,j}^K,z_{k,j}^K]\times(z_{m-1,l}^K,z_{m,l}^K])}{p_{\{j,l\}}((z_{k-1,j}^K,z_{k,j}^K]\times(z_{m-1,l}^K,z_{m,l}^K])}\\
	   & =\mathbb{E}[\Delta_f^{\{j,l\}}(K,k,m;\mathbf{X}_{\backslash {\{j,l\}}})|X_j\in(z_{k-1,j}^K,z_{k,j}^K],X_l\in(z_{m-1,l}^K,z_{m,l}^K]].
\end{align*}}
Again, by the almost sure convergence of the finite sum of  almost-surely convergent random variables, as $n\to\infty$, 
\begin{align}\label{consistency thm eq 2nd-3}
    \begin{split}
     &\quad \hat{h}_{\{j,l\},ALE,K,n}(x_j, x_l)=\\
    &\sum_{k=1}^{ k_j^K(x_j)}\sum_{m = 1}^{k_l^K(x_l)}\frac{\sum_{i=1}^{n}\mathbb{I}(X_{i,j}\in(z_{k-1,j}^K,z_{k,j}^K],X_{i,l}\in(z_{m-1,l}^K,z_{m,l}^K])\Delta_f^{\{j,l\}}(K, k, m; \mathbf{X}_{i,\backslash {\{j,l\}}})} {\sum_{i=1}^{n}\mathbb{I}(X_{i,j}\in(z_{k-1,j}^K,z_{k,j}^K],X_{i,l}\in(z_{m-1,l}^K,z_{m,l}^K])}\\
   &\overset{\mathrm{a.s.}}{\to}\sum_{k=1}^{ k_j^K(x_j)}\sum_{m = 1}^{k_l^K(x_l)}\mathbb{E}[\Delta_f^{\{j,l\}}(K,k,m;\mathbf{X}_{\backslash {\{j,l\}}})|X_j\in(z_{k-1,j}^K,z_{k,j}^K],X_l\in(z_{m-1,l}^K,z_{m,l}^K]]\\
   &=h_{\{j,l\},ALE,K}(x_j, x_l).
    \end{split}
\end{align}
By definition of $h_{\{j,l\},ALE}(x_j, x_l)$ in \eqref{ALE uncentered second def as limit}, for any $\varepsilon>0$ there exists some sufficiently large $K_1 = K_1(\varepsilon)$ such that 
\begin{equation}\label{consistency thm eq 2nd-4}
\left|h_{\{j,l\},ALE,K}(x_j, x_l)- h_{\{j,l\},ALE}(x_j, x_l)\right| < \frac{\varepsilon}{2} \quad\forall K > K_1.
\end{equation}
Eqs. (\eqref{consistency thm eq 2nd-3}) and (\eqref{consistency thm eq 2nd-4}) together imply the strong consistency claim.

Using the definition of convergence in probability, which is implied from the almost sure convergence in \eqref{consistency thm eq 2nd-3}, we have for each $\varepsilon >0$,
\begin{equation}\label{consistency thm eq 2nd-5}
    \lim_{n \to \infty}P\left\{\left|\hat{h}_{\{j,l\},ALE,K,n}(x_j,x_l) - h_{\{j,l\},ALE,K}(x_j,x_l)\right| > \frac{\varepsilon}{2} \right\}=0.
\end{equation}
Thus, for each $K$, each $\varepsilon > 0$, and each $\delta > 0$ there exists an integer $n_K = n_K(\varepsilon, \delta)$ such that for all $n \geq n_K$,
\begin{equation}\label{consistency thm eq 2nd-6}
    P\left\{\left|\hat{h}_{\{j,l\},ALE,K,n}(x_j,x_l) - h_{\{j,l\},ALE,K}(x_j,x_l)\right| > \frac{\varepsilon}{2} \right\}<\delta.
\end{equation}

Since
\begin{align*}
   & \left|\hat{h}_{\{j,l\},ALE,K,n_K}(x_j,x_l) - h_{\{j,l\},ALE}(x_j,x_l)\right| \\
   &\leq \left|\hat{h}_{\{j,l\},ALE,K,n_K}(x_j,x_l) - h_{\{j,l\},ALE,K}(x_j,x_l)\right| + \left|h_{\{j,l\},ALE,K}(x_j,x_l) - h_{\{j,l\},ALE}(x_j,x_l)\right|,
\end{align*}
for $K > K_1$ in \eqref{consistency thm eq 2nd-4} we have 
\begin{equation}\label{consistency thm eq 2nd-7}
    \begin{split}
    &\quad P\left\{\left|\hat{h}_{\{j,l\},ALE,K,n_K}(x_j,x_l) - h_{\{j,l\},ALE}(x_j,x_l)\right| > \varepsilon \right\}\\
    &\leq P\left\{\left|\hat{h}_{\{j,l\},ALE,K,n_K}(x_j,x_l) - h_{\{j,l\},ALE,K}(x_j,x_l)\right| + \left|h_{\{j,l\},ALE,K}(x_j,x_l) - h_{\{j,l\},ALE}(x_j,x_l)\right| > \varepsilon \right\}\\
    &\leq P\left\{\left|\hat{h}_{\{j,l\},ALE,K,n_K}(x_j,x_l) - h_{\{j,l\},ALE,K}(x_j,x_l)\right| + \frac{\varepsilon}{2} > \varepsilon \right\}\\
    &= P\left\{\left|\hat{h}_{\{j,l\},ALE,K,n_K}(x_j,x_l) - h_{\{j,l\},ALE,K}(x_j,x_l)\right| > \frac{\varepsilon}{2} \right\}<\delta \qquad(\text{using } \eqref{consistency thm eq 2nd-6}).
    \end{split}
\end{equation}
Since $\delta$ and $\varepsilon$ were arbitrary, \eqref{consistency thm eq 2nd-7} implies that $\hat{h}_{\{j,l\},ALE,K,n_K}(x_j,x_l)$ converges in probability to $h_{\{j,l\},ALE}(x_j,x_l)$ as $K \to \infty$, as claimed.

We now prove that for any fixed $K$, the following convergence in mean holds:
\begin{equation}\label{consistency thm eq 2nd-8}
    \lim_{n \to \infty}\mathbb{E}\left[\left|\hat{h}_{\{j,l\},ALE,K,n}(x_j,x_l) - h_{\{j,l\},ALE,K}(x_j,x_l)\right|\right]=0.
\end{equation}
By similar arguments as in the proof of Theorem \ref{Theorem3}, it suffices to show uniform (over all $n$) integrability of $\hat{h}_{\{j,l\},ALE,K,n}(x_j,x_l)$, which follows if we can show that $\hat{h}_{\{j,l\},ALE,K,n}(x_j,x_l)$ is bounded for all $n$ and for any fixed $K$. The boundedness follows since $|f|$ is bounded by some $M < \infty$, so that for every $n$, 
\begin{equation}\label{consistency thm eq 2nd-9}
    \begin{split}
    &\left|\hat{h}_{\{j,l\},ALE,K,n}(x_j,x_l)\right| \leq \\
    &\sum_{k=1}^{ k_j^K(x_j)}\sum_{m = 1}^{k_l^K(x_l)}\frac{\sum_{i=1}^{n}\mathbb{I}(X_{i,j}\in(z_{k-1,j}^K,z_{k,j}^K],X_{i,l}\in(z_{m-1,l}^K,z_{m,l}^K])\left|\Delta_f^{\{j,l\}}(K, k, m; \mathbf{X}_{i,\backslash {\{j,l\}}})\right|} {\sum_{i=1}^{n}\mathbb{I}(X_{i,j}\in(z_{k-1,j}^K,z_{k,j}^K],X_{i,l}\in(z_{m-1,l}^K,z_{m,l}^K])}\\
    &\leq 4M\sum_{k=1}^{ k_j^K(x_j)}\sum_{m = 1}^{k_l^K(x_l)}\frac{\sum_{i=1}^{n}\mathbb{I}(X_{i,j}\in(z_{k-1,j}^K,z_{k,j}^K],X_{i,l}\in(z_{m-1,l}^K,z_{m,l}^K]} {\sum_{i=1}^{n}\mathbb{I}(X_{i,j}\in(z_{k-1,j}^K,z_{k,j}^K],X_{i,l}\in(z_{m-1,l}^K,z_{m,l}^K])}\\
    &\leq 4MK^2.
    \end{split}
\end{equation}
Thus, we have proven \eqref{consistency thm eq 2nd-8}, which in turn implies that for each fixed $K$ and each $\varepsilon > 0$, there exists an integer $n_K = n_K(\varepsilon)$ such that for all $n \geq n_K$, 
\begin{equation}\label{consistency thm eq 2nd-10}
    \mathbb{E}\left[\left|\hat{h}_{\{j,l\},ALE,K,n}(x_j,x_l) - h_{\{j,l\},ALE,K}(x_j,x_l)\right|\right] < \frac{\varepsilon}{2}.
\end{equation}
Hence, for $K > K_1$ in \eqref{consistency thm eq 2nd-4}, we have
\begin{equation}\label{consistency thm eq 2nd-11}
    \begin{split}
    &\mathbb{E}\left[\left|\hat{h}_{\{j,l\},ALE,K,n_K}(x_j,x_l) - h_{\{j,l\},ALE}(x_j,x_l)\right|\right] \\
    &\leq \mathbb{E}\left[\left|\hat{h}_{\{j,l\},ALE,K,n_K}(x_j,x_l) - h_{\{j,l\},ALE,K}(x_j,x_l)\right|\right] + \mathbb{E}\left[\left|h_{\{j,l\},ALE,K}(x_j,x_l) - h_{\{j,l\},ALE}(x_j,x_l)\right|\right]\\
    &< \frac{\varepsilon}{2} + \frac{\varepsilon}{2} = \varepsilon.
    \end{split}
\end{equation}
Since $\varepsilon$ was arbitrary, this proves the claimed convergence in mean
\begin{equation*}
    \lim_{K \to \infty}\mathbb{E}\left[\left|\hat{h}_{\{j,l\},ALE,K,n_K}(x_j,x_l) - h_{\{j,l\},ALE}(x_j,x_l)\right|\right]=0.
\end{equation*}
\end{proof}
\section{ALE Plot Definition for Higher-Order Effects}\label{def of higher-order effects}
\par 
Although we do not envision ALE plots being commonly used to visualize third-and-higher order effects, the notion of higher-order ALE effects is needed to derive the ALE decomposition theorem in Appendix \ref{ale decomposition theorem and properties of L} and the additive recovery properties discussed in Section \ref{paired diff and add recovery}. For this reason and for completeness, we define the ALE higher-order effects and their sample estimators in this appendix and Appendix \ref{estimation of higher-order effects}, respectively. Here, we use the notations defined in Section \ref{def of main and second} and make the same assumptions for $p_j$, $\mathcal{S}_j$, and $\mathcal{P}_{j}^K$ (for each $j \in \{1, \ldots, d\}$, $K = 1, 2, \ldots$) as in Definitions 1 and 2. Again, $\delta_{j,K}$ represents the fineness of the partition $\mathcal{P}_{j}^K$ with $\lim_{K\to\infty} \delta_{j,K} = 0$ for each $j$, and $k_j^K(x)$ denotes the index of the interval of $\mathcal{P}_{j}^K$ into which $x$ falls.  

\par 
Consider any index subset $J\subseteq D\equiv \{1,2,\ldots,d\}$ and the corresponding subset of predictors $\mathbf{X}_J=(X_j: j\in J)$ and their complement $\mathbf{X}_{\backslash J}$. In order to define the $|J|$-order ALE effect of $\mathbf{X}_J$, we first extend the definition of the uncentered ALE effects $g_{j,ALE}(x_j)$ and $h_{\{j,l\},ALE}(x_j,x_l)$ to general $J\subseteq D$, for which it will be convenient to introduce the following operator notation. Let $g:\mathbb{R}^d \to \mathbb{R}$ be any function and consider the operator $\mathcal{L}_J$ that maps $g$ onto another function $\mathcal{L}_J(g):\mathbb{R}^d\to \mathbb{R}$ defined via
\begin{equation}\label{ALE uncentered J def as limit}
\mathcal{L}_J(g)(\mathbf{x}_J) \equiv \lim_{K\to\infty} \sum_{\{\mathbf{k}: 1\leq k_j \leq k^K_j(x_j), j\in J\}}\mathbb{E}[\Delta_g^{J}(K,\mathbf{k};\mathbf{X}_{\backslash J})|X_j\in (z^K_{k_j-1,j}, z^K_{k_j,j}], \forall j \in J],
\end{equation}
where the notation is as follows. For each $K$, the $\mathbf{X}_J$-space is partitioned into the $|J|$-dimensional grid of rectangular cells that is the Cartesian product of $\{\mathcal{P}_{j}^K: j\in J\}$. The $|J|$-length vector $\mathbf{k}=(k_j: j\in J)$ (with each $k_j$ an integer between $1$ and $K$) indicates a specific cell in this grid, i.e., cell-$\mathbf{k}$ is the Cartesian product of  $\{(z_{k_j-1,j}^K,z_{k_j,j}^K]: j\in J\}$. $\Delta_g^{J}(K,\mathbf{k};\mathbf{x}_{\backslash J})$ denotes the $|J|$-order finite difference of $g(\mathbf{x}) = g(\mathbf{x}_J, \mathbf{x}_{\backslash J})$ with respect to $\mathbf{x}_J=(x_j: j\in J)$ across cell-$\mathbf{k}$. For example, for $J=1$ and $\mathbf{k}=k$, $\Delta_g^{J} (K,\mathbf{k};\mathbf{x}_{\backslash J})$ is the difference $[g(z^K_{k,1},\mathbf{x}_{\backslash 1})-g(z^K_{k-1,1},\mathbf{x}_{\backslash 1})]$. For $J=\{1,3\}$ and $\mathbf{k}=(k,m)$, $\Delta_g^{J} (K,\mathbf{k};\mathbf{x}_{\backslash J})$ is the difference of the difference $[g(z^K_{k,1},z^K_{m,3},\mathbf{x}_{\backslash \{1,3\}} )-g(z^K_{k-1,1},z^K_{m,3},\mathbf{x}_{\backslash\{1,3\}} )]-[g(z^K_{k,1},z^K_{m-1,3},\mathbf{x}_{\backslash\{1,3\}})-g(z^K_{k-1,1},z^K_{m-1,3},\mathbf{x}_{\backslash \{1,3\}})]$. For general $J$, $\Delta_g^{J} (K,\mathbf{k};\mathbf{x}_{\backslash J})$ is the difference of the difference of the difference $\ldots$ ($|J|$ times).

\par
Note that if we substitute $g=f$ in \eqref{ALE uncentered J def as limit} for the special case $J=\{j,l\}$, \eqref{ALE uncentered J def as limit} reduces to $h_{\{j,l\},ALE}(x_j,x_l)$ in \eqref{ALE uncentered second def as limit}. In \eqref{ALE uncentered J def as limit} we have written $\mathcal{L}_J(g)$ as a function of only $\mathbf{x}_J$ to make explicit the fact that it only depends on $\mathbf{x}_J$. However, it could be viewed as a function of $\mathbf{x}\in \mathbb{R}^d$, if we take it to be its extension from $\mathbb{R}^{|J|}$  to $\mathbb{R}^d$. For $J=\emptyset$ (the empty set of indices), $\mathcal{L}_{\emptyset} (g)$ is defined as $\mathbb{E}[g(\mathbf{X})]=\int p_D(\mathbf{x})g(\mathbf{x})d\mathbf{x}$, the marginal mean of $g(\mathbf{X})$.
\par 
We will define the ALE $|J|$-order effect of $\mathbf{X}_J$ on $f$, which we denote by $f_{J,ALE}$, as a centered version of the function $\mathcal{L}_J(f)$, analogous to how $f_{j,ALE}(x_j)$ and $f_{\{j,l\},ALE} (x_j,x_l)$ were obtained by centering $g_{j,ALE}(x_j)$ and $h_{\{j,l\},ALE}(x_j,x_l)$ in Section \ref{def of main and second}.  For general $J$, $\mathcal{L}_J(f)$ is comprised of the desired $f_{J,ALE}$, plus lower-order effect functions that are related to $f$ evaluated at the lower boundaries of the rectangular summation region in \eqref{ALE uncentered J def as limit}. Broadly speaking, our strategy is to sequentially subtract the lower-order ALE effects from $\mathcal{L}_J(f)$ to obtain $f_{J,ALE}$.
\par 
More formally, define $\mathcal{H}_{J}(f): f \to f_{J,ALE}$ as the operator that maps a function $f$ to its $|J|$-order ALE effect $f_{J,ALE}$, and let the symbol $\circ$ denote the composition of two operators. For $|J|=0$ (i.e., $J=\emptyset$), we define the zero-order ALE effect for $f(\cdot)$ as \begin{equation}\label{ALE zero def} f_{\emptyset,ALE}\equiv \mathcal{H}_{\emptyset}(f)\equiv \mathcal{L}_{\emptyset}(f),\end{equation} a constant that does not depend on $\mathbf{x}$ and that represents the marginal mean $\mathbb{E}[f(\mathbf{X})]$ of the function $f(\mathbf{X})$. For $1\leq |J|<d$, we define the $|J|$-order effect of $\mathbf{X}_J$ on $f$ as \begin{equation}\label{ALE J def} \begin{split}
f_{J,ALE}(\mathbf{x}_J)\equiv \mathcal{H}_J(f)(\mathbf{x}_J)&\equiv [(I-\mathcal{L}_{\emptyset})\circ(I-\sum_{v\subset J,|v|=1}\mathcal{L}_v)\circ(I-\sum_{v\subset J,|v|=2}\mathcal{L}_v)\circ\\ &\ldots \circ (I-\sum_{v\subset J,|v|=|J|-1}\mathcal{L}_v )\circ \mathcal{L}_J ](f)(\mathbf{x}_J ),\end{split}\end{equation}	
where $I$ denotes the identity operator, i.e., $I(g)=g$ for a function $g:\mathbb{R}^d\to\mathbb{R}$.  The rightmost term in the composite operator $\mathcal{H}_J$ defined in \eqref{ALE J def} is just $\mathcal{L}_J$; the next rightmost term $(I-\sum_{v\subset J,|v|=J-1}\mathcal{L}_v)$ serves to subtract all of the interactions effects of order $|J|-1$ from the result $\mathcal{L}_J(f)$ of the previous operation; the next rightmost term $(I-\sum_{v\subset J,|v|=J-2}\mathcal{L}_v)$ serves to subtract all of the interaction terms of order $|J|-2$ from the result $(I-\sum_{v\subset J,|v|=J-1}\mathcal{L}_v)\circ \mathcal{L}_J(f)$ of the previous operation; and so on. In other words, proceeding from right-to-left, \eqref{ALE J def} iteratively subtracts the effects of smaller and smaller order, until the final operator $(I-\mathcal{L}_\emptyset )$ is encountered, which subtracts the zero-order effect from the result of the previous operation. Collectively, these composite operations serve to properly (in the sense of the decomposition theorem in Appendix \ref{ale decomposition theorem and properties of L}) subtract from $\mathcal{L}_J(f)$ all lower-order effects when forming $f_{J,ALE}$. Finally, for $J=D$, we define $f_{D,ALE}(\mathbf{x})$ as
\begin{equation}\label{ALE D def}
f_{D,ALE}(\mathbf{x})\equiv \mathcal{H}_D(f)(\mathbf{x})\equiv [I-\sum_{v\subset D} \mathcal{H}_v ](f)(\mathbf{x}).\end{equation}
\par 
For the special cases $J=j$ ($|J|=1$) and $J=\{j,l\}$ $(|J|=2)$, \eqref{ALE J def} reduces to $f_{j,ALE} (x_j)$ and $f_{\{j,l\},ALE}(x_j,x_l)$ from \eqref{ALE centered main def} and \eqref{ALE centered second def}, respectively. That is, for $J=j$,
\begin{equation*} \begin{split}	f_{j,ALE}(x_j)&=[(I-\mathcal{L}_\emptyset )\circ\mathcal{L}_j ](f)(x_j )=\mathcal{L}_j (f)(x_j )-\mathcal{L}_\emptyset\circ\mathcal{L}_j (f)(x_j )\\
		&=\mathcal{L}_j (f)(x_j )-\mathbb{E}[\mathcal{L}_j (f)(X_j )]=g_{j,ALE} (x_j )-\mathbb{E}[g_{j,ALE} (X_j )],		\end{split}\end{equation*}
which is the same as \eqref{ALE centered main def}; and for $J=\{j,l\}$,
\begin{equation*}\begin{split}f_{\{j,l\},ALE} (x_j,x_l )&=[(I-\mathcal{L}_\emptyset )\circ(I-\mathcal{L}_j-\mathcal{L}_l )\circ\mathcal{L}_{\{j,l\}} ](f)(x_j,x_l )\\
	&=(I-\mathcal{L}_\emptyset )\circ[\mathcal{L}_{\{j,l\}}  (f)(x_j,x_l )-\mathcal{L}_j\circ\mathcal{L}_{\{j,l\}}  (f)(x_j )-\mathcal{L}_l\circ\mathcal{L}_{\{j,l\}}  (f)(x_l )]\\
	&=\mathcal{L}_{\{j,l\}}  (f)(x_j,x_l )-\mathcal{L}_j\circ\mathcal{L}_{\{j,l\}}  (f)(x_j )-\mathcal{L}_l\circ\mathcal{L}_{\{j,l\}}  (f)(x_l )\\
	&-\mathbb{E}[\mathcal{L}_{\{j,l\}}  (f)(X_j,X_l )-\mathcal{L}_j\circ\mathcal{L}_{\{j,l\}}  (f)(X_j )-\mathcal{L}_l\circ\mathcal{L}_{\{j,l\}}  (f)(X_l )],	\end{split}\end{equation*}
which is the same as \eqref{ALE centered second def}. 
\section{ALE Decomposition Theorem and Some Properties of  $\mathcal{L}_J$ and $\mathcal{H}_J$}\label{ale decomposition theorem and properties of L}
\par 
We first state some properties of $\mathcal{L}_J$ and $\mathcal{H}_J$, which will be used in the proof of the main result in this appendix. The main result is the ALE decomposition theorem, which states that, in some sense, the ALE plots are estimating the correct quantities. 
\par 
\textbf{Properties of $\mathcal{L}_J$ and $\mathcal{H}_J$}:  For any two sets of indices $u\subseteq D$ and $J\subseteq D$, we have:
\begin{enumerate}[i]
\item $\mathcal{L}_u\circ\mathcal{L}_u=\mathcal{L}_u.$
\item $\mathcal{L}_u\circ\mathcal{L}_J=0$ if $u\not\subseteq J$, i.e., if $u$ contains at least one index that is not in $J$.
\item $\mathcal{L}_J$ is a linear operator, i.e., $\mathcal{L}_J (a_1 g_1+a_2 g_2 )=a_1 \mathcal{L}_J (g_1 )+a_2 \mathcal{L}_J (g_2 )$ for functions $g_1$ and $g_2$ in the domain of $\mathcal{L}_J$ and constants $a_1$ and $a_2$.
\item $\mathcal{L}_u\circ\mathcal{H}_J=0$, for $u\neq J$.
\item $\mathcal{L}_u\circ\mathcal{H}_u=\mathcal{L}_u$.
\item $\mathcal{H}_u\circ\mathcal{H}_J=0$, for $u\neq J$.
\item $\mathcal{H}_u\circ\mathcal{H}_u=\mathcal{H}_u$
\end{enumerate}

\par  
The statement $\mathcal{L}_u\circ\mathcal{L}_J=0$ is an abbreviation for $\mathcal{L}_u\circ\mathcal{L}_J(g)(\mathbf{x})=0$ for all $g$ and for all $\mathbf{x}$, and likewise for similar statements. The preceding properties are mostly obvious by inspection of the definitions of $\mathcal{L}_J$ in \eqref{ALE uncentered J def as limit} and $\mathcal{H}_J$ in \eqref{ALE zero def}—\eqref{ALE D def}. Property (\textit{i}) follows because applying $\mathcal{L}_u$ to a function $g(\mathbf{x}_u)$ that does not depend on $\mathbf{x}_{\backslash u}$ returns the same function $g(\mathbf{x}_u)$ plus lower-order functions of proper subsets of elements of $\mathbf{x}_u$. Hence, $\mathcal{L}_u\circ\mathcal{L}_u (g)(\mathbf{x}_u )=\mathcal{L}_u (g)(\mathbf{x}_u )$ plus lower order functions, but the sum of these lower-order functions must be identically zero because of the boundary conditions that $\mathcal{L}_u\circ\mathcal{L}_u (g)(\mathbf{x}_u )=\mathcal{L}_u (g)(\mathbf{x}_u )=0$ when any element of $\mathbf{x}_u$ (say $x_j$) is at its lower boundary value $x_{\min,j}$ over the integration region in \eqref{ALE uncentered J def as limit}. Properties (\textit{ii}) and (\textit{iii}) are obvious. Regarding Property (\textit{iv}), if $u\neq J$, we must have either $u\not \subseteq J$ or $u\subset J$. Property (\textit{iv}) is obvious for $u\not \subseteq J$, i.e., if $u$ contains at least one index that is not in $J$. For $u\subset J$, Property (\textit{iv}) follows by noting that, when applying $\mathcal{L}_u$ to \eqref{ALE J def} from left to right, $\mathcal{L}_u\circ(I-\mathcal{L}_\emptyset)\circ\ldots \circ(I-\sum_{v\subset J,|v|=|u|-1}\mathcal{L}_v )=\mathcal{L}_u$ (by Properties (\textit{ii}) and (\textit{iii})), so that $\mathcal{L}_u\circ(I-\mathcal{L}_\emptyset)\circ\ldots \circ(I-\sum_{v\subset J,|v|=|u|}\mathcal{L}_v )=\mathcal{L}_u\circ(I-\mathcal{L}_u-\sum_{v\subset J,|v|=|u|,v\neq u}\mathcal{L}_v )=\mathcal{L}_u-\mathcal{L}_u-0=0$ (by Properties (\textit{i}), (\textit{ii}), and (\textit{iii})). Property (\textit{v}) follows similarly. Properties (\textit{vi}) and (\textit{vii}) follow immediately from Properties (\textit{iv}) and (\textit{v}), respectively.

The next theorem follows trivially from the above definitions and properties, although we formally state it for completeness.  
\par  \textbf{ALE Decomposition Theorem:} A function $f(\cdot)$ can be decomposed as\\ $f(\mathbf{x})=\sum_{J\subseteq D}f_{J,ALE} (\mathbf{x}_J )$, where each ALE component function $f_{J,ALE}$ represents the $|J|$-order effect of $\mathbf{X}_J$ on $f(\cdot)$ and is directly constructed via $f_{J,ALE}=\mathcal{H}_J(f)$. Moreover, the ALE component functions have the following orthogonality-like property. For all $J\subseteq D$, we have $\mathcal{H}_J(f_{J,ALE} )=f_{J,ALE}$, and $\mathcal{H}_u(f_{J,ALE} )=0$ for all $u\subseteq D$ with $u\neq J$. That is, for each $J\subseteq D$, the $|J|$-order effect of $\mathbf{X}_J$ on $f_{J,ALE}$ is $f_{J,ALE}$ itself, and all other effects on $f_{J,ALE}$ are identically zero. The ALE decomposition is unique in that for any decomposition $f=\sum_{J\subseteq D}f_J$ with $\{f_J: J\subseteq D\}$ having this orthogonality-like property (i.e., with $\mathcal{H}_J (f_J )=f_J$ and $\mathcal{H}_u (f_J )=0$ for $u\neq J$), it must be the case that $f_J=f_{J,ALE}$. 
\par  
\begin{proof}That $f(\cdot)$ can be decomposed as $f(\mathbf{x})=\sum_{J\subseteq D}f_{J,ALE} (\mathbf{x}_J )$ follows directly from the definition \eqref{ALE D def} of $f_{D,ALE}$. The orthogonality-like property follows directly from Properties (vi) and (vii). The uniqueness of the ALE decomposition follows directly from Property (iii).
\end{proof}
\section{Estimation of $f_{J,ALE} (\mathbf{x}_J)$ for Higher-Order Effects}\label{estimation of higher-order effects}
\par 
Estimation of $f_{J,ALE}$ for $|J|=1$ and $|J|=2$ is described in Section \ref{estimation of main and second}. Here we describe estimation of $f_{J,ALE}$ for general $J\subseteq D\equiv \{1,2,\ldots,d\}$. We compute the estimate $\hat{f}_{J,ALE}$ by computing estimates of the quantities in the composite expression \eqref{ALE J def} from right-to-left. Although the notation necessary to formally define $\hat{f}_{J,ALE}$ for general $J$ is tedious, the concept is straightforward: To estimate $\mathcal{L}_J(f)(\mathbf{x}_J)$, we make the following replacements in \eqref{ALE uncentered J def as limit}. We replace the sequence of $|J|$-dimensional Cartesian product partitions in \eqref{ALE uncentered J def as limit} by the single $|J|$-dimensional Cartesian product of some appropriate fixed partitions of the sample ranges of $\{\mathbf{x}_{i, j}: i = 1,\ldots,n\}$ (for $j \in J$), and we replace the conditional expectation in \eqref{ALE uncentered J def as limit} by the sample average across all $\{\mathbf{x}_{i,\backslash J}:i=1,2,\ldots,n\}$, conditioned on $\mathbf{x}_{i,J}$ falling into
the corresponding cell of the partition. 
\par 
More precisely, for any function $g:\mathbb{R}^d\to \mathbb{R}$, we estimate $\mathcal{L}_J (g)(\mathbf{x}_J )$ for $J\subset D$ via \begin{equation}\label{Lu estimate}
\hat{\mathcal{L}_J}(g)(\mathbf{x}_J) \equiv \sum_{\{\mathbf{k}: 1\leq k_j \leq k_j(x_j), j\in J\}} \frac{1}{n_J(\mathbf{k})}\sum_{\{i: \mathbf{x}_{i,J} \in N_J(\mathbf{k})\}} \Delta_g^{J} (K,\mathbf{k};\mathbf{x}_{i, \backslash J}),
\end{equation}
where the notation is as follows. The $|J|$-order finite difference $\Delta_g^{J} (K,\mathbf{k};\mathbf{x}_{i, \backslash u})$ and the index vector $\mathbf{k}=( k_j: j\in J)$ of cell-$\mathbf{k}$ of the partition grid are the same as in Appendix \ref{def of higher-order effects}. Let $\{N_j(k)=(z_{k-1,j},z_{k,j}]:k=1,2,\ldots,K\}$ denote a partition of the sample range of $\{x_{i,j}: i=1,2,\ldots,n\}$ as in Section \ref{estimation of main and second}. We partition the $|J|$-dimensional range of $\mathbf{x}_J$ into a grid of $K^{|J|}$  rectangular cells obtained as the cross product of the individual one-dimensional partitions. Let $N_J(\mathbf{k})$ denote cell-$\mathbf{k}$ of the grid, i.e., the Cartesian product of the intervals $\{(z_{k_j-1,j},z_{k_j,j}]:j\in J\}$, and let $n_J(\mathbf{k})$ denote the number of training observations that fall into $N_J(\mathbf{k})$, so that the sum of $n_J(\mathbf{k})$ over all $K^{|J|}$  rectangles is $\sum_{\{\mathbf{k}: 1\leq k_j\leq K,j\in J\}} n_J(\mathbf{k})=n$. For each $j \in D$ and each $x$, let $k_j(x)$ denote the index of the $X_j$ partition interval into which $x$ falls, i.e., $x \in N_j(k_j(x))$.  
\par 
The estimator $\hat{f}_{J,ALE}$ is obtained by substituting the estimator \eqref{Lu estimate} for each term of the form $\mathcal{L}_u(g)(\mathbf{x}_u)$ in \eqref{ALE J def}. Eqs. \eqref{ALE centered main est} and \eqref{ALE centered second est} in Section \ref{estimation of main and second} are special cases of $\hat{f}_{J,ALE}$ for $|J|=1$ and $|J|=2$. 
\section{Some Implementation Details}\label{implementation details}
\textbf{Handling categorical predictors.} 
It is often desirable to visualize the effect of any categorical predictor $X_j$ on $f(\cdot)$, and our package \textbf{\texttt{ALEPlot}} includes such functionality. Recall that for the estimation of the ALE effects (for example, in \eqref{ALE uncentered main est} and \eqref{ALE uncentered second est}), we need to take the difference of $f(\cdot)$ across neighboring values of $X_j$. It is therefore important to come up with a reasonable ordering of the levels of $X_j$. The main consideration for a ``reasonable'' ordering is that function extrapolation outside the data envelope should be avoided as much as possible when neighboring values of $x_j$ are plugged into $f(x_j,\mathbf{x}_{\backslash j})$.
\par 
To accomplish this, we order the levels of $X_j$ based on how dissimilar the sample values $\{\mathbf{x}_{i,\backslash j}: i=1,2, \ldots, n\}$ are across the levels of $X_j$. More specifically, we set the number of bins $K$ to the number of nonempty levels of $X_j$. We then calculate a $K\times K$ dissimilarity matrix, the $(k,l)$th component of which accumulates (over the other predictors $X_{j'}$ with $j' \in \{1,2, \ldots, d\}\backslash j$) the distances (e.g., Kolmogorov-Smirnov distance for continuous predictors) between the univariate conditional distributions of $\{x_{i,j'}: i=1,2, \ldots, n; x_{i,j} = \text{level-}k\}$ and $\{x_{i,j'}: i=1,2, \ldots, n; x_{i,j} = \text{level-}l\}$. We then order the levels of $X_j$ according to the ordering of the first coordinate from applying multidimensional scaling (MDS) to the dissimilarity matrix. 
\begin{figure}
\centering
    \includegraphics[trim={0cm 0cm 0cm 0cm},clip,width = \textwidth]{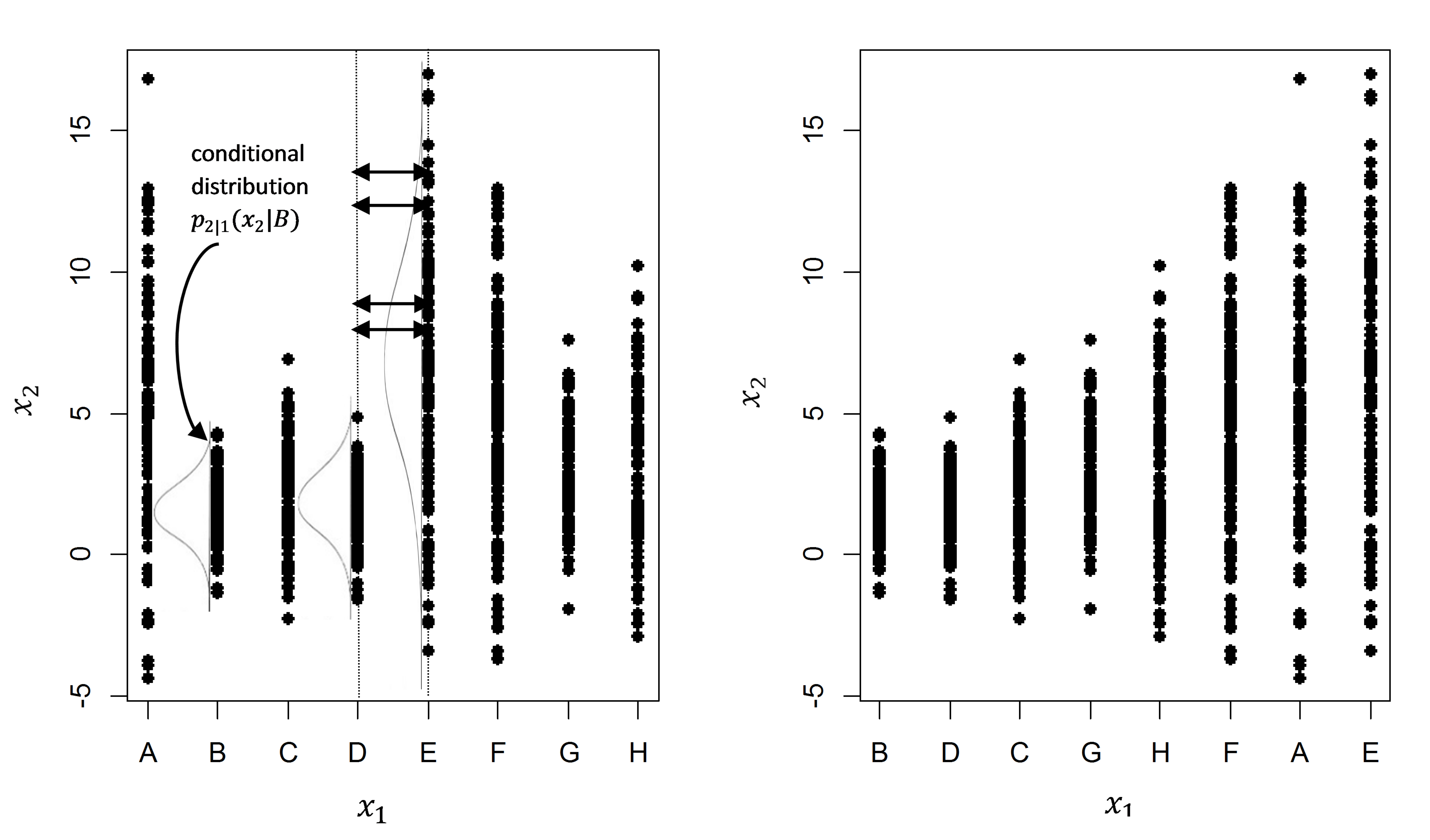}
    \caption{Illustration of the method of ordering the levels of a categorical predictor $X_j$ ($X_1$, having levels ``A'', ``B", $\ldots$, ``H") when calculating its ALE main effect. The left panel is a scatter plot of $\mathbf{X}_{\backslash j}$ ($X_2$, numerical) vs. $X_1$ in the original alphabetical ordering, along with the conditional distributions $p_{2|1}(x_2|x_1)$ for $x_1 = B,D,E$. The right panel is the same but after reordering the levels of $X_1$ using the MDS method. The horizontal black arrows in the left panel indicate where severe extrapolation is required if the levels are not reordered.}  \label{categorical plot}
\end{figure}
\par 
Figure \ref{categorical plot} uses a toy example to illustrate the logic of this ordering and how it helps to avoid function extrapolation. Here, there are $d = 2$ predictors $X_1$ (categorical with levels ``A'', ``B'', $\ldots$, ``H'') and $X_2$ (numerical), and we are interested in the main effect of $X_1$. A scatterplot of $X_2$ vs.\ $X_1$, along with several conditional distributions $p_{2|1}(x_2|x_1)$ for $x_1 = B, D, E$, are shown in the left panel where levels of $X_1$ are alphabetically ordered. If this ordering is kept, we must evaluate $f(x_1=E,x_{i,2}) - f(x_1 = D, x_{i,2})$ for many observations $i$ that require extrapolating $f$ outside the effective range of $p_{2|1}(x_2|x_1=D)$ and $p_{2|1}(x_2|x_1=E)$ (as indicated by the dark horizongal arrows in Figure \ref{categorical plot}). In contrast, the right panel of Figure \ref{categorical plot} shows the same data after ordering the levels of $X_1$ according to MDS as described above. In this case, much less function extrapolation is required when we evaluate $f(x_1,x_{i,2}) - f(x'_1, x_{i,2})$ for $x_1$ and $x'_1$ in neighboring levels after reordering. Our \textbf{\texttt{ALEPlot}} package uses the method of ordering depicted in Figure \ref{categorical plot}, except that for $d>2$ the distance between distributions is summed over all $d-1$ components of $\mathbf{X}_{\backslash j}$.
\par 
\textbf{Vectorization.}  It is important to note that computation of $\hat{f}_{J,ALE}(\mathbf{x}_J)$ is easily vectorizable. In \texttt{R} for example, to produce the $2^{|J|}\times n$ evaluations of $f(\mathbf{x})$ in \eqref{Lu estimate} (or in \eqref{ALE centered main est} or \eqref{ALE centered second est} for ALE main and second-order effects), we can construct a predictor variable array with $2^{|J|}\times n$ rows and d columns and then call the \textbf{\texttt{predict}} command (which is available for most supervised learning models) a single time. This is typically orders of magnitude faster than using a \textbf{\texttt{for}} loop to call the \textbf{\texttt{predict}} command $2^{|J|}\times n$ times. Similarly, the averaging and summation operations in \eqref{Lu estimate}, \eqref{ALE centered main est}, or \eqref{ALE centered second est} can be vectorized without the need for a \textbf{\texttt{for}} loop. Our \texttt{R} package \textbf{\texttt{ALEPlot}} uses this vectorization.
\par
\begin{figure}[!ht]
    \centering
    \includegraphics[trim={0cm 0cm 0cm 0cm},clip,width = \textwidth]{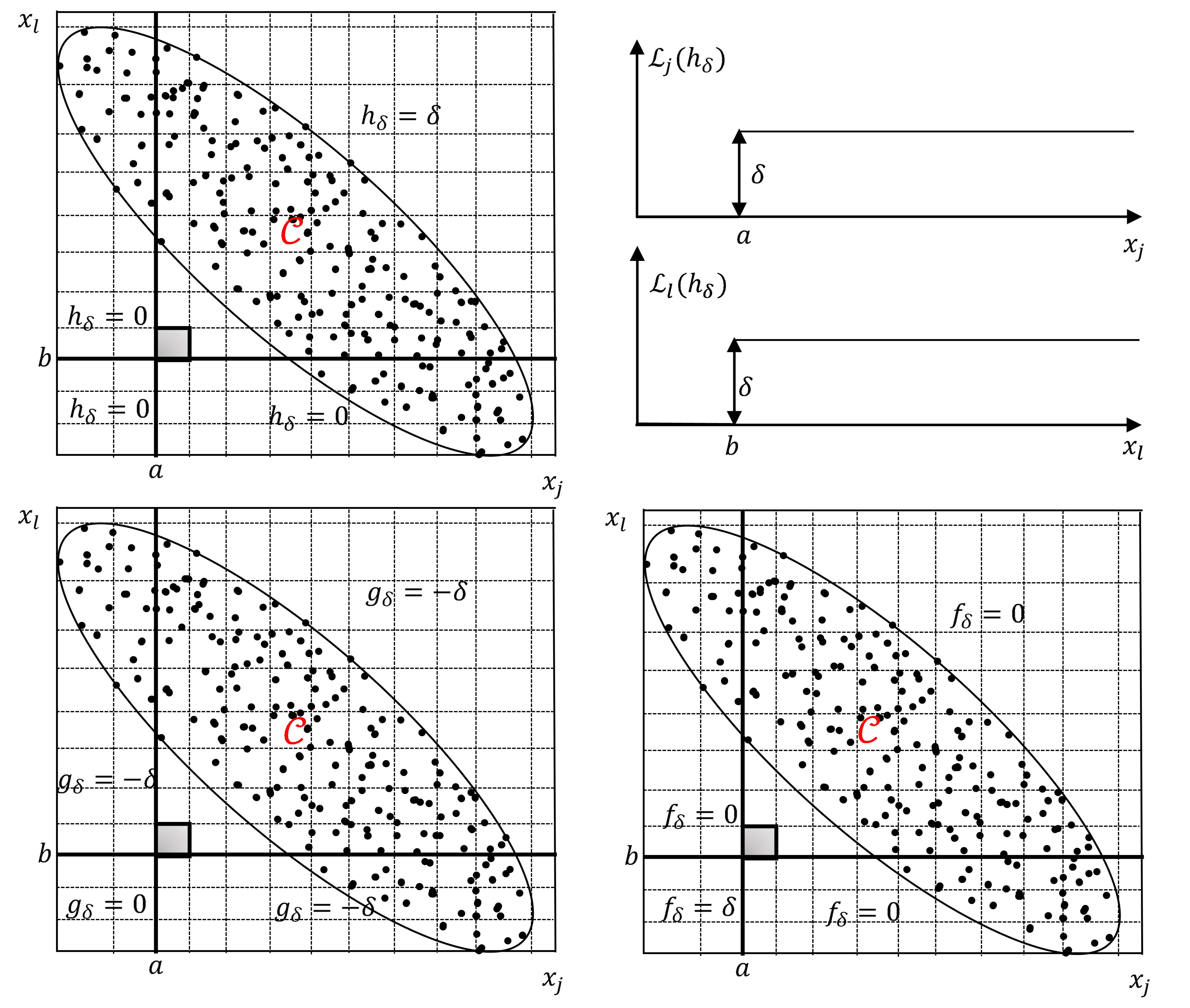}
    \caption{Illustration of the ideas behind the proof of Theorem \ref{Theorem: Interaction2}. All panels except the top-right show sample data for two negatively correlated predictors $(X_j, X_l)$, the approximate convex hull $\mathcal{C}$ (shown as an ellipse) of the bivariate training sample, and the given partition of the space into a 2-D grid. The shaded rectangle with lower left corner at $(a,b)$ is an empty cell (indexed by $(k^*,m^*)$) that lies outside $\mathcal{C}$. The heavy horizontal and vertical lines passing through the point $(a,b)$ divide the grid into four quadrants. The top-left panel shows the values of $h_{\delta}$ in each quadrant. The top-right panel plots $\mathcal{L}_j(h_{\delta})$ and $\mathcal{L}_l(h_{\delta})$ as functions of $x_j$ and $x_l$ respectively. The bottom-left and bottom-right panels show the values of $g_{\delta}$ and $f_{\delta}$, respectively, in each quadrant.}
    \label{fig:2ndch proof}
\end{figure}

 
\textbf{Dealing with empty cells in second-order interaction effect ALE plots.} For a second-order interaction effect ALE plot with $J=\{j,l\}$, we discretize the $(x_j,x_l)$ sample-space into the $K^2$ rectangular cells by taking the cross product of the quantile-based discretizations of the sample spaces of $X_j$ and $X_l$ individually. If $(X_j,X_l)$ are correlated, this will often result in some cells that contain few or no observations, as illustrated in Fig. \ref{fig:2ndch proof}. In this case, it may be useful to add scatter plot bullets to the second-order interaction effect ALE plots to indicate the bivariate training sample values $\{(x_{i,j},x_{i,l}):i=1,2,\ldots,n\}$. This would help to identify empty cells (i.e., cells with $n_{\{j,l\}}(k,m)=0)$ and, more generally, to identify regions of the $(x_j,x_l)$-space in which $f_{J,ALE}(\mathbf{x}_J)$ may not be reliable due to lack of data in that region. This is not necessary for main effect ALE plots $(J=j)$, because our quantile-based discretization always results in the same number $\frac{n}{K}$ of observations in each region. In our \textbf{\texttt{ALEPlot}} package, rather than adding scatter plot bullets on the ALE second-order effect plot, we allow black rectangles to be plotted to indicate empty cells in the chosen grid of partition.
\par 
Denote the average local effect associated with any cell $(k,m)$ (i.e., the summand in \eqref{ALE uncentered second est}) by 
\begin{equation}\label{avg local effect}
\Bar{\Delta}(k,m) \equiv \frac{1}{n_{\{j,l\}}(k,m)} \sum_{\{i: \mathbf{x}_{i,\{j,l\}} \in N_{\{j,l\}}(k,m)\}}\Delta_f^{\{j,l\}}(K, k, m;\mathbf{x}_{i,\backslash\{j,l\}}),
\end{equation} 
where for notational simplicity, we have omitted the dependence of $\Bar{\Delta}(k,m)$ on $\{j,l\}$ and $K$. Note that the quantity in \eqref{avg local effect} is not defined for any cell $(k,m)$ that is empty. On the surface, this may appear to cause a problem when calculating $\hat{h}_{\{j,l\},ALE}(x_j,x_l)$ via \eqref{ALE uncentered second est}, since the outer two summations in \eqref{ALE uncentered second est} are over a rectangular array of cells. If an empty cell is within this array, we need to replace \eqref{avg local effect} by some appropriate value in order to allow the outer two summations in \eqref{ALE uncentered second est} to be calculated. 
\par 
This problem of empty cells is fundamentally different from the extrapolation problem illustrated in Figure \ref{PD and M difference}(a). If $J \subseteq \{1,2,...,d\}$ denotes the predictor indices, the former involves extrapolation in $\mathbf{X}_J$, whereas the latter involves extrapolation in $\mathbf{X}_{\backslash J}$. For the latter, ALE plots themselves are an effective solution, as they are designed to avoid the extrapolation. However, for the former, there is no way around substituting some value for \eqref{avg local effect} for empty cells. For this, we recommend the following strategy, which we use in our \textbf{\texttt{ALEPlot}} package. For any empty cells, we replace \eqref{avg local effect} by the average of the $\Bar{\Delta}(k,m)$ values for the $M$ nearest non-empty cells, weighted by the number of training observations in each of these non-empty cells. We take $M = \min\{10, M_{0.1}\}$, where $M_{0.1}$ is the smallest number of nearest non-empty cells that together contain at least 10 percent of the training observations. 
\par
The $\Bar{\Delta}(k,m)$ value that we substitute for empty cells is somewhat arbitrary, and substituting a different value will alter $\hat{f}_{\{j,l\},ALE}(x_j,x_l)$ in general. However, for empty cells that lie outside the convex hull of the bivariate training values $\{(x_{i,j},x_{i,l}), i = 1, \ldots, n\}$ (which is where empty cells are more likely to occur, as illustrated in Fig. \ref{fig:2ndch proof}), the following theorem implies that the $\Bar{\Delta}(k,m)$ values that we choose do not alter $\hat{f}_{\{j,l\},ALE}(x_j,x_l)$ where it matters, which is in cells that are not empty.

\begin{thm}\label{Theorem: Interaction2}
Consider a supervised learning model $f(x_1,x_2, \ldots, x_d)$ and two predictors of interest $X_j$ and $X_l$. Let $\mathcal{C}$ denote the convex hull of the bivariate training values $\{(x_{i,j}, x_{i,l}): i = 1, \ldots, n\}$ and $(k^*, m^*)$ denote the index of an empty cell outside $\mathcal{C}$ in a given partition of the $(X_j, X_l)$ sample space into a 2-D grid (see Fig. \ref{fig:2ndch proof}). Suppose that we substitute some value $\Bar{\Delta}(k^*,m^*)$ for the quantity in \eqref{avg local effect} for the empty cell when calculating the functions $\hat{h}_{\{j,l\}, ALE}$, $\hat{g}_{\{j,l\}, ALE}$, and $\hat{f}_{\{j,l\}, ALE}$ via \eqref{ALE uncentered second est}, \eqref{ALE second centering 1 est}, and \eqref{ALE centered second est}. We claim that the value chosen for $\Bar{\Delta}(k^*,m^*)$ is arbitrary and does not affect  $\hat{f}_{\{j,l\}, ALE}$ inside $\mathcal{C}$ (i.e., in the region for which we have data), in the sense that changing $\Bar{\Delta}(k^*,m^*)$ to $\Bar{\Delta}(k^*,m^*) + \delta$ for any constant $\delta$ has no effect on $\hat{f}_{\{j,l\}, ALE}(x_j,x_l)$ for $(x_j, x_l) \in \mathcal{C}$.
\end{thm}
\begin{proof} To avoid obscuring the simple logic behind this claim with tedious arguments and cumbersome notation, we present only a sketch of proof and only for the case in which $X_j$ and $X_l$ are negatively correlated and the sample data are as shown in Figure \ref{fig:2ndch proof}. Denote the lower-left corner of empty cell $(k^*,m^*)$ by $(a,b)$, as depicted in Figure \ref{fig:2ndch proof}. Let $\hat{h}_{\{j,l\}, ALE}$, $\hat{g}_{\{j,l\}, ALE}$, and $\hat{f}_{\{j,l\}, ALE}$ denote the functions computed in \eqref{ALE uncentered second est}, \eqref{ALE second centering 1 est}, and \eqref{ALE centered second est} when we use the value $\Bar{\Delta}(k^*,m^*)$ for \eqref{avg local effect} for the empty cell, let $\widetilde{h}_{\{j,l\}, ALE}$, $\widetilde{g}_{\{j,l\}, ALE}$, and $\widetilde{f}_{\{j,l\}, ALE}$ denote the corresponding functions computed when we use the value $\Bar{\Delta}(k^*,m^*) + \delta$, instead of $\Bar{\Delta}(k^*,m^*)$, and let $h_{\delta} = \widetilde{h}_{\{j,l\}, ALE}-\hat{h}_{\{j,l\}, ALE}$, $g_{\delta} = \widetilde{g}_{\{j,l\}, ALE}-\hat{g}_{\{j,l\}, ALE}$, and $f_{\delta} = \widetilde{f}_{\{j,l\}, ALE}-\hat{f}_{\{j,l\}, ALE}$ denote their differences.  

The goal is to show that $f_{\delta}(x_j,x_l)=0$ for $(x_j, x_l) \in \mathcal{C}$. Beginning with the uncentered ALE effect functions $\widetilde{h}_{\{j,l\}, ALE}$ and $\hat{h}_{\{j,l\}, ALE}$, adding $\delta$ for cell $(k^*,m^*)$ will only affect the accumulation of local effects in \eqref{ALE uncentered second est} in cells that are above and to the right of $(k^*,m^*)$. Hence
\begin{equation}\label{thm-interaction eq h}
    h_{\delta}(x_j,x_l)= 
\begin{cases}
    \delta,& \text{if } x_j> a, x_l >b\\
    0, & \text{otherwise}
\end{cases} = 0 + \delta \cdot\mathbb{I}(x_j > a, x_l >b),
\end{equation}
which is shown in the top-left panel of Figure \ref{fig:2ndch proof}. 

To relate $\widetilde{g}_{\{j,l\}, ALE}$ and $\hat{g}_{\{j,l\}, ALE}$, using the operator notation defined in Appendix \ref{estimation of higher-order effects}, \eqref{ALE second centering 1 est} becomes
\begin{align}\label{thm-interaction eq 1}
\begin{split}
\widetilde{g}_{\{j,l\},ALE} &\equiv \widetilde{h}_{\{j,l\},ALE} -\hat{\mathcal{L}}_j(\widetilde{h}_{\{j,l\},ALE}) -   \hat{\mathcal{L}}_l(\widetilde{h}_{\{j,l\},ALE})\\
& = \left[\hat{h}_{\{j,l\},ALE} - \hat{\mathcal{L}}_j(\hat{h}_{\{j,l\},ALE}) - \hat{\mathcal{L}}_l(\hat{h}_{\{j,l\},ALE})\right]+ \left[h_{\delta} -\hat{\mathcal{L}}_j(h_{\delta}) -   \hat{\mathcal{L}}_l(h_{\delta})\right]\\
& = \hat{g}_{\{j,l\},ALE} + \left[h_{\delta} -\hat{\mathcal{L}}_j(h_{\delta}) -   \hat{\mathcal{L}}_l(h_{\delta})\right]\\
& = \hat{g}_{\{j,l\},ALE} + g_{\delta}.
\end{split}
\end{align}
From \eqref{thm-interaction eq h}, the uncentered main effects of $h_{\delta}$ are $\hat{\mathcal{L}}_j(h_{\delta})(x_j) = 0 + \delta \cdot\mathbb{I}(x_j > a)$ and $\hat{\mathcal{L}}_l(h_{\delta})(x_l) = 0 + \delta \cdot\mathbb{I}(x_l > b)$, which are the step functions shown in the top-right panel of Figure \ref{fig:2ndch proof}. Using this in \eqref{thm-interaction eq 1} gives 
\begin{equation}\label{thm-interaction eq 2}
\qquad g_{\delta}(x_j,x_l) = h_{\delta}(x_j,x_l)-\hat{\mathcal{L}}_j(h_{\delta})(x_j,x_l)-\hat{\mathcal{L}}_l(h_{\delta})(x_j,x_l)= -\delta \cdot\mathbb{I}(x_j > a \text{ or } x_l > b),
\end{equation}
which is plotted in the bottom-left panel of Figure \ref{fig:2ndch proof}. 

Finally, using \eqref{thm-interaction eq 2}, the centered second-order ALE effect from \eqref{ALE centered second est} becomes
\begin{align*}
 \widetilde{f}_{\{j,l\},ALE}(x_j,x_l)&\equiv \widetilde{g}_{\{j,l\},ALE}(x_j,x_l) - \hat{\mathcal{L}}_{\emptyset}\left(\widetilde{g}_{\{j,l\},ALE}\right)(x_j,x_l) \\
&=\hat{g}_{\{j,l\},ALE}(x_j,x_l) + g_{\delta}(x_j, x_l) - \hat{\mathcal{L}}_{\emptyset}\left(\hat{g}_{\{j,l\},ALE} + g_{\delta}\right)(x_j,x_l) \\
&= \hat{g}_{\{j,l\},ALE}(x_j,x_l) - \hat{\mathcal{L}}_{\emptyset}\left(\hat{g}_{\{j,l\},ALE}\right)(x_j,x_l) + g_{\delta}(x_j,x_l) - \hat{\mathcal{L}}_{\emptyset}\left(g_{\delta}\right)(x_j,x_l) \\
& = \hat{f}_{\{j,l\},ALE}(x_j,x_l) -\delta \cdot\mathbb{I}(x_j > a \text{ or } x_l > b) - ( -\delta)\\
& = \hat{f}_{\{j,l\},ALE}(x_j,x_l) + \delta \cdot \mathbb{I}(x_j \leq a, x_l \leq b)\\
& = \hat{f}_{\{j,l\},ALE}(x_j,x_l) + f_{\delta}(x_j,x_l).
\end{align*}

Hence, $f_{\delta}(x_j,x_l) = \delta \cdot \mathbb{I}(x_j \leq a, x_l \leq b) = 0$ for $(x_j,x_l) \in \mathcal{C}$, which proves the claim.
\end{proof}

\end{appendices}

\end{document}